\documentclass[aps,superscriptaddress,floatfix,bm,braket,notitlepage,onecolumn, eqsecnum]{revtex4-1}
\usepackage{amsfonts}
\usepackage{amssymb}
\usepackage{amsmath}
\usepackage{times}
\usepackage{graphicx}
\usepackage{color}
\usepackage{bm}
\usepackage{braket}
\usepackage{mathtools}
\usepackage{mathcomp}
\usepackage[caption=false,justification=raggedright,position=top,singlelinecheck=off]{subfig}
\usepackage{soul}
\usepackage{tabularx}
\usepackage{calrsfs}
\usepackage[vcentermath]{youngtab}

\newcolumntype{L}[1]{>{\raggedright\let\newline\\\arraybackslash\hspace{0pt}}m{#1}}
\newcolumntype{C}[1]{>{\centering\let\newline\\\arraybackslash\hspace{0pt}}m{#1}}
\newcolumntype{R}[1]{>{\raggedleft\let\newline\\\arraybackslash\hspace{0pt}}m{#1}}

\newcommand{\enbe}{\begin{equation}}
\newcommand{\enee}{\end{equation}}
\newcommand{\enba}{\begin{align}}
\newcommand{\enea}{\end{align}}

\begin{document}

\title{Kondo effect due to a hydrogen impurity in graphene: a multichannel Kondo
problem with diverging hybridization}
\author{Zheng Shi}
\affiliation{Dahlem Center for Complex Quantum Systems and Physics Department, Freie
Universit\"{a}t Berlin, Arnimallee 14, 14195 Berlin, Germany}
\author{Emilian M. Nica}
\affiliation{Department of Physics, Arizona State University, Box 871504, Tempe, Arizona
85287-1504, USA}
\author{Ian Affleck}
\affiliation{Department of Physics and Astronomy and Stewart Blusson Quantum Matter
Institute, University of British Columbia, Vancouver, B.C., Canada, V6T1Z1}

\begin{abstract}
We consider the Kondo effect arising from a hydrogen impurity in graphene.
As a first approximation, the strong covalent bond to a carbon atom removes that carbon
atom without breaking the $C_{3}$ rotation symmetry, and we only retain the Hubbard interaction on the three nearest neighbors of the removed carbon atom which then behave as magnetic impurities. These three impurity spins are coupled to three conduction channels with definite helicity, two of which support a diverging local density of states (LDOS) $\propto 1/%
\left[ \left\vert \omega \right\vert \ln ^{2}\left( \Lambda /\left\vert \omega \right\vert \right) \right] $ near the Dirac point $\omega \rightarrow 0$ even though the bulk density of states vanishes linearly. We study the resulting 3-impurity multi-channel Kondo model using the numerical renormalization group method. For weak potential scattering, the ground state of the Kondo model is a particle-hole symmetric spin-$1/2$ doublet, with ferromagnetic coupling between the three impurity spins; for moderate potential scattering, the ground state becomes a particle-hole asymmetric spin singlet, with antiferromagnetic coupling between the three impurity spins. This behavior is inherited by the Anderson model containing the hydrogen impurity and all four carbon atoms in its vicinity.
\end{abstract}

\date{\today }
\maketitle

\section{Introduction}

Interest in graphene magnetism and its potential applications in spintronics started to grow soon after the isolation and characterization of this two-dimensional material\cite{Science.306.666,Nature.438.197,RevModPhys.81.109,RepProgPhys.73.056501,Nature.448.571,PhysRevLett.100.047209,PhysRevB.81.165409,PhysRevB.97.195425,NatNanotechnol.9.794}. One particularly intensively explored approach to making graphene magnetic is through point defects\cite{PhysRevB.75.125408,JPSJ.76.064713,PhysRevLett.101.037203,JPhysCondensMatter.21.196002,PhysRevB.83.241408,JPhysCondensMatter.27.156001,PhysRevB.91.035132}, such as adsorbing adatoms\cite{ApplPhysLett.93.082504,PhysRevB.84.245446,PhysRevB.87.174435,PhysRevLett.101.026805,PhysRevB.81.125433,NewJPhys.12.053012,Nanotechnology.21.505202,PhysRevB.89.115428} and vacancies\cite{PhysRevB.77.195428,PhysicaE.41.80,PhysRevB.86.165438,Nanoscale.6.8814,PhysRevB.90.014401,PhysRevB.90.201406,PhysRevLett.117.166801,PhysRevB.96.125431,JPhysChemC.121.8653,PhysRevB.99.075102}. While transition metal adatoms with $d$ or $f$ electrons constitute an obvious option, rather amazingly, it has been shown both theoretically and experimentally that hydrogen impurities in graphene are also capable of inducing local magnetic moments\cite{PhysRevB.77.035427,PhysRevB.78.085417,PhysRevLett.109.186604,PhysRevB.85.115405,2DMater.2.022002,Science.352.437,PhysRevB.95.060408,PhysRevB.96.024403}, of the order of one Bohr magneton per defect. An intuitive explanation follows from the strong coupling between the hydrogen impurity and the carbon atom directly below it\cite{PhysRevLett.101.196803,PhysRevLett.105.056802}. In the limit of this coupling going to infinity, the large energy cost of transferring an electron from or to the hydrogen-carbon pair effectively removes the $p_{z}$ orbital of the carbon atom from the graphene sheet. If we approximate graphene as a Hubbard model defined on the bipartite honeycomb lattice, then Lieb's theorem predicts that the total spin of the ground state should be $1/2$ after the removal of the $p_{z}$ orbital\cite{PhysRevLett.62.1201}.

In metals with dilute magnetic impurities, conduction electrons screen the impurity magnetic moments at low temperatures, forming many-body singlets in the famous Kondo effect\cite{ProgTheorPhys.32.37,hewson1997kondo}. Now understood in great detail, the Kondo effect is frequently employed in various unconventional materials in order to locally probe the bulk properties of the conducting host. From a theoretical perspective, graphene is predicted to support many exotic variants of the Kondo effect\cite{[][{, and references therein.}]RepProgPhys.76.032501}, thanks to the Dirac cones in its electronic structure and a diversity of possible impurity locations in the unit cell\cite{JStatMech.2010.P01007,EurophysLett.90.67001}.

On the experimental side, it has been found early on that irradiation induced carbon vacancies in single-layer graphene produce a resistivity minimum versus temperature\cite{NatPhys.7.535}, which is consistent with a high Kondo temperature ($\sim 70$K) if attributed to Kondo effect. However, subsequent magnetization measurements in irradiated thick graphite laminates suggest paramagnetism down to the lowest accessed temperatures\cite{NatPhys.8.199}. While alternative explanations of the resistivity minimum based on weak localization or electron-electron interactions have been proposed\cite{PhysRevB.88.155412,CurrApplPhys.17.474,PhysicaE.100.40}, the apparent contradiction between resistivity and magnetization measurements is eventually resolved by scanning tunneling spectroscopy\cite{NatCommun.9.2349,PhysRevB.97.155419}. The Kondo screening of vacancy-induced moments generally takes place when the graphene layer is not locally perfectly flat, and the Kondo temperature depends sensitively on the local curvature. In corrugated graphene samples, vacancies come with different local curvatures and are subjected to varying degrees of screening, but the distribution of Kondo temperatures is not fully captured by either resistivity or magnetization measurements, with the former probing screened moments and the latter probing unscreened ones.

Unlike a carbon vacancy which is subject to the Jahn-Teller distortion, a hydrogen impurity preserves the $C_{3}$ rotational symmetry of the graphene lattice around the carbon atom directly below it (henceforth referred to as the ``central site''). The induced magnetic moment predominantly resides on the carbon sublattice where the impurity is not adsorbed. Despite extending many lattice constants, the magnetization is the strongest on the three nearest neighbors of the central site\cite{PhysRevB.78.085417,PhysRevB.83.241408,Science.352.437,PhysRevB.85.115405,PhysRevB.96.024403}. This has motivated Ref.~\onlinecite{PhysRevB.85.115405} to examine, among other models, a reduced Hamiltonian where the Hubbard interactions are taken into account only on the 5-atom cluster including the hydrogen impurity, the central site and its three nearest neighbors. To study the Kondo effect, Ref.~\onlinecite{PhysRevB.85.115405} replaces the rest of the system by a non-interacting bath of Dirac electrons; the cluster hybridizes with the bath via a local density of states (LDOS) that vanishes linearly as a function of energy near the Dirac points, as is characteristic of the density of states in bulk graphene\cite{RevModPhys.81.109}.

Upon closer inspection, however, one realizes that the system with the 5-atom cluster removed cannot be equivalent to bulk graphene, but rather comes with a 4-site vacancy: one site is removed from one of the sublattices and three sites are removed from the other. In the nearest-neighbor tight-binding model, it is known that a sublattice site number imbalance produces the same number of zero energy states (or ``zero modes'') dwelling exclusively on the sublattice with more sites\cite{PhysRevB.77.115109}. In the simpler case of a single-site vacancy, the single zero mode produced by the vacancy cannot be normalized, because its wave function decays as $1/r$ away from the vacancy\cite{PhysRevLett.96.036801,PhysRevB.79.155442,NewJPhys.14.083004,PhysRevB.88.075413}. This is intimately related to a strongly enhanced LDOS around the vacancy\cite{PhysRevB.77.115109}, which diverges as $1/\left[ \left\vert \omega \right\vert \ln ^{2}\left( \Lambda /\left\vert \omega \right\vert \right) \right] $ ($\Lambda$ is a high-energy cutoff) near the Dirac point $\omega \rightarrow 0$. (We refer to this as a ``logarithmic divergence'' in the following.) When an impurity magnetic moment located at the vacancy is coupled to the rest of the graphene sheet, the divergent LDOS has a profound impact on the ensuing single-channel Kondo effect at half filling\cite{PhysRevB.88.075104,2012arXiv1207.3135C,PhysRevB.90.195109,JPhysConfSer.603.012013,PhysRevB.95.165412}: in stark contrast to the linear-LDOS case\cite{PhysRevLett.64.1835,PhysRevB.53.15079,PhysRevB.57.14254,PhysRevB.70.214427,JPhysCommun.2.085001}, both potential scattering and Kondo scattering perturbations become strongly relevant in the renormalization group (RG) sense, leading to a high Kondo temperature, and the low-energy behavior of the system is always controlled by a strong-coupling fixed point. Similar impurity-related LDOS enhancement mechanisms and their effects on Kondo screening have been discussed in the context of $d$-wave superconductors\cite{PhysRevB.51.15547,PhysRevB.63.020506,*PhysRevB.64.060501}.

In this paper, we apply the above considerations to the hydrogen impurity problem under the 5-atom cluster approximation. The non-interacting bath with the 4-site vacancy allows $3-1=2$ zero modes, both of which are non-normalizable. Correspondingly, we show that two conduction channels with a diverging hybridization appear in the Kondo problem. There is a third conduction channel with a vanishing LDOS at low energies, but the perturbations associated with it are strongly irrelevant; its importance is therefore diminished by the other two channels.

As a first approximation, we consider the limit of infinite coupling between the hydrogen impurity and the central site. The central site is essentially eliminated from the low-energy theory in this limit. Consequently, only the three nearest neighbors of the central site are left to host the magnetic moment, mapping to a 3-impurity Kondo problem where the impurities are symmetric under $Z_{3}$ permutations. In a metallic host, this problem is known to yield a rich phase diagram\cite{cond-mat/9607190, PhysRevLett.95.257204}, and many insights are carried over to our case of a diverging hybridization.

To tackle the 3-impurity problem, we first construct an auxiliary problem: a single impurity spin of arbitrary size $S$ coupled to two conduction channels through the same diverging hybridization. This is analyzed with the aid of the numerical renormalization group (NRG) algorithm\cite{RevModPhys.47.773,RevModPhys.80.395,NRGLjubljana}. Systematic studies have been performed on the pseudogap case where the hybridization vanishes at zero energy\cite{PhysRevB.84.125139}. However, as in the single-channel case, the diverging hybridization makes a qualitative difference. We find that all low-energy fixed points are strong-coupling and Fermi-liquid like; in particular, in the presence of particle-hole (p-h) symmetry, the low-energy fixed point involves the impurity spin screened by conduction electrons from both channels, forming a residual spin of size $\left\vert S-1\right\vert $ for any $S$, including $S=1/2$ together with two phase-shifted conduction channels. This is very different from the 2-channel spin-$1/2$ Kondo problem with a constant hybridization, whose ground state is a non-Fermi liquid\cite{PhysRevB.48.7297,PhysRevLett.67.161,PhysRevLett.52.364,JStatPhys.38.125}.

Returning to the 3-impurity Kondo model, we map out the phase diagram by NRG and study the thermodynamics and the impurity spin correlations in each phase. The most important coupling constants are the relevant ones associated with potential and Kondo scattering in the two conduction channels with a diverging hybridization. For weak and intermediate potential scattering, we find two stable low-energy fixed points: a p-h symmetric spin-$1/2$ fixed point with the Kondo effect taking place in the spin sector which we label as K-S, and a p-h asymmetric spin-singlet fixed point which we label as AF-ASC. The latter is connected to an unstable fixed point of the Kondo effect taking place in the isospin sector. K-S is favored by weak potential scattering at the impurities and ferromagnetic Ruderman-Kittel-Kasuya-Yosida (RKKY) coupling between the impurities, and exhibits ferromagnetic impurity spin correlations, whereas AF-ASC exhibits antiferromagnetic impurity spin correlations; the transition between the K-S phase and the AF-ASC phase is shown to be a simple level crossing. On the other hand, we find that very strong potential scattering can overwhelm the Kondo scattering and suppress the Kondo effect, in which case the impurity spins couple to form a magnetic moment decoupled from the conduction electrons. Finally, the divergence of the hybridization is inevitably cut off at low energies in more realistic models of graphene\cite{PhysRevB.77.115109}, and we examine the consequences of such a cutoff on a phenomenological level.

We then proceed to analyze with NRG the Anderson model with Hubbard interactions on the 5-atom cluster. The system flows to the p-h symmetric spin-$1/2$ K-S fixed point when p-h symmetry breaking perturbation is weak, and to the p-h asymmetric spin-singlet AF-ASC fixed point otherwise; the two phases are again respectively characterized by ferromagnetic and antiferromagnetic impurity spin correlations. We conclude that the Kondo effect occurs at the hydrogen impurity both in the 3-impurity Kondo model and the 5-atom cluster Anderson model.

The rest of this paper is organized as follows. In Sec.~\ref{sec:model}, we introduce the 3-impurity Kondo model of a hydrogen impurity in the infinite hydrogen-carbon coupling limit, highlighting the diverging hybridization between the magnetic impurities and the two conduction channels. Sec.~\ref{sec:Kondo} is devoted to the 3-impurity Kondo model: we give the scaling behavior of various perturbations and the RKKY interactions in the weak-coupling limit in Sec.~\ref{sec:weak}, then discuss the NRG results on the 2-channel spin-$S$ Kondo model with a diverging hybridization in Sec.~\ref{sec:S2CK}. The numerical results on the 3-impurity Kondo model with a diverging hybridization are analyzed in depth in Sec.~\ref{sec:3I2CK}. Sec.~\ref{sec:LDOScutoff} closes our discussion of the 3-impurity Kondo model by demonstrating the effects of a low-energy cutoff on the divergent LDOS. In Sec.~\ref{sec:Anderson} we interpret our numerical results for the 5-atom cluster Anderson model with a diverging hybridization. Sec.~\ref{sec:conclusions} concludes the paper and discusses some open problems. Appendix~\ref{sec:app4vscatbas} contains a derivation of the divergent LDOS from the 4-site vacancy lattice model, and the corresponding zero mode solutions of this model are discussed in Appendix~\ref{sec:appzero}. Finally, in Appendix~\ref{sec:appRKKY}, we calculate the RKKY interaction in the 3-impurity Kondo model to the second order in Kondo couplings. The abbreviations used in this paper are summarized in Table~\ref{listabbr}.

\begin{table}[htb]
\caption{Abbreviations used in this paper.} \label{listabbr} 
\begin{tabular}{c p{6cm}}
\hline\hline \\[-1em]
\text{Abbreviation} & \text{Meaning}\\ \\[-1em]
\hline \\[-0.5em]
    LDOS & local density of states\\  \\[-0.5em]  
    RG & renormalization group\\  \\[-0.5em]  
    NRG & numerical renormalization group\\  \\[-0.5em]  
    p-h & particle-hole\\  \\[-0.5em]  
    RKKY & Ruderman-Kittel-Kasuya-Yosida\\  \\[-0.5em]  

    ALM & p-h asymmetric local moment fixed point of the 2-channel spin-$S$ Kondo problem with a logarithmically divergent LDOS\\  \\[-0.5em]  
    SSC & p-h symmetric strong-coupling fixed point of the 2-channel spin-$S$ Kondo problem with a logarithmically divergent LDOS\\  \\[-0.5em]  
    LM & free-spin p-h symmetric local moment fixed point\\  \\[-0.5em]  
    free-ALM & free-spin p-h asymmetric local moment fixed point\\  \\[-0.5em]  
    K-S & ferromagnetic p-h symmetric Kondo fixed point\\  \\[-0.5em]  
    F-ALM & ferromagnetic p-h asymmetric local moment fixed point\\  \\[-0.5em]  
    K-I & antiferromagnetic p-h symmetric isospin Kondo fixed point\\  \\[-0.5em]  
    AF-ASC & antiferromagnetic p-h asymmetric strong-coupling fixed point\\  \\[-0.5em]  
    AF-ALM & antiferromagnetic p-h asymmetric local moment fixed point\\  \\[-0.5em]  
    F-ASC & ferromagnetic p-h asymmetric strong-coupling fixed point\\  \\[-0.5em]  
    
\hline\hline
\end{tabular}
\end{table}

\section{Model\label{sec:model}}

We consider the nearest-neighbor tight-binding model of a graphene layer defined on a bipartite honeycomb lattice with two sublattices A and B. The ``central site'', to which the hydrogen impurity is coupled, is assumed to be on the A sublattice. Throughout the paper, we follow Ref.~\onlinecite{PhysRevB.85.115405} and only retain the local Hubbard interactions on the 5-atom cluster, composed of the hydrogen impurity, the central A site and its three nearest-neighbor B sites. To highlight the Kondo physics, in this section we furthermore approximate the hydrogen-carbon coupling as infinity, which leaves both the hydrogen and the central site decoupled from the rest of the system. In the limit of strong Hubbard interactions, we effectively localize the electrons on the three nearest-neighboring B sites. The effect of the hopping from the 3 B sites to the 6 outer A sites can be considered to leading order in perturbation theory. The associated Anderson model corresponds to 3 interacting impurities hybridizing with a non-interacting bath through the outer A electrons:

\begin{equation}
H=H_{\text{vac}}+H_{\text{hyb}}+H_{\text{imp}}.  \label{Anderson0}
\end{equation}

\begin{figure}[!h]
\includegraphics[width=0.5\columnwidth]{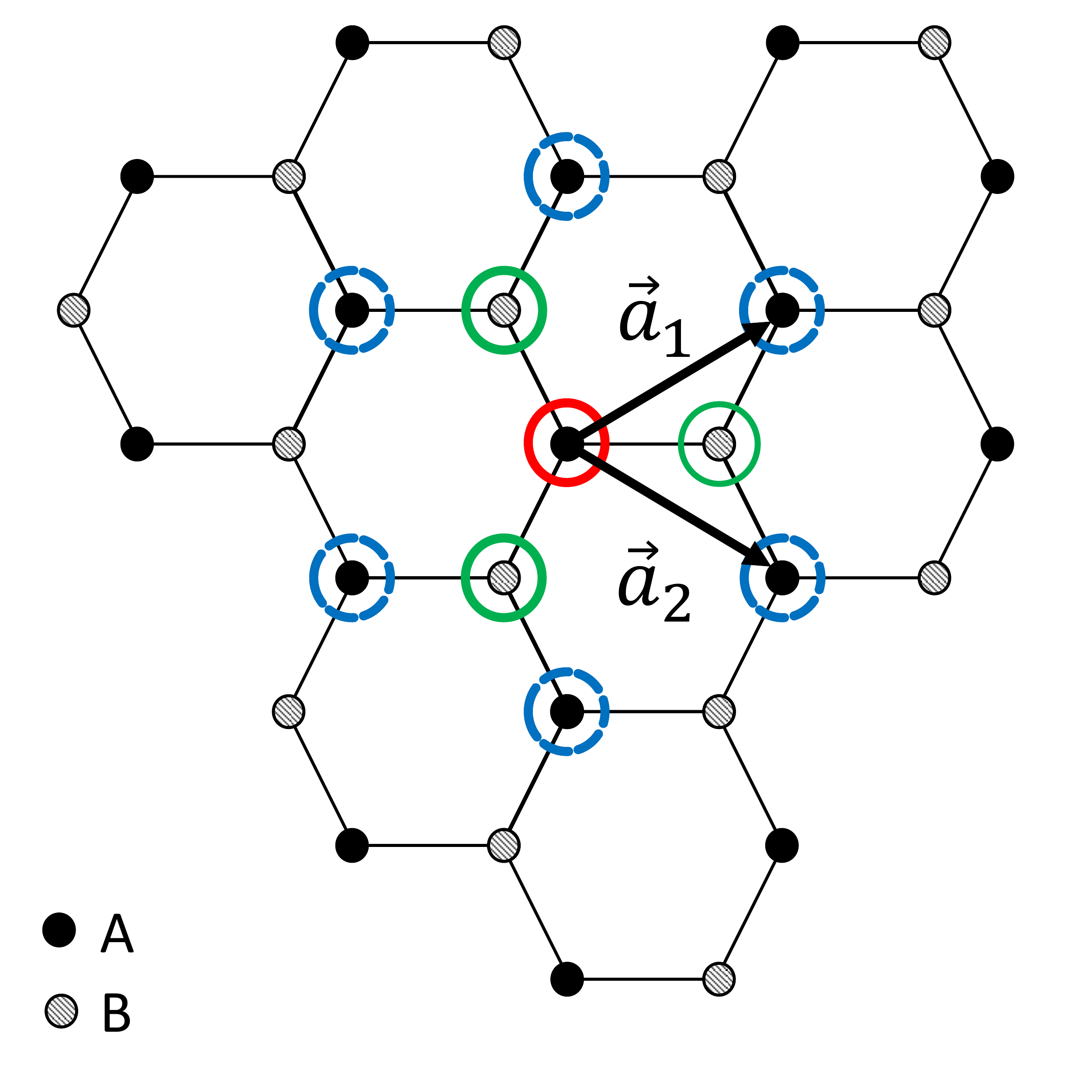} 
\caption{Schematic representation of the effective model for a hydrogen impurity in graphene. The central site on the A sublattice, which is directly coupled to the hydrogen impurity, is marked by a solid red circle. Its three nearest neighbors on the B sublattice are marked by solid green circles. The repulsive Hubbard interactions only reside on these four sites and the hydrogen impurity. The outer 6 next-nearest neighbors on the A sublattice are marked by dashed blue circles. Each pair of these hybridizes with one of the nearest-neighboring (green) B sites. When the hydrogen-carbon coupling goes to infinity, the central (red) A site is removed from the model along with the hydrogen impurity itself, and the nearest-neighboring (green) B sites become the effective impurities.} \label{modelsketch}
\end{figure}

The non-interacting bath part of the effective 3-impurity Hamiltonian describes a graphene sheet with a 4-site vacancy

\begin{equation}
H_{\text{vac}}=-t\sum\nolimits_{\vec{R}}^{\prime }\left\{ \left[ b^{\dag
}\left( \vec{R}\right) +b^{\dag }\left( \vec{R}-\vec{a}_{2}\right) +b^{\dag
}\left( \vec{R}-\vec{a}_{1}\right) \right] a\left( \vec{R}\right) +\text{h.c.%
}\right\} ,  \label{bathH}
\end{equation}%
where the summation excludes the central A site and its three nearest
neighbor B sites, as shown in Fig.~\ref{modelsketch}. Here, $\vec{R}=n\vec{a}_{1}+m\vec{a}_{2}$ are the Bravais lattice vectors with

\begin{equation}
\vec{a}_{1} =\frac{a}{2}\left( \sqrt{3},1\right), \vec{a}_{2} =\frac{a}{2}\left( \sqrt{3},-1\right) ,
\end{equation}%
where $a$ is the lattice constant, and the B site labeled by $\vec{R}$ is
displaced by $(a/\sqrt{3},0)$ from the corresponding A site. The
four sites removed from $H_{\text{vac}}$ are then labeled by $a(\vec{0})$, $%
b(\vec{0})$, $b\left( -\vec{a}_{1}\right) $ and $b\left( -\vec{a}_{2}\right) 
$.

The impurity-bath hybridization, given by

\begin{equation}
H_{\text{hyb}}=-\sqrt{2}t\sum_{j=1}^{3}a_{j}^{\dag }b_{j}+\text{h.c.},
\label{hybH}
\end{equation}%
is invariant under $C_{3}$ rotations. Here we have relabeled the 3
impurity B sites as

\begin{equation}
b(\vec{0})=b_{1},b\left( -\vec{a}_{2}\right) =b_{2},b\left( -\vec{a}%
_{1}\right) =b_{3},
\end{equation}%
and defined the symmetric linear combinations of pairs of the neighboring $%
a^{\dagger }(\vec{R})$ electrons as

\begin{subequations}
\begin{align}
a_{1}^{\dagger }& =\frac{1}{\sqrt{2}}\left[ a^{\dagger }\left( \vec{a}%
_{1}\right) +a^{\dagger }\left( \vec{a}_{2}\right) \right] , \\
a_{2}^{\dagger }& =\frac{1}{\sqrt{2}}\left[ a^{\dagger }\left( -\vec{a}%
_{2}\right) +a^{\dagger }\left( \vec{a}_{1}-\vec{a}_{2}\right) \right] , \\
a_{3}^{\dagger }& =\frac{1}{\sqrt{2}}\left[ a^{\dagger }\left( -\vec{a}%
_{1}\right) +a^{\dagger }\left( -\vec{a}_{1}+\vec{a}_{2}\right) \right] .
\end{align}
\end{subequations}

Finally, the Hamiltonian of the 3 nearest-neighbor B sites consists of a
local Hubbard interaction term and a local on-site potential term:

\begin{equation}
H_{\text{imp}}=\sum_{j=1}^{3}\left( \epsilon _{b}n_{b,j}+Un_{b,j\uparrow
}n_{b,j\downarrow }\right)
\end{equation}%
where $n_{b,j\alpha }=b_{j\alpha }^{\dag }b_{j\alpha }$, and $%
n_{b,j}=n_{b,j\uparrow }+n_{b,j\downarrow }$ is the number operator for B
electrons at site $j$. This model is p-h symmetric when $\epsilon _{b}=-U/2$
and the chemical potential $\mu =0$.

The lattice with a 4-site vacancy inherits the $C_{3}$ symmetry of the
pristine lattice. Hence, we can construct helicity eigenstates $h=0,1,\bar{1}$ from the $a_{1,2,3}^{\dagger }$ states:

\begin{subequations}
\begin{align}
c_{h=0}^{\dagger }& =\frac{1}{\sqrt{3}}\left( a_{1}^{\dagger
}+a_{2}^{\dagger }+a_{3}^{\dagger }\right)  \label{Eq:Dfnt_hlct} \\
c_{h=1}^{\dagger }& =\frac{1}{\sqrt{3}}\left( a_{1}^{\dagger }+e^{-i\frac{%
2\pi }{3}}a_{2}^{\dagger }+e^{i\frac{2\pi }{3}}a_{3}^{\dagger }\right) \\
c_{h=\bar{1}}^{\dagger }& =\frac{1}{\sqrt{3}}\left( a_{1}^{\dagger }+e^{i%
\frac{2\pi }{3}}a_{2}^{\dagger }+e^{-i\frac{2\pi }{3}}a_{3}^{\dagger
}\right) .
\end{align}
\end{subequations}
A counter-clockwise $2\pi /3$ rotation about the central A site acts
as a permutation of the three $a_{1,2,3}^{\dagger }$ states. Then, under
this rotation we have $c_{h}^{\dagger }\rightarrow e^{i2\pi
h/3}c_{h}^{\dagger }$.

In the limit $U\sim \left\vert \epsilon _{b}\right\vert \gg t$, by applying
a Schrieffer-Wolff projection as in Ref.~\onlinecite{cond-mat/9607190}, we
obtain an effective Kondo model which includes potential scattering and Kondo
interactions:

\begin{align}
H=& H_{\text{vac}}+V_{0}n_{0}+V_{1}\left( n_{1}+n_{\bar{1}}\right) +J_{00}%
\bm{s}_{00}\cdot \mathcal{S}_{0}
\notag \\
& +J_{11}\left( \bm{s}_{11}+\bm{s}_{\bar{1}%
\bar{1}}\right) \cdot \mathcal{S}_{0}+J_{1\bar{1}}\left( \bm{s}_{1\bar{1}%
}\cdot \mathcal{S}_{1}+\bm{s}_{\bar{1}1}\cdot \mathcal{S}_{\bar{1}}\right) 
\notag \\
& +J_{01}\bigg[\left( \bm{s}_{01}+\bm{s}_{\bar{1}0}\right) \cdot \mathcal{S}%
_{1}+\left( \bm{s}_{10}+\bm{s}_{0\bar{1}}\right) \cdot \mathcal{S}_{\bar{1}}%
\bigg],  \label{3I3CK}
\end{align}%
where the local moment operators of definite helicities are\cite{cond-mat/9607190,PhysRevLett.95.257204}

\begin{equation}
\mathcal{S}_{h}=\sum_{j=1}^{3}e^{-ih2\pi \left( j-1\right) /3}\bm{S}_{j},~%
\bm{S}_{j}=\frac{1}{2}\sum_{\alpha \beta }b_{j\alpha }^{\dagger }\bm{\sigma}%
_{\alpha \beta }b_{j\beta },~n_{b,j}=1.
\end{equation}%
We note that $\mathcal{S}_{h=0}$ is simply the total impurity spin operator, and the total impurity spin quantum number can be $S=3/2$ (one quartet) or $S=1/2$ (two doublets); the two $S=1/2$ doublets can be distinguished by their behavior under the $Z_{3}$ permutation of impurity spins. We have also defined the particle number operators for $c_{h}$, $n_{h}=\sum_{\alpha }c_{h\alpha }^{\dagger }c_{h\alpha }$, and the spin operators

\begin{equation}
\bm{s}_{hh^{\prime }}=\frac{1}{2}\sum_{\alpha \beta }c_{h\alpha }^{\dagger }%
\bm{\sigma}_{\alpha \beta }c_{h^{\prime }\beta }
\end{equation}%
which can involve conduction electrons of different helicities. For our
particular microscopic model, the unrenormalized couplings are

\begin{equation}
J_{hh^{\prime }}\approx \frac{4}{3}t^{2}\left( \frac{1}{U+\epsilon _{b}}+%
\frac{1}{-\epsilon _{b}}\right) ,V_{h}\approx t^{2}\left( \frac{1}{-\epsilon
_{b}}-\frac{1}{U+\epsilon _{b}}\right)
\end{equation}%
Generally $\left\vert V_{h}\right\vert /J_{hh^{\prime }}\leq 3/4$. Note that 
$V_{h}$ will be generated by breaking the p-h symmetry. This can occur not
only when $\epsilon _{b}\neq -U/2$, but also when we move away from the
charge neutrality point or take into consideration second neighbor hopping\cite{PhysRevB.82.165426}.

The scaling dimensions of the various couplings in $H$ are determined by the LDOS of the $c_{h}$ conduction channels for the
4-site-vacancy graphene. A detailed solution of this non-interacting
problem in Appendix~\ref{sec:app4vscatbas} gives the following leading
contributions to the $c_{h}$ channels:

\begin{subequations}
\begin{align}
c_{0}^{\dagger }& \approx -\frac{3^{\frac{3}{4}}a}{4\pi ^{\frac{1}{2}}}%
\int_{-\infty }^{\infty }dk\sqrt{\left\vert k\right\vert }\left( \tilde{\phi}%
_{\vec{K},0,k}^{\dagger }+\tilde{\phi}_{\vec{K}^{\prime },0,k}^{\dagger
}\right)  \label{c0tilde} \\
c_{1}^{\dagger }& \approx -i\frac{\pi ^{\frac{1}{2}}}{3^{\frac{1}{4}}}%
\int_{-\infty }^{\infty }dk\frac{\operatorname{sgn}k}{\sqrt{\left\vert k\right\vert }%
\left( \ln \frac{\Lambda ^{2}}{\left( v_{F} k\right) ^{2}}-i\pi \operatorname{sgn}k\right) }\tilde{\phi}_{%
\vec{K},-1,k}^{\dagger }  \label{c1tilde} \\
c_{\bar{1}}^{\dagger }& \approx -i\frac{\pi ^{\frac{1}{2}}}{3^{\frac{1}{4}}}%
\int_{-\infty }^{\infty }dk\frac{\operatorname{sgn}k}{\sqrt{\left\vert k\right\vert }%
\left( \ln \frac{\Lambda ^{2}}{\left( v_{F} k\right) ^{2}}-i\pi \operatorname{sgn}k\right) }\tilde{\phi}_{%
\vec{K}^{\prime },1,k}^{\dagger }\text{.}  \label{c1bartilde}
\end{align}
\end{subequations}
where $\Lambda \sim t$ is an ultraviolet energy cutoff, $v_{F}=\sqrt{3}ta/2$ is the Fermi velocity, and $\tilde{\phi}_{\vec{K}/\vec{K}^{\prime },m,k}^{\dagger }$ creates an electron in the
eigenstate of $H_{\text{vac}}$ in valley $\vec{K}$ or $\vec{K}^{\prime
}=\left( \sqrt{3},\pm 1\right) \left( 2\pi /3a\right) $, with angular
momentum $m$ and momentum amplitude $k$. (The low energy Dirac theory has
full rotational symmetry so that eigenstates can be labeled by $m$, the
2-dimensional angular momentum quantum number\cite{PhysRevB.53.15079}.) Note that the low-energy spectrum is determined from $\epsilon _{\vec{K}+\vec{k}}\approx v_{F}k$. Due to the additional factor of $a$ in Eq.~(\ref{c0tilde}), $c_{0}$ should become less and less important compared to $c_{1}$ and $c_{\bar{1}}$ at low energy scales, as will be confirmed in Sec.~\ref{sec:Kondo}. From this, is it straightforward to determine the leading contributions to the
LDOS for $c_{h}$ in the low-energy limit:

\begin{equation}
\rho _{h=0}\left( \omega \right) =\frac{3\sqrt{3}a^{2}}{8\pi v_{F}^{2}}%
\left\vert \omega \right\vert \text{,}
\end{equation}

\begin{equation}
\rho _{h=1,\bar{1}}\left( \omega \right) \approx \frac{\pi }{\sqrt{3}\left\vert
\omega \right\vert \ln ^{2}\frac{\Lambda ^{2}}{
\omega ^{2}}} \text{.}  \label{logLDOS}
\end{equation}%
While the helicity-0 channel has a behavior similar to pristine graphene,
helicities $1$ and $\bar{1}$ show a logarithmic divergence in their LDOS. We
attribute such a divergence to the presence of two non-normalizable zero modes
in the 4-site-vacancy graphene, whose wave functions behave as $1/r$ when the
distance to the vacancy $r$ is large\cite%
{PhysRevB.79.155442,PhysRevLett.96.036801,PhysRevB.88.075413}; see Appendix~\ref{sec:appzero} for details. Because $N_{A}=1$ A site and $N_{B}=3$ B
sites are removed, there are $\left\vert N_{A}-N_{B}\right\vert =2$ zero
modes (for each spin) living on the A sublattice; due to the $C_{3}$
symmetry in this case, these two zero modes can be chosen as helicity
eigenstates\cite{PhysRevB.77.115109}. Even though these zero modes are not
true eigenstates in an infinite system, they hybridize strongly with the
low-energy itinerant states in pristine graphene, forming low-energy
scattering states which are true eigenstates of $H_{\text{vac}}$ and
contribute to the divergent LDOS\cite{PhysRevB.88.075413}.

\section{Kondo model\label{sec:Kondo}}

In this section, we establish the phase diagram of the 3-impurity 3-channel
Kondo model Eq.~(\ref{3I3CK}), using a combination of analytical arguments
and NRG.

\subsection{Scaling and RKKY interactions at weak coupling\label{sec:weak}}

It is instructive to begin by analyzing the weak-coupling fixed point. To
find the scaling behavior of various coupling constants, we first define
dimensionless couplings at the running energy cutoff $D$:

\begin{equation}
v_{h}\left( D\right) \equiv \rho _{h}\left( D\right) V_{h}\left( D\right) 
\text{,}
\end{equation}
and

\begin{equation}
j_{hh^{\prime }}\left( D\right) \equiv \sqrt{\rho _{h}\left( D\right) \rho
_{h^{\prime }}\left( D\right) }J_{hh^{\prime }}\left( D\right) \text{.}
\end{equation}%
The second-order weak-coupling RG equations then read

\begin{subequations}
\begin{equation}
-\frac{dj_{11}}{d\ln D}=\left( 1-\frac{2}{\ln \frac{\Lambda }{D}}\right)
j_{11}+\left( j_{11}^{2}+j_{1\bar{1}}^{2}+j_{01}^{2}\right) \text{,}
\end{equation}

\begin{equation}
-\frac{dj_{1\bar{1}}}{d\ln D}=\left( 1-\frac{2}{\ln \frac{\Lambda }{D}}%
\right) j_{1\bar{1}}+\left( 2j_{11}j_{1\bar{1}}+j_{01}^{2}\right) \text{,}
\end{equation}

\begin{equation}
-\frac{dj_{00}}{d\ln D}=-j_{00}+j_{00}^{2}+2j_{01}^{2}\text{,}
\end{equation}

\begin{equation}
-\frac{dj_{01}}{d\ln D}=-\frac{1}{\ln \frac{\Lambda }{D}}j_{01}+j_{01}\left(
j_{11}+j_{1\bar{1}}+j_{00}\right) \text{,}
\end{equation}

\begin{equation}
-\frac{dv_{0}}{d\ln D}=-v_{0}\text{,}
\end{equation}

\begin{equation}
-\frac{dv_{1}}{d\ln D}=\left( 1-\frac{2}{\ln \frac{\Lambda }{D}}\right) v_{1}%
\text{.}
\end{equation}%
From these equations we can also see that the relation $J_{11}=J_{1\bar{1}}$%
, if true for the bare couplings, is preserved along the RG flow.

Due to the singular LDOS for helicities $1$ and $\bar{1}$, $J_{11}$ and $J_{1%
\bar{1}}$ are \emph{relevant} at low energies ($D\ll \Lambda $), in analogy
to the single-channel problem with a divergent LDOS discussed in Ref.~%
\onlinecite{PhysRevB.88.075104}. This leads to a greatly enhanced Kondo
temperature for the corresponding Kondo couplings:

\end{subequations}
\begin{equation}
T_{K}\propto J_{K}/\ln ^{2}\left( \Lambda /J_{K}\right) \text{,}
\end{equation}%
where $J_{K}$ is either $J_{11}\left( D_{0}\right)$ or $J_{1\bar{1}}\left( D_{0}\right)$, $D_{0}$ being the initial semi-bandwidth of the Kondo model. The potential scattering
term $V_{1}$ is likewise relevant, and has its own characteristic energy
scale

\begin{equation}
T_{P}\propto V_{1}\left( D_{0}\right)/\ln ^{2}\left( \Lambda /V_{1}\left( D_{0}\right)\right) \text{,}
\end{equation}%
at which it flows to strong-coupling. On the other hand, $J_{01}$ is weakly
irrelevant, becoming almost marginal only at very low energies, even though
it generates relevant couplings $J_{11}$ and $J_{1\bar{1}}$ at the second
order. Finally, the linear LDOS of the helicity-0 channel renders $V_{0}$
and $J_{00}$ strongly irrelevant.

It is also possible to consider the RKKY interactions between magnetic impurities at weak coupling mediated by conduction electrons\cite{PhysRevLett.58.843,*PhysRevLett.61.125,*PhysRevB.40.324,cond-mat/9607190}; this gives us some intuition on possible magnetic orders of the impurities. It should be clarified that these interactions are only introduced to help us understand the Kondo model; they are not part of the NRG input (as the Kondo couplings are), and we make no a priori assumptions about the associated RKKY energy scale in our NRG calculations. As in Refs.~\onlinecite{cond-mat/9607190,PhysRevLett.95.257204}, the RKKY interactions are of the form

\begin{equation}
H_{\text{RKKY}}=I\sum_{i<j}\bm{S}_{i}\cdot \bm{S}_{j},  \label{HRKKY}
\end{equation}
where we labeled a generic RKKY interaction by $I$ in order to avoid
confusion with the valley-momenta $\vec{K},\vec{K}^{\prime }$. This
expression can be re-cast as

\begin{equation}
H_{\text{RKKY}}=\frac{I}{2}\left( \mathcal{S}_{h=0}^{2}-\bm{S}_{1}^{2}-\bm{S}%
_{2}^{2}-\bm{S}_{3}^{2}\right) .
\end{equation}%
$H_{\text{RKKY}}$ takes the value $3I/4$ in the $S=3/2$ multiplet state, and 
$-3I/8$ otherwise. Hence, strong antiferromagnetic RKKY interactions ($I>0$)
project onto the $S=1/2$ manifold, while strong ferromagnetic RKKY ($I<0$)
prefers the $S=3/2$ configuration.

The RKKY coupling strength $I$ has been evaluated in bulk graphene\cite%
{PhysRevB.76.184430,PhysRevB.80.153414,PhysRevB.81.205416,PhysRevB.83.165425,PhysRevB.84.115119}%
, under the assumption that each magnetic impurity interacts with one carbon
atom but does not disrupt the graphene lattice (e.g. by introducing
vacancies). In that case, it has been shown to be ferromagnetic between
impurities on the same sublattice, and antiferromagnetic between impurities
on different sublattices. Nevertheless, antiferromagnetic RKKY interactions
have also been reported between same-sublattice magnetic impurities when
other non-magnetic impurities are present\cite{PhysRevB.86.205427}, or when a large on-site potential energy is associated with the magnetic impurities\cite{PhysRevB.95.075411}.

In Appendix \ref{sec:appRKKY}, we analyze the RKKY interaction between the
three effective magnetic impurities $b_{1,2,3}$ generated by Kondo couplings
to $O\left( J_{hh^{\prime }}^{2}\right) $, carefully including the effects
of the vacancy. At low temperatures, we find $I\propto J_{1\bar{1}%
}^{2}-2J_{11}^{2}$; thus $I$ is ferromagnetic in a model with $J_{11}$ only,
and antiferromagnetic in a model with $J_{1\bar{1}}$ only. In the Kondo
model obtained through the Schrieffer-Wolff transformation, where $%
J_{11}=J_{1\bar{1}}$, $I$ is expected to be ferromagnetic. The RG flow of
the RKKY interaction is controlled by

\begin{equation}
-\frac{d\left( j_{1\bar{1}}^{2}-2j_{11}^{2}\right) }{d\ln D}=\left( 2-\frac{4%
}{\ln \frac{\Lambda }{D}}\right) \left( j_{1\bar{1}}^{2}-2j_{11}^{2}\right)
+2\left( j_{1\bar{1}}-2j_{11}\right) j_{01}^{2}-4j_{11}^{3}\text{.}
\end{equation}%
In other words, the RKKY interaction does not change sign along the RG flow
near the weak coupling fixed point.

\subsection{Auxiliary model: 2-channel spin-$S$ Kondo model with a logarithmically divergent LDOS\label%
{sec:S2CK}}

The weak-coupling analysis in Sec.~\ref{sec:weak} shows that all
coupling constants associated with the helicity-0 conduction channel are
irrelevant, and the collective state of the impurity spins can be either an $%
S=3/2$, helicity-0 multiplet or an $S=1/2$, helicity-$\pm 1$ multiplet. We
are therefore motivated to study the 2-channel spin-$S$ Kondo model where
both conduction channels $1$ and $\bar{1}$ are characterized by the
logarithmically divergent LDOS Eq.~(\ref{logLDOS}):

\begin{equation}
H=H_{\text{vac}}+V\left( n_{1}+n_{\bar{1}}\right) +J\left( \bm{s}_{11}+\bm{s}%
_{\bar{1}\bar{1}}\right) \cdot \mathbf{S}  \label{log2CK}
\end{equation}%
with an antiferromagnetic Kondo coupling $J>0$. Although this model is interesting in its own right, it is quite different from Eq.~(\ref{3I3CK}) even after ignoring the helicity-0 conduction channel [see Eq.~(\ref{3I2CK})]. In Eq.~(\ref{3I3CK}) we have spin operators of conduction electrons that are diagonal ($\bm{s}_{11}$, $\bm{s}_{\bar{1}\bar{1}}$) and off-diagonal ($\bm{s}_{1\bar{1}}$, $\bm{s}_{\bar{1}1}$) in the helicity basis, while in Eq.~(\ref{log2CK}) we only have diagonal operators in the helicity basis; also, the impurity in Eq.~(\ref{3I3CK}) comprises three $S=1/2$ spins appearing in different total spin sectors and different helicity combinations, whereas the impurity in Eq.~(\ref{log2CK}) is a single spin of rigid size $S$. Nevertheless, as we will show in Sec.~\ref{sec:3I2CK}, the auxiliary model Eq.~(\ref{log2CK}) provides helpful intuitions for understanding the limiting cases of the 3-impurity model, where the impurity can be effectively viewed as a rigid spin. The auxiliary model also serves as an excellent benchmark for the impurity contributions to thermodynamic quantities, which we will discuss later in this section.

Let us briefly review the case where both channels have a constant LDOS. In
this case, at the weak-coupling fixed point, $V$ is exactly marginal and $J$
is marginally relevant due to the dynamics of the impurity spin. It is well
known that the low-energy behavior of this problem depends on the size of
the impurity spin\cite{JPhysFrance.41.193,ActaPhysPolB.26.1869}. When $S>1$ (underscreened) or $%
S=1$ (exactly screened), the low-energy fixed point is simply the
strong-coupling one. This involves a decoupled residual impurity spin of
size $S-1$ (when $S>1$) and a Fermi liquid theory for the conduction
electrons, with a phase shift $\pi /2+\delta $ ($\delta $ being an odd
function of $V$) for each conduction channel taking part in the screening
of the impurity. On the other hand, in the overscreened case $S=1/2$, the
strong-coupling fixed point is no longer stable. If we consider a naive $%
J\rightarrow +\infty $ theory on a nearest-neighbor tight-binding lattice
with hopping amplitude $t\ll J$, then the electrons $c_{1}$, $c_{\bar{1}}$
are strongly bound to the impurity spin, forming an effective spin-1/2. The
effective spin is in turn coupled to the lattice with $c_{1}$ and $c_{\bar{1}%
}$ removed (which we name as the \textquotedblleft strong-coupling
lattice\textquotedblright ). This $O\left( t^{2}/J\right) $ coupling
constant is antiferromagnetic, which we already know to be marginally
relevant. In contrast, the underscreened strong-coupling fixed point is
stable because the large-$J$ effective model is a Kondo model with a \emph{%
ferromagnetic} Kondo coupling, which is marginally irrelevant (and turns the system into a singular Fermi liquid\cite{PhysRevB.72.014430}). The stable
low-energy fixed point in the overscreened case is therefore an
intermediate-coupling, non-Fermi-liquid one. Its properties can be obtained
by a range of theoretical methods, including boundary conformal field theory%
\cite{PhysRevB.48.7297}, Bethe ansatz\cite%
{PhysRevLett.52.364,JStatPhys.38.125} and Abelian bosonization\cite%
{PhysRevB.46.10812}; in particular an impurity entropy of $\frac{1}{2}\ln 2$
and an impurity magnetic susceptibility that depends logarithmically on
temperature have been predicted and numerically verified.

We return to the 2-channel Kondo model Eq.~(\ref{log2CK}) with a divergent LDOS Eq.~(\ref{logLDOS}). As discussed in Sec.~\ref{sec:weak}, at the
weak-coupling, p-h symmetric local moment fixed point, both $J$ and $V$ are
relevant on account of the energy dependence of the LDOS. It is therefore
natural to investigate the corresponding strong-coupling theories on a
lattice with hopping amplitude $t\ll J$. The $J\rightarrow +\infty $ lattice
theory as before gives rise to an effective impurity spin of size $%
\left\vert S-1\right\vert $ formed by the impurity spin and the electrons $%
c_{1}$ and $c_{\bar{1}}$, which is coupled to the remaining
\textquotedblleft strong-coupling lattice\textquotedblright\ with an either
ferromagnetic (when $S>1$) or antiferromagnetic (when $S=1/2$) coupling
constant of $O\left( t^{2}/J\right) $. In the $V\rightarrow \pm \infty $
lattice theory, on the other hand, no Kondo physics takes place; the $c_{1}$
and $c_{\bar{1}}$ states become nevertheless inaccessible to other
conduction electrons, being either empty or occupied depending on the sign
of $V$, and the original spin $S$ is coupled to the remaining
\textquotedblleft strong-coupling lattice\textquotedblright .

At this point we should accentuate the crucial difference between the
logarithmically divergent LDOS and the constant LDOS. In the constant LDOS case, removing $%
c_{1}$ and $c_{\bar{1}}$ from the original lattice does not change the LDOS
at the sites which the effective impurity can couple to, because the
original lattice and the \textquotedblleft strong-coupling
lattice\textquotedblright\ are both composed of one-dimensional chains. In
graphene with a 4-site vacancy, however, projecting out $c_{1}$ and $c_{\bar{1}}$
will remove the logarithmic divergence in the LDOS, and the leading
contribution to the LDOS becomes linear near the Dirac point as in the bulk.
This can be shown explicitly using the method outlined in Appendix~\ref%
{sec:app4vscatbas}. On a more intuitive level, we can also explain the
disappearance of the logarithmic divergence through the removal of the two
non-normalizable zero modes. In Appendix~\ref{sec:appzero}, we see that the
wave functions associated with $c_{1}$ and $c_{\bar{1}}$ cannot be
simultaneously zero for a nontrivial zero-energy solution of the lattice Schr%
\"{o}dinger equation; in other words, if we project out $c_{1}$ and $c_{\bar{%
1}}$ from the lattice, i.e. demand the wave functions associated with $c_{1}$
and $c_{\bar{1}}$ should vanish, then no zero mode exists.

Because of the linear LDOS in the \textquotedblleft strong-coupling
lattice\textquotedblright , any perturbation at the naive strong-coupling
fixed point-- Kondo coupling or potential scattering-- is strongly irrelevant%
\cite{PhysRevB.70.214427}. It is therefore reasonable to conclude that the
large-$J$ and large-$V$ strong-coupling fixed points are \emph{stable} in
the 2-channel Kondo model Eq.~(\ref{log2CK}) with a divergent LDOS Eq.~(%
\ref{logLDOS}), irrespective of the size of the impurity spin $S$.

\begin{figure}[!h]
\includegraphics[width=0.5\columnwidth]{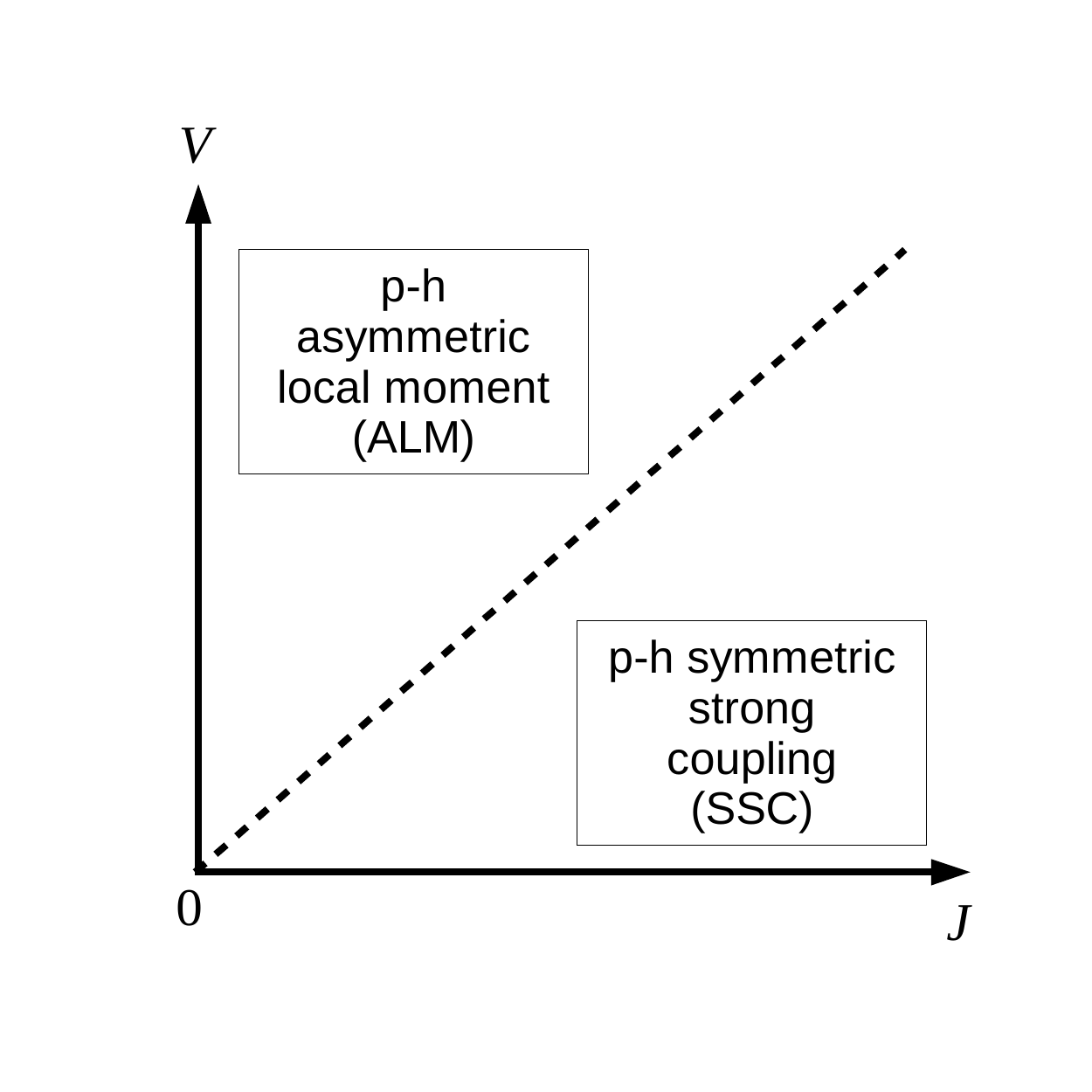} 
\caption{Schematic phase diagram of the 2-channel Kondo model Eq.~(\ref{log2CK})
with a single spin-$S$ impurity and the logarithmically divergent LDOS Eq.~(\ref{logLDOS}) on the $J$-$V$
plane. There are two phases, the p-h symmetric strong-coupling (SSC) phase and the p-h asymmetric local-moment
(ALM) phase, characterized by $J\to \infty$ and $V\to \infty$ respectively. The residual spin is of size
$\left| S-1\right|$ in the maximally screened SSC phase, and $S$ in the unscreened ALM phase. The two phases
are separated by a second-order phase transition, as in the single-channel case of
Ref.~\onlinecite{PhysRevB.88.075104}. Note that these results are independent of the impurity spin size $S$.}
\label{NRGphasediagS}
\end{figure}

This picture is verified by NRG calculations, which we perform with the
\textquotedblleft NRG Ljubljana\textquotedblright\ code\cite{NRGLjubljana}.
The schematic phase diagram is shown in Fig.~\ref{NRGphasediagS}. As
conjectured, the two regimes are controlled by the large-$J$ p-h symmetric
strong-coupling (SSC) fixed point and the large-$V$ p-h asymmetric local
moment (ALM) fixed point, with residual spin sizes $\left\vert
S-1\right\vert $ and $S$ respectively. At both fixed points it is possible
to construct the entire finite-size spectrum from single-particle
excitations, in contrast to the non-Fermi-liquid overscreened Kondo fixed
point in the case of constant LDOS. As in the single-channel spin-1/2 case
discussed in Ref.~\onlinecite{PhysRevB.88.075104}, the two phases are
separated by a second-order transition.

To shed further light on the nature of the low-energy fixed points, let us
examine their thermodynamic properties. The impurity contribution to any
quantity $\Omega $ in a quantum impurity system is defined as $\Omega_{\text{imp}}=\Omega -\Omega _{0}$, the difference between this quantity
evaluated in the entire impurity system and in the reference ``clean'' system without the
impurity. In our case, there are two possibilities for the reference system: the pristine graphene, and the non-interacting bath of a 4-site-vacancy graphene lattice with a logarithmically divergent LDOS. Following Ref.~\onlinecite{PhysRevB.88.075104}, we choose the 4-site-vacancy graphene as the reference system while presenting our numerical results, but impurity quantities measured with respect to pristine graphene will also be discussed because they are experimentally directly accessible (see Table~\ref{listSTchi}). We focus on the impurity entropy $S_{\text{imp}}=-\partial \mathcal{F_{\text{imp}}}/\partial T$ ($\mathcal{F}$ being the
free energy) and the impurity magnetic susceptibility $\chi_{\text{imp}} =\left\langle
S_{z}^{2}\right\rangle _{\text{imp}} /T$.

It is useful to first discuss the effect of non-normalizable zero modes. In
short, with respect to pristine graphene, each non-normalizable zero mode
contributes $\ln 4$ to the zero-temperature impurity entropy, and $1/\left(
8T\right) $ to the magnetic susceptibility\cite{PhysRevB.88.075104}. We can
directly derive these results by viewing pristine graphene as a
(non-interacting) resonant-level model with a logarithmically divergent LDOS, as has been
done in Appendix B of Ref.~\onlinecite{PhysRevB.88.075104}. Alternatively,
when we calculate the impurity-induced density of states (which appears in
the Friedel sum rule)\cite{hewson1997kondo} in the single-vacancy graphene
lattice, we find a $\delta $-function peak at zero energy\cite%
{PhysRevB.79.155442,PhysRevLett.96.036801,PhysRevB.88.075413}; this $\delta $%
-function is what contributes to the zero-temperature impurity entropy and
the impurity susceptibility, as if it were a real spin-degenerate
single-particle eigenstate of the system. When the non-normalizable zero
mode is removed, for instance at strong coupling, the impurity contributions
vanish correspondingly. This applies equally to the 4-site-vacancy graphene
lattice, except there are now two channels with a logarithmically divergent LDOS, so that
the strength of the $\delta $-function peak is doubled together with the
impurity entropy and impurity magnetic susceptibility.

We can now determine the limiting behavior of the thermodynamic properties
at the fixed points of the 2-channel spin-$S$ Kondo model Eq.~(\ref{log2CK}) with a divergent LDOS Eq.~(\ref{logLDOS}). At the ALM fixed point, the
non-normalizable zero modes are removed by potential scattering; the local
moment then results in a zero-temperature impurity entropy of $S_{\text{imp}%
}\left( T=0\right) =\ln \left( 2S+1\right) $ and an impurity susceptibility
of $\chi _{\text{imp}}=S\left( S+1\right) /\left( 3T\right) $ relative to
pristine graphene. At the SSC fixed point, the zero modes are removed by
strong Kondo screening, so the residual spin $\left\vert S-1\right\vert $
yields $S_{\text{imp}}\left( T=0\right) =\ln \left( 2\left\vert
S-1\right\vert +1\right) $ and $T\chi _{\text{imp}}=\left\vert
S-1\right\vert \left( \left\vert S-1\right\vert +1\right) /3$. Finally, at
the p-h symmetric local moment fixed point, the zero modes together with the
local moment give $S_{\text{imp}}\left( T=0\right) =\ln \left( 2S+1\right)
+2\ln 4$ and $T\chi _{\text{imp}}=1/4+S\left( S+1\right) /3$.

\begin{figure}[!h]
\includegraphics[width=1\columnwidth]{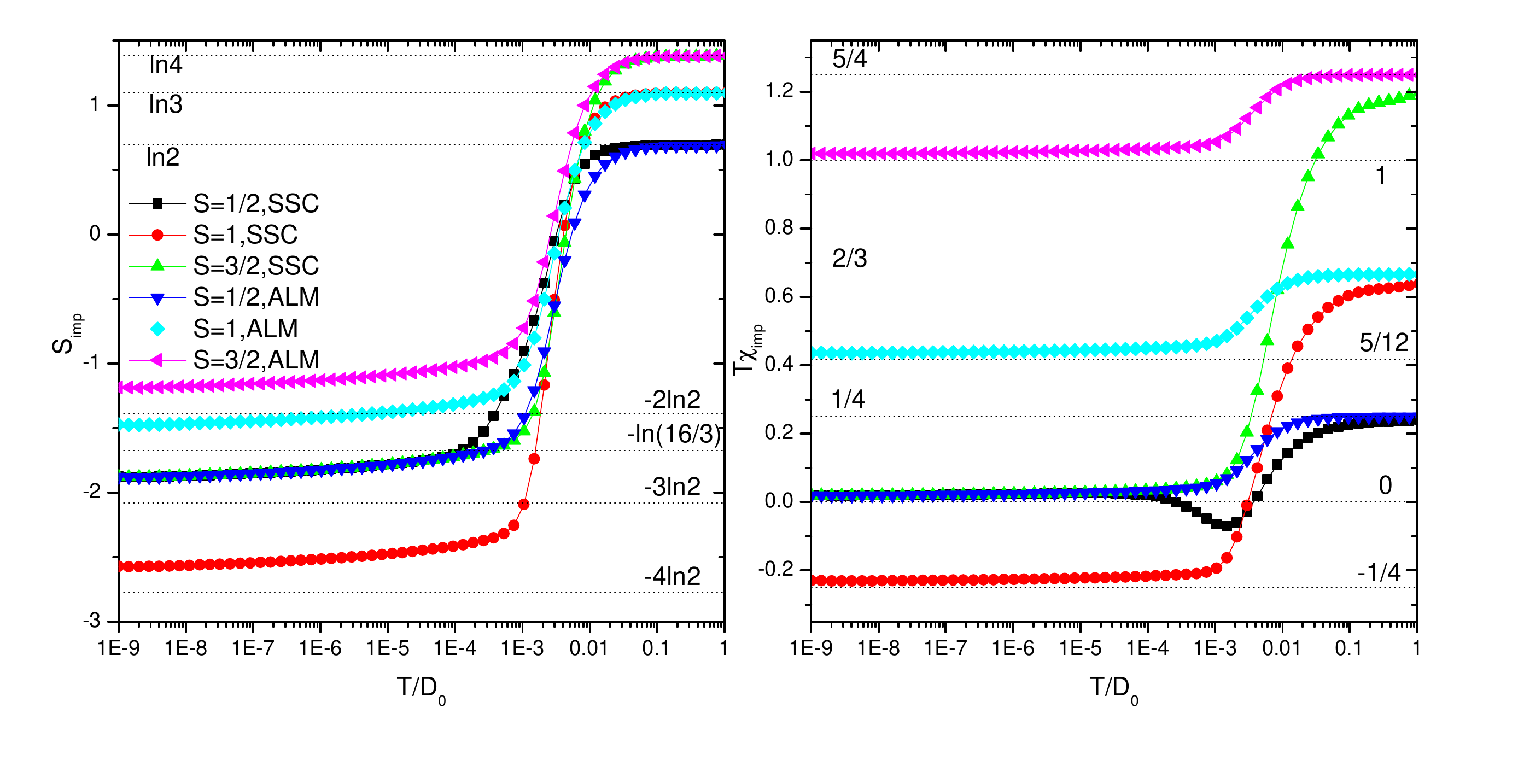} 
\caption{Impurity entropy $S_{\text{imp}}$ and impurity magnetic susceptibility multiplied by temperature,
$T \chi_{\text{imp}}$, versus temperature $T$ in the 2-channel Kondo model Eq.~(\ref{log2CK})
with a single spin-$S$ impurity and a logarithmically divergent LDOS Eq.~(\ref{logLDOS}). Note that the reference ``clean'' system is taken to be 4-site-vacancy graphene rather than pristine graphene. We have chosen $\Lambda=1.5D_{0}$ in Eq.~(\ref{logLDOS}), where $D_{0}$ is the initial semi-bandwidth of the Kondo model, and have used $J=0.1D_{0}$, $V=0$ for the SSC curves, and $J=0$, $V=0.1D_{0}$ for the ALM curves.}
\label{STchiS}
\end{figure}

The high- to low-temperature crossover of $S_{\text{imp}}$ and $T\chi _{%
\text{imp}}$ is plotted in Fig.~\ref{STchiS}, with the reference system
chosen as the 4-site-vacancy graphene (where $S_{\text{imp}}\left( T=0\right)
=2\ln 4$ and $T\chi _{\text{imp}}=2\cdot 1/8=1/4$). In the $S=1/2$ case, the
crossover between the unscreened $S=1/2$ and the overscreened $S=1/2$ is
clearly visible from the non-monotonicity of $T\chi _{\text{imp}}$. At low temperatures, $S_{\text{imp}}$ and $T\chi _{\text{imp}}$ have logarithmic corrections in the form of $1/\ln \left(\Lambda /T\right)$ near all fixed points\cite{PhysRevB.88.075104}. However, unlike in a singular Fermi liquid, this logarithmic behavior does not originate from a marginally irrelevant operator, because the allowed operators at any of these fixed points are strongly irrelevant irrespective of the spin size. We can verify that the logarithmic corrections vanish when the reference system is pristine graphene (in the form of a non-interacting 4-site cluster coupled to the 4-site-vacancy graphene); therefore, the logarithmic behavior in Fig.~\ref{STchiS} can be fully attributed to the non-normalizable zero modes of the reference system.

\subsection{Phases of the 3-impurity model\label{sec:3I2CK}}

We are in the position to present the NRG results on the 3-impurity
3-channel Kondo model Eq.~(\ref{3I3CK}). Again, as a first approximation, we
neglect the helicity-0 conduction channel entirely on the grounds that all
couplings $J_{00}$, $J_{01}$ and $V_{0}$ associated with it are irrelevant;
we shall see later that this approximation is usually justified. This leaves
us with a 3-impurity 2-channel Kondo model,

\begin{equation}
H=H_{\text{vac}}+V_{1}\left( n_{1}+n_{\bar{1}}\right) +J_{11}\left( \bm{s}%
_{11}+\bm{s}_{\bar{1}\bar{1}}\right) \cdot \mathcal{S}_{0}+J_{1\bar{1}%
}\left( \bm{s}_{1\bar{1}}\cdot \mathcal{S}_{1}+\bm{s}_{\bar{1}1}\cdot 
\mathcal{S}_{\bar{1}}\right) \text{,}  \label{3I2CK}
\end{equation}%
with two relevant Kondo couplings $J_{11}$ and $J_{1\bar{1}}$ as well as a
relevant potential scattering $V_{1}$. As in Refs.~\onlinecite{cond-mat/9607190, PhysRevLett.95.257204}, one may measure the relative strengths of $J_{11}$ and $J_{1\bar{1}}$ with the dimensionless RKKY coupling strength $\tilde{I}\equiv \left( J_{1\bar{1}}^{2}-2J_{11}^{2}\right) /\left( 2J_{11}^{2}+2J_{1\bar{1}}^{2}\right) $. $J_{11}=J_{1\bar{1}}=V_{1}=0$ marks
the unstable p-h symmetric local moment (\textquotedblleft
LM\textquotedblright ) fixed point, which has three decoupled impurity spins and two
non-normalizable zero modes; therefore, at the LM fixed point, $S_{%
\text{imp}}\left( T=0\right) =3\ln 2+2\ln 4=7\ln 2$, and $T\chi _{\text{imp}%
}=3\cdot 1/4+2\cdot 1/8=1$ relative to pristine graphene.

\begin{figure}[!h]
\includegraphics[width=1\columnwidth]{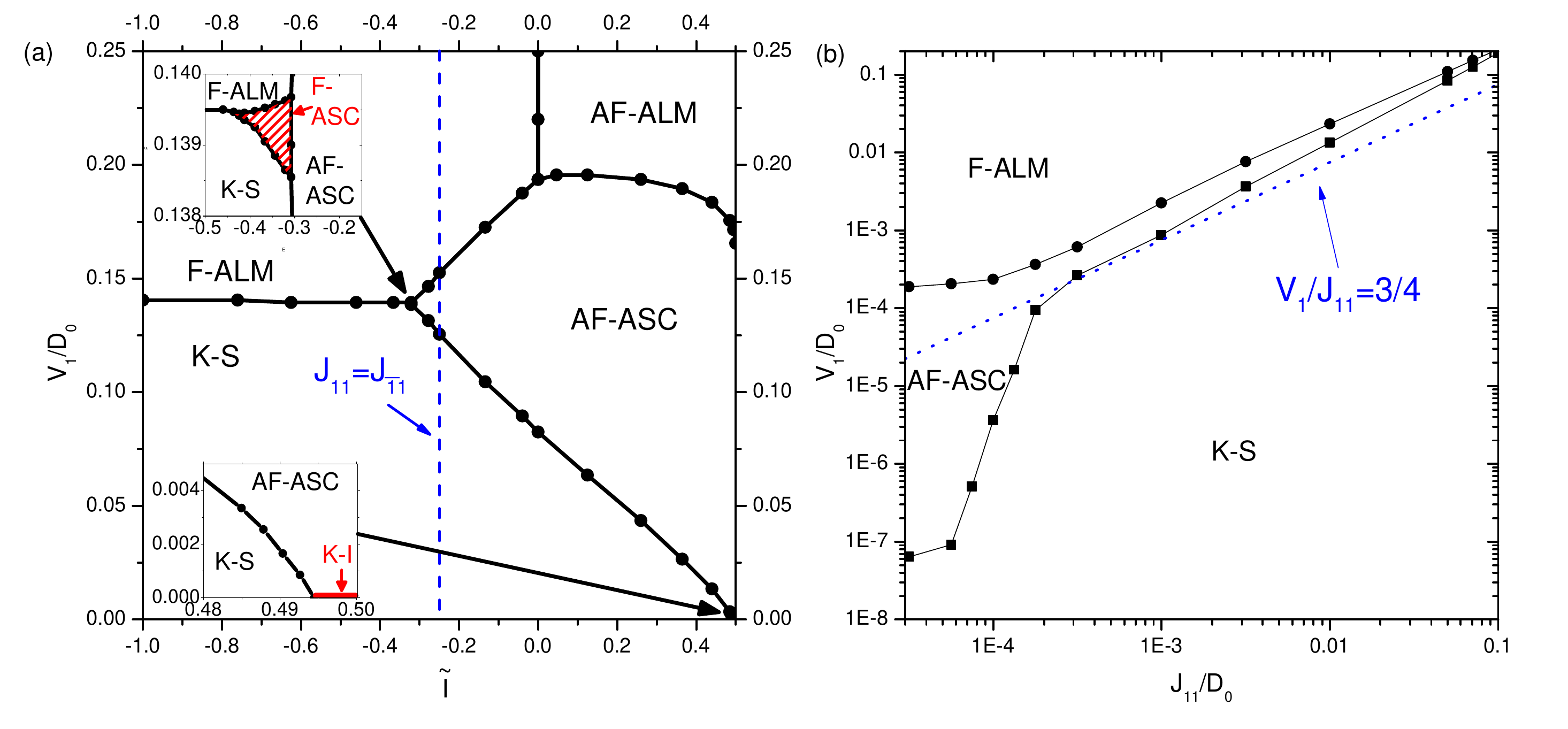} 
\caption{Phase diagram of the 3-impurity 2-channel Kondo model Eq.~(\ref{3I2CK}) with a logarithmically divergent LDOS given by Eq.~(\ref{logLDOS}). We have again chosen the ultraviolet cutoff as $\Lambda=1.5D_{0}$.
(a): phase diagram on the $\tilde{I}$-$V_{1}$ plane, where $\tilde{I} \equiv \left( J^2_{1\bar{1}}-2J^2_{11}\right) / \left( 2J^2_{11} + 2J^2_{1\bar{1}} \right) $ is the dimensionless RKKY interaction. We fix $J^2_{11} + J^2_{1\bar{1}} =\left( 0.1D_{0} \right) ^{2}$. When $V_{1}=0$, we find an underscreened Kondo strong-coupling phase (K-S) with an effective impurity spin $S=3/2$ in the ferromagnetic RKKY limit ($\tilde{I}=-1$), and a spin-singlet isospin-doublet phase (K-I) with an isospin $1/2$ in the antiferromagnetic RKKY limit ($\tilde{I}=1/2$). 
For $\tilde{I}=-1$, sufficiently strong potential scattering $V_{1}$ will overcome the strong-coupling phase at
$V_{1}/J_{11}\approx O(1)$, resulting in a p-h asymmetric $S=3/2$ local-moment (F-ALM) phase. 
For $\tilde{I}=1/2$, even an infinitesimal $V_{1}$ drives the system into a p-h asymmetric, exactly screened strong-coupling phase (AF-ASC) with $S=0$ (lower inset), characterized by antiferromagnetic impurity spin correlations. A larger $V_{1}$ comparable to $J_{1\bar{1}}$ leads to a p-h asymmetric local-moment phase (AF-ALM) with $S=1/2$.
Finally, a p-h asymmetric strong coupling phase with $S=1$ and ferromagnetic impurity spin correlations (F-ASC) exists in a small region of the parameter space, and separates the three phases K-S, F-ALM and AF-ASC (upper inset).
(b): phase diagram on the $J_{11}$-$V_{1}$ plane for $J_{11} = J_{1\bar{1}}$ (i.e. $\tilde{I}=-1/4$). We find that the critical value of $V_{1}/J_{11}$ at the K-S/AF-ASC transition becomes smaller as $J_{11}/D_{0}$ is reduced.} \label{NRGphasediag}
\end{figure}

\begin{figure}[!h]
\includegraphics[width=1\columnwidth]{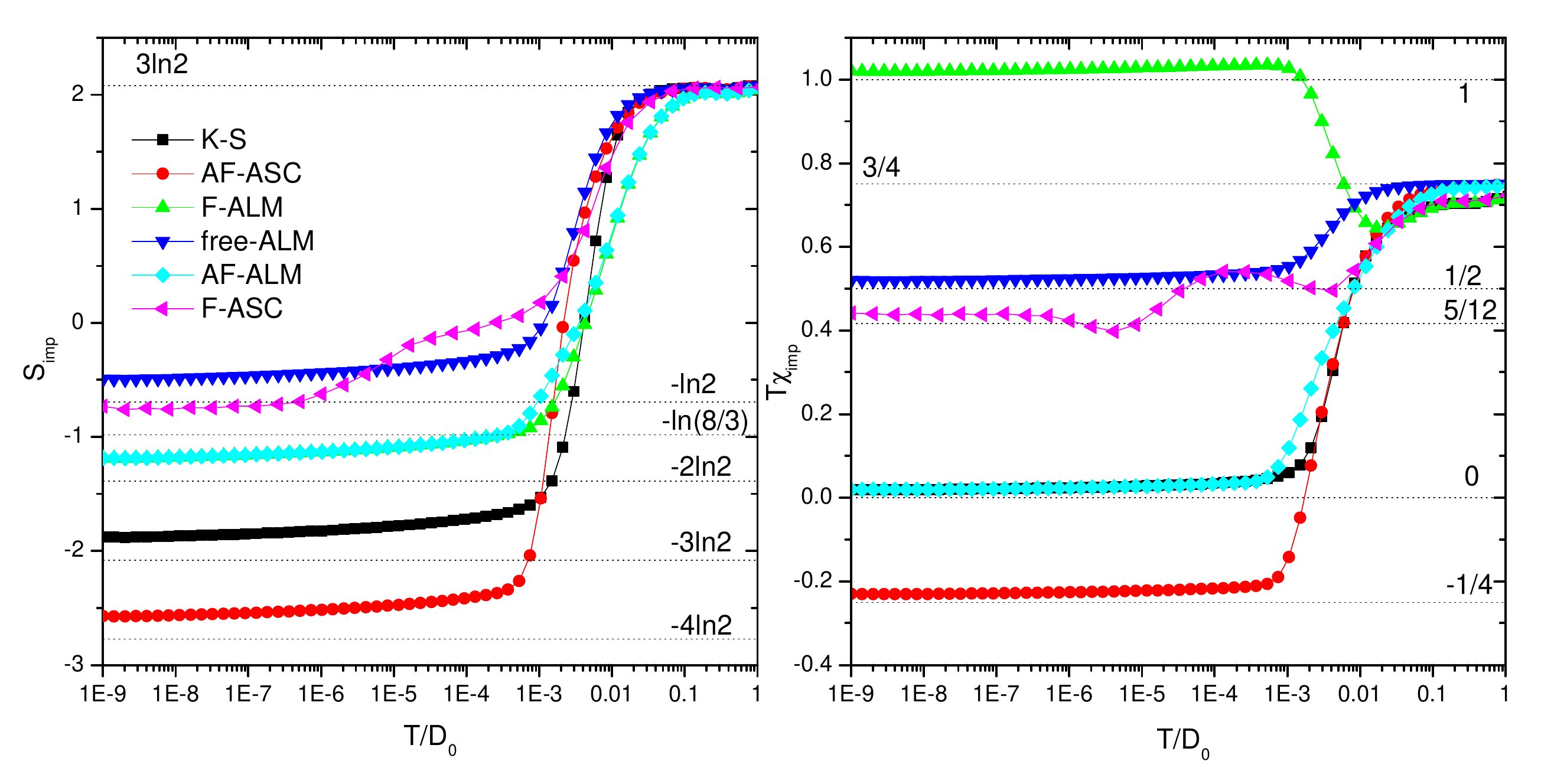} 
\caption{$S_{\text{imp}}$ and $T \chi_{\text{imp}}$ versus $T$ in various phases of the 3-impurity 2-channel Kondo model Eq.~(\ref{3I2CK}) with a logarithmically divergent LDOS given by Eq.~(\ref{logLDOS}). Note again that the reference
``clean'' system is taken to be graphene with a 4-site vacancy rather than pristine graphene. $\Lambda=1.5D_{0}$; $\left( J_{11}, J_{1\bar{1}},
V_{1} \right) /D_{0}=\left( 0.1, 0, 0\right) $ for K-S, $\left( 0, 0.1, 0.1\right)$ for AF-ASC, $\left( 0.1, 0, 0.3\right) $ for F-ALM, $\left( 0, 0, 0.1\right) $ for free-ALM, $\left( 0, 0.1, 0.3\right) $ for AF-ALM, and $\left( 0.075, 0.0661438, 0.139\right) $ for F-ASC.  Data for F-ASC is not $z$-averaged\cite{PhysRevB.33.7871,*PhysRevB.49.11986} and therefore contains spurious oscillations. The K-I fixed point is shown separately in Fig.~\ref{STchiKI}.}
\label{STchi3I2CK}
\end{figure}

\begin{figure}[!h]
\includegraphics[width=1\columnwidth]{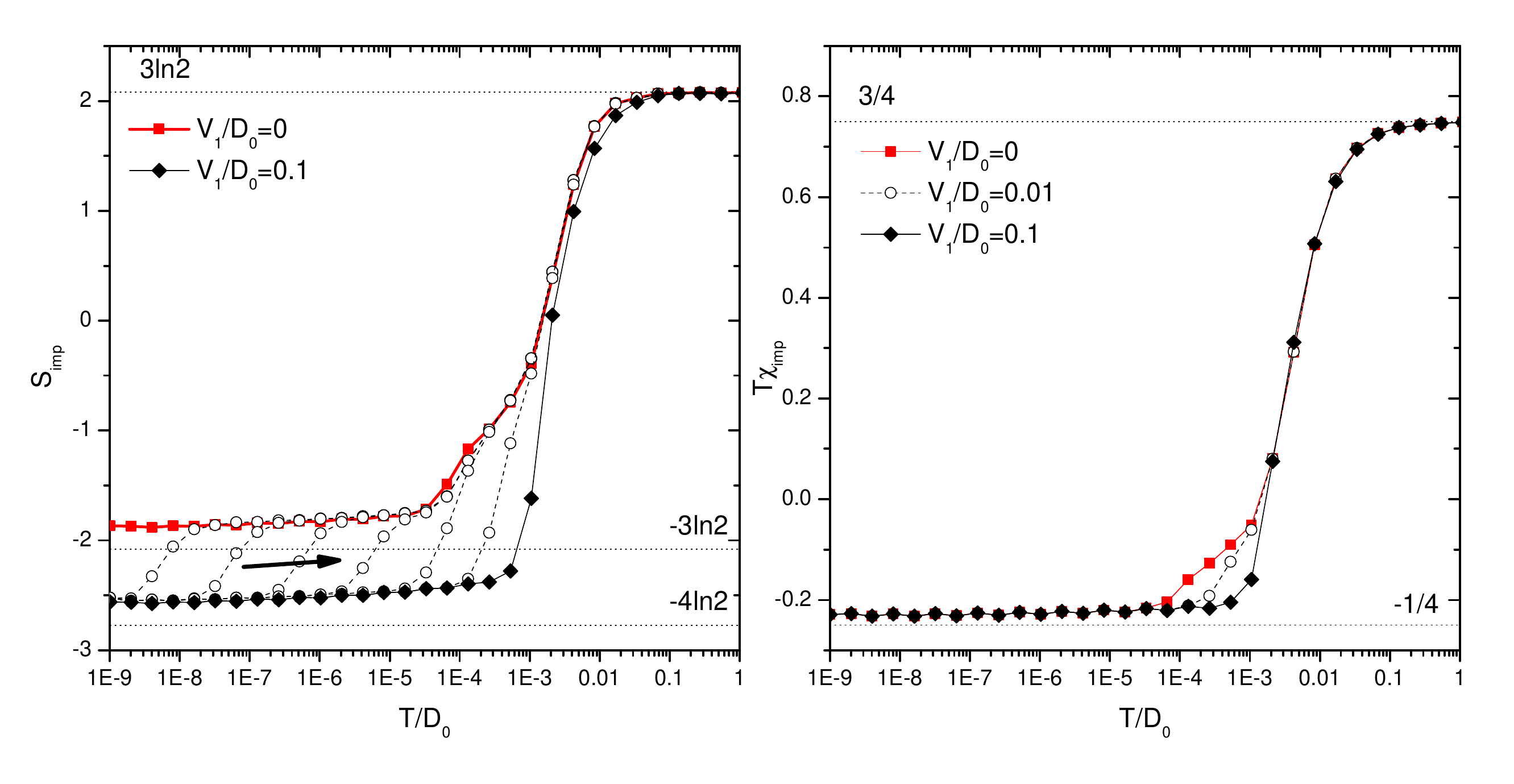} 
\caption{$S_{\text{imp}}$ and $T \chi_{\text{imp}}$ versus $T$ in a crossover from the unstable K-I fixed point to the stable AF-ASC fixed point as the p-h symmetry breaking potential scattering $V_{1}$ is increased. $\Lambda=1.5D_{0}$, $J_{11} = 0$, $J_{1\bar{1}} = 0.1D_{0}$, and different curves correspond to different values
of $V_{1}/D_{0}$. Solid red squares represent the K-I fixed point $V_{1}=0$, and solid black diamonds represent $V_{1}=0.1D_{0}$. In the $S_{\text{imp}}$ plot, the open black circles correspond to $V_{1}/D_{0}=10^{-7}$, $10^{-6}$, $10^{-5}$,$10^{-4}$, $10^{-3}$ and $0.01$ in the direction of the arrow; in the $T \chi_{\text{imp}}$ plot, only $V_{1}=0.01D_{0}$ is shown in open black circles for clarity. Data in these figures is not $z$-averaged and therefore contains spurious oscillations.}
\label{STchiKI}
\end{figure}

In Fig.~\ref{NRGphasediag} we present the NRG phase diagram of the 3-impurity 2-channel Kondo model Eq.~(\ref{3I2CK}). Panel (a) is the phase diagram on the $\tilde{I}$-$V_{1}$ plane when $J_{11}^{2}+J_{1\bar{1}}^{2}=\left( 0.1D_{0}\right) ^{2}$, and panel (b) is the phase diagram on the $J_{11}$-$V_{1}$ plane for $J_{11}=J_{1\bar{1}}$. We also show the high- to low-temperature crossover of $S_{\text{imp}}$ and $T\chi _{\text{imp}}$ in different phases in Figs.~\ref{STchi3I2CK} and \ref{STchiKI}.

In the following we discuss the different regimes in the phase diagram. We
begin by considering each of the two relevant Kondo couplings $J_{11}$ and $%
J_{1\bar{1}}$ separately.

\emph{(i) $J_{11}>0,J_{1\bar{1}}=0$:}

The RKKY interaction is ferromagnetic ($\tilde{I}=-1$), favoring $S=3/2$;
therefore it is plausible that, at low energies, the model is reduced to a 2-channel problem with a logarithmically divergent LDOS and a single $S=3/2$ impurity. This picture for $J_{1\bar{1}}=0$ is confirmed by NRG, and the coupling between the two spin sectors $S=3/2$ and $S=1/2$ turns out to be irrelevant.

Applying our results from Sec.~\ref{sec:S2CK}, we find two different phases:
a p-h symmetric strong-coupling phase and a p-h asymmetric local moment
phase. In the symmetric phase, the effective $S=3/2$ impurity is strongly
screened by both $c_{1}$ and $c_{\bar{1}}$, and a residual $S=1/2$ impurity
emerges; thus we name it \textquotedblleft K-S\textquotedblright\ after the
Kondo screening taking place in the spin sector. In the asymmetric phase,
the effective $S=3/2$ local moment is unscreened, while both $c_{1}$ and $c_{%
\bar{1}}$ are blocked locally by strong potential scattering; we call this phase
\textquotedblleft F-ALM\textquotedblright\ after the ferromagnetic RKKY
interaction. K-S and F-ALM are both stable fixed points as discussed in Sec.~%
\ref{sec:S2CK}; they are separated by a critical boundary marked by $%
\left\vert V_{1}\right\vert /J_{11}\sim O\left( 1\right) $, in analogy to
the single-channel case\cite{PhysRevB.88.075104}. The low-temperature
thermodynamic properties of K-S (F-ALM) is identical to those of the SSC
(ALM) fixed point in the 2-channel $S=3/2$ problem with a logarithmically divergent LDOS. Therefore, at K-S, $S_{\text{imp}}\left( T=0\right) =\ln 2$ and $T\chi _{\text{imp}}=1/4$ relative
to pristine graphene; at F-ALM, $S_{\text{imp}}\left( T=0\right) =\ln 4$ and 
$T\chi _{\text{imp}}=5/4$ relative to pristine graphene.

\begin{figure}[!h]
\includegraphics[width=0.6\columnwidth]{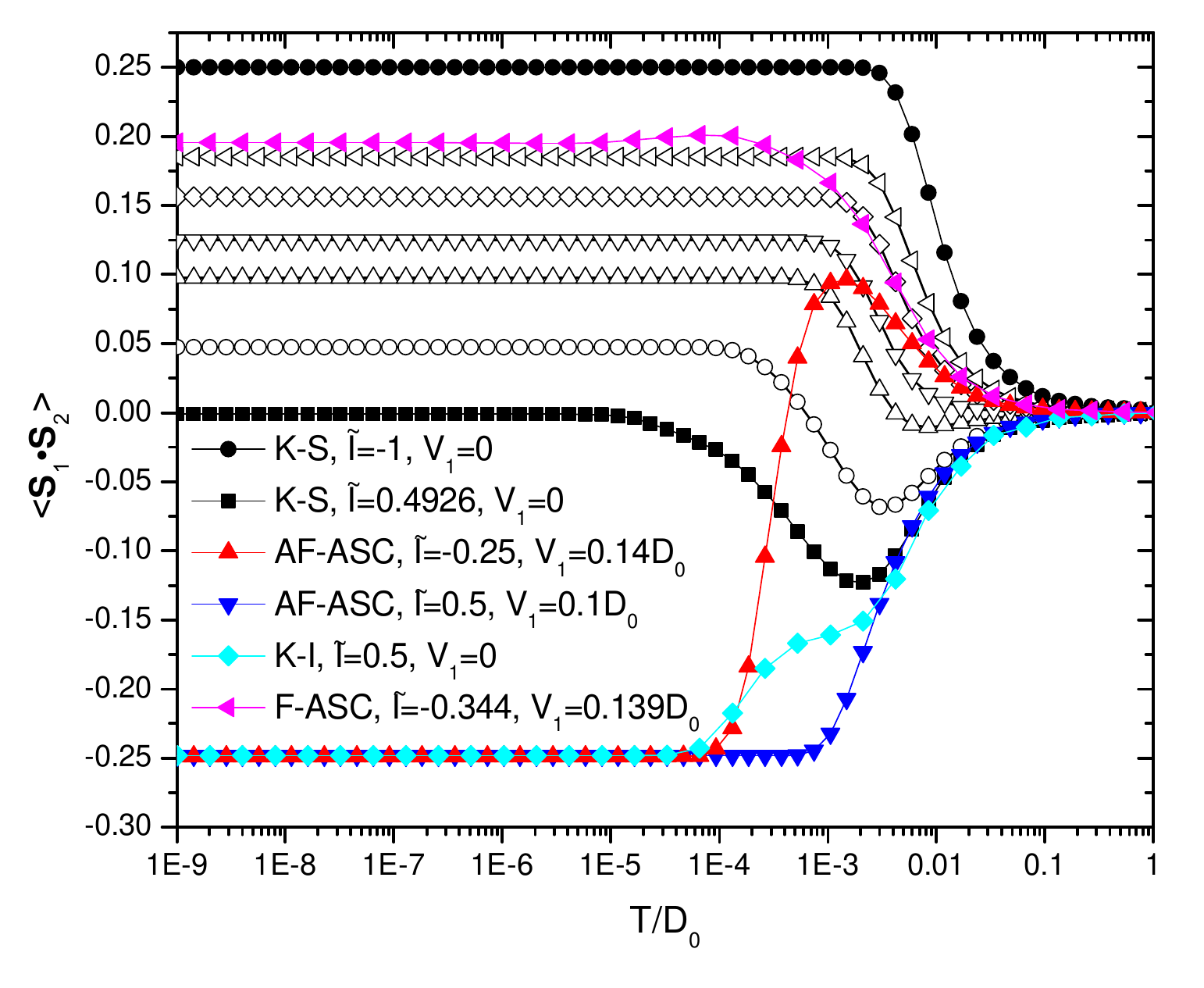} 
\caption{Equal-time impurity spin correlation $\left< \bm{S}_1 \cdot \bm{S}_2 \right>$ in the Kondo model
Eq.~(\ref{3I2CK}) with a logarithmically divergent LDOS given by Eq.~(\ref{logLDOS}).
$\Lambda=1.5D_{0}$ and $J^2_{11} + J^2_{1\bar{1}} =\left( 0.1D_{0} \right) ^{2}$.
The topmost solid black circles represent a K-S system in the maximally ferromagnetic RKKY limit, with $\tilde{I}=-1$ and $V_{1}=0$; solid black squares correspond to a K-S system close to the K-S/K-I transition, with $\tilde{I}=0.4926$ and $V_{1}=0$; the open black symbols represent K-S systems between these two limiting cases, with $V_{1}=0.02 D_{0}$, and $\tilde{I}=-0.46$, $-0.25$, $-0.04$, $0.125$ and $0.365$ from top to bottom. Solid red up-pointing triangles represent an AF-ASC system with $\tilde{I}=-0.25$ (i.e. $J_{11}=J_{1\bar{1}}$) and $V_{1}=0.14 D_{0}$, solid blue down-pointing triangles represent an AF-ASC system with $\tilde{I}=0.5$ and $V_{1}=0.1 D_{0}$, solid cyan diamonds represent a K-I system with $\tilde{I}=0.5$ and $V_{1}=0$, and solid pink left-pointing triangles represent an F-ASC system with $\tilde{I}=-0.344$ and $V_{1}=0.139 D_{0}$. F-ALM systems behave qualitatively similarly to K-S systems with $\tilde{I}<0$, and AF-ALM systems behave qualitatively similarly to AF-ASC systems with $\tilde{I}>0$.}
\label{spinspin3I2CK}
\end{figure}

While our heuristic picture correctly predicts the thermodynamic quantities and the finite-size spectrum, it would only be strictly true had we introduced by hand a strong ferromagnetic RKKY interaction Eq.~(\ref{HRKKY}) into the Hamiltonian, assuming that the RKKY strength $I$ were much greater in magnitude than any other energy scale in the problem. In reality, there might be no clear separation between the RKKY energy scale, at which the effective $S=3/2$ impurity forms, and the Kondo temperature $T_{K}$ (or $T_{P}$), at which $J_{11}$ (or $V_{1}$) flows to strong coupling. These energy scales must be extracted numerically, e.g. from thermodynamics (Fig.~\ref{STchi3I2CK}) and impurity spin correlations (Fig.~\ref{spinspin3I2CK} which we will discuss later).

\emph{(ii) $J_{1\bar{1}}>0,J_{11}=0$:}

The RKKY interaction is now antiferromagnetic ($\tilde{I}=1/2$), and favors
the $S=1/2$ state for the magnetic impurities. While one might assume that
the low-energy physics would be captured by the 2-channel $S=1/2$ Kondo model with a logarithmically divergent LDOS, we must also take note of the additional helicity degeneracy of the $S=1/2$
subspace, and the fact that $J_{1\bar{1}}$ is not a conventional Kondo
coupling.

We can again glean some insight from the constant-LDOS version of this
problem\cite{cond-mat/9607190,PhysRevLett.95.257204}. In the constant-LDOS
3-impurity 3-channel Kondo problem, when the RKKY interactions are
antiferromagnetic and $J_{1\bar{1}}$ overwhelms $J_{01}$, it has been
reported that the helicity-0 channel decouples, and the low-energy effective
model is a 2-channel spin-1/2 Kondo model with spin and isospin sectors
interchanged. (The isospin for helicities $1$ and $\bar{1}$ is defined as
usual by $\hat{I}^{z}=\frac{1}{2}\sum_{h=\pm 1,\alpha }\left( c_{h\alpha }^{\dag
}c_{h\alpha } -1/2\right)$ and $\hat{I}^{+}=\frac{1}{2}\sum_{h=\pm 1,\alpha \beta
}\epsilon _{\alpha \beta }c_{h\alpha }^{\dag }c_{\bar{h}\beta }^{\dag }$,
where $\epsilon _{\alpha \beta }$ is the Levi-Civita symbol.) To be more
concrete, the effective model involves an isospin-1/2 impurity screened by
two conduction channels, one spin-up and the other spin-down. We emphasize that such a fictitious isospin-1/2 impurity is only invoked to describe the isospin state of the conduction electrons, since the impurities themselves always possess p-h symmetry and have isospin 0 by construction. The resulting
low-energy non-Fermi-liquid fixed point, dubbed ``isospin Kondo'' in Ref.~\onlinecite{cond-mat/9607190}, is unstable against an infinitesimal
p-h symmetry breaking perturbation $V_{1}$, which plays the role of a
magnetic field in the isospin sector; the system flows to a Fermi-liquid
state in the presence of $V_{1}$.

In our problem with a logarithmically divergent LDOS, the condition of $J_{1\bar{1}}$
dominating over $J_{01}$ is always satisfied when the bare couplings are
weak, since $J_{1\bar{1}}$ is relevant while $J_{01}$ is irrelevant. We find
that, as in the constant-LDOS case, the low-energy physics is governed by a
2-channel spin-1/2 Kondo model with spin and isospin sectors interchanged;
however, as shown in Sec.~\ref{sec:S2CK}, the logarithmically divergent LDOS dictates that
the low-energy fixed point for $V_{1}=0$ should be located at strong
coupling rather than intermediate coupling. We label this strong-coupling
fixed point as \textquotedblleft K-I\textquotedblright\ due to the Kondo
effect taking place in the isospin sector. The ground state of K-I has spin
zero and isospin $1/2$, i.e. one electron is either removed from or added to half filling. This yields $S_{\text{imp}}\left( T=0\right) =\ln 2$ and $%
T\chi _{\text{imp}}\rightarrow 0$ at the K-I fixed point relative to
pristine graphene. As an intuitive picture, in the lattice version of K-I, the impurity spins form an effective spin-$1/2$, which is in turn strongly coupled to (i.e. screened by) one of the conduction channels; the other conduction channel can be either empty or doubly occupied at the lattice site closest to the impurity.

K-I is unstable against an infinitesimal $V_{1}$, which picks out a
preferred isospin state from the doublet (i.e. one electron removed from or
added to half filling). We call the resulting p-h asymmetric strong-coupling
fixed point \textquotedblleft AF-ASC\textquotedblright . In the lattice
version of AF-ASC, one conduction channel forms a spin singlet with the
impurities, and the other channel takes advantage of the local potential
scattering to lower the ground state energy; the remaining conduction
electrons are essentially free apart from constraints imposed by the Pauli
principle. With the isospin symmetry broken and the ground state a spin
singlet, we simply have $S_{\text{imp}}\left( T=0\right) =0$ and $T\chi _{%
\text{imp}}\rightarrow 0$ at AF-ASC relative to pristine graphene. Fig.~\ref{STchiKI} shows how the system flows from K-I to AF-ASC for $V_{1}/D_{0}$ ranging from from $10^{-7}$ to $0.1$.

Increasing $V_{1}$ further, for sufficiently large $\left\vert
V_{1}\right\vert /J_{11}\sim O\left( 1\right) $, we eventually encounter a
second transition to the large-$V_{1}$ fixed point, where both $c_{1}$ and $%
c_{\bar{1}}$ become blocked by strong potential scattering and the ground state electric charge differs from
half-filling by two. The antiferromagnetic RKKY interactions remain in
effect even though the impurity spins are already decoupled from the
conduction electrons, so we are left with an $S=1/2$ local moment with an
additional helicity degeneracy $h=\pm 1$. Therefore, at this fixed point
which we call \textquotedblleft AF-ALM\textquotedblright , $S_{\text{imp}%
}\left( T=0\right) =\ln 4$ and $T\chi _{\text{imp}}=1/4$ relative to
pristine graphene.

In addition to cases (i) and (ii), it is also worth mentioning that taking $J_{11}=J_{1\bar{1}}=0$ but $V_{1}\neq 0$ will lead to
another free-spin p-h asymmetric local moment fixed point (\textquotedblleft
free-ALM\textquotedblright ), where the impurity spins completely decouple and the non-normalizable zero modes vanish. This is an unstable fixed point, because even an infinitesimal RKKY interaction induced by $J_{11}$ or $J_{1\bar{1}}$ drives the impurity spins into the $S=3/2$ or the $S=1/2$ state.
Obviously, due to the three impurity spins, the free-ALM fixed point has $S_{%
\text{imp}}\left( T=0\right) =3\ln 2$ and $T\chi _{\text{imp}}=3/4$ relative
to pristine graphene. The three fixed points F-ALM, AF-ALM and free-ALM
differ only in their impurity spin states.

Cases (i) and (ii) represent the limits of maximally ferromagnetic and
maximally antiferromagnetic RKKY interaction respectively. As shown in Fig.~%
\ref{NRGphasediag}, when $J_{11}^{2}+J_{1\bar{1}}^{2}=\left( 0.1D_{0}\right)
^{2}$ and $\Lambda /D_{0}=1.5$, the K-S fixed point controls a large region
of the parameter space where $J_{11}/J_{1\bar{1}}$ is not too small and $%
V_{1}$ is not too large. The K-I phase only occurs when the p-h symmetry is
preserved and $J_{11}/J_{1\bar{1}}$ is very small (i.e. $\tilde{I}>\tilde{I}_{c0}$ where $\tilde{I}_{c0}$ is close to $1/2$)\footnote{We have obtained the phase diagram Fig.~\ref{NRGphasediag} using the NRG parameters $\Lambda_{\text{NRG}}=4$ and $z=1$ in the conduction band discretization scheme of Ref.~\onlinecite{PhysRevB.33.7871,*PhysRevB.49.11986}. With these parameters we find that the system flows to the K-I fixed point when $J_{11}/J_{1\bar{1}}<0.0060$, or $\tilde{I}>\tilde{I}_{c0}\approx 0.494$. Changing $\Lambda_{\text{NRG}}$ and $z$ shifts the phase boundaries only slightly; however, where (and indeed, whether) the K-I fixed point is numerically attainable depends sensitively on the value of $z$. In the continuum limit $\Lambda_{\text{NRG}} \to 1$, it is not clear to us whether $\tilde{I}_{c0}$ is exactly $1/2$, i.e. whether the K-I phase exists only in the limiting case $J_{11}=0$.}, but its direct
descendant-- the AF-ASC phase-- becomes progressively more important at
intermediate values of $V_{1}$ when the RKKY interactions are not too
strongly ferromagnetic. Finally, at very large values of $V_{1}$, the F-ALM
and AF-ALM phases come into play, separated approximately by the $\tilde{I}=0
$ line.

While the phase boundaries between K-S, AF-ASC and F-ALM seemingly meet at a single tricritical point on the $\tilde{I}$-$V_{1}$ plane, a more careful survey of the parameter space reveals the presence of another phase in a small area separating these three phases. The ground state in this phase is again p-h asymmetric with one electron removed from or added to half filling. However, in contrast to the AF-ASC phase, the ground state has a residual $S=1$ impurity, consistent with the ferromagnetic RKKY interaction ($\tilde{I}<0$); thus we name this phase \textquotedblleft F-ASC\textquotedblright . Because one of the two conduction channels couples to the impurity spins and the other is blocked by potential scattering, the $S=1$ residual impurity in the ground state can form in two distinct helicities. Combined with the threefold degeneracy of the spin state, this helicity degeneracy gives an impurity entropy of $S_{\text{imp}}\left( T=0\right) =\ln 6$ relative to pristine graphene. The impurity susceptibility from the residual impurity is simply $T\chi _{\text{imp}}=2/3$.

It is also enlightening to look at the correlation between the impurity
spins. In Fig.~\ref{spinspin3I2CK} we plot the expectation value $%
\left\langle \bm{S}_{1}\cdot \bm{S}_{2}\right\rangle $ as a function of
temperature at various points in the phase diagram. At high temperatures its
sign is simply opposite to that of $\tilde{I}$, as perturbation theory
predicts. At low temperatures, $\left\langle \bm{S}_{1}\cdot \bm{S}%
_{2}\right\rangle $ takes the minimum possible value $-1/4$ in the K-I, AF-ASC
and AF-ALM phases, and the maximum possible value $1/4$ in the F-ALM phase;
these results are consistent with our previous analysis that the impurity
spins form a spin-$1/2$ in the K-I, AF-ASC and AF-ALM phases, and a spin-$3/2$ in
the F-ALM phase. The low-temperature spin correlation in the K-S phase is
more interesting: $\left\langle \bm{S}_{1}\cdot \bm{S}_{2}\right\rangle $
varies smoothly from $1/4$ to $0$ as $\tilde{I}$ increases from $-1$ to $%
\tilde{I}_{c0}$ (recall that the K-I phase takes over for $\tilde{I}>\tilde{I%
}_{c0}$ and $V_{1}=0$), and is only weakly dependent on $V_{1}$ as long as
the system remains in the K-S phase. Therefore, away from the strongly
ferromagnetic limit $\tilde{I}=-1$, the impurity spins generally form a
superposition of spin-$3/2$ and spin-$1/2$ states in the K-S phase, even
though the residual spin is always $1/2$, with two electrons participating
in screening. A similar statement can be made for the F-ASC phase: the impurity spins form a superposition of spin-$3/2$ and spin-$1/2$ states, which couples to one conduction electron to produce a residual spin-$1$.

In the Kondo model obtained by Schrieffer-Wolff transforming the 3-impurity
Anderson model, $J_{11}=J_{1\bar{1}}$ and $\tilde{I}=-1/4$; in this case it
is clear from Fig.~\ref{NRGphasediag} that K-S and F-ALM control small- and
large-$V_{1}$ physics respectively as in the strongly ferromagnetic RKKY
limit, whereas the AF-ASC phase sets in for intermediate values of $V_{1}$
as in the strongly antiferromagnetic RKKY limit. For $J_{11}=J_{1\bar{1}%
}=0.1D_{0}/\sqrt{2}$, the AF-ASC phase occurs for $0.126<\left\vert
V_{1}\right\vert /D_{0}<0.152$; this cannot be realized by a
Schrieffer-Wolff transformation which requires $\left\vert V_{1}\right\vert
/J_{11}\lesssim 3/4$. However, panel (b) of Fig.~\ref{NRGphasediag} indicates that when $%
J_{11}=J_{1\bar{1}}\lesssim 10^{-4}D_{0}$, the critical value of $\left\vert
V_{1}\right\vert /J_{11}$ at the K-S/AF-ASC transition can be reduced
dramatically, well below $3/4$. This strongly suggests that both K-S and AF-ASC (or at least their generalizations) are accessible in an Anderson model, although F-ALM and AF-ALM may still be out of reach. We will confirm this picture in Sec.~\ref{sec:Anderson}.

\begin{figure}[!h]
\includegraphics[width=1\columnwidth]{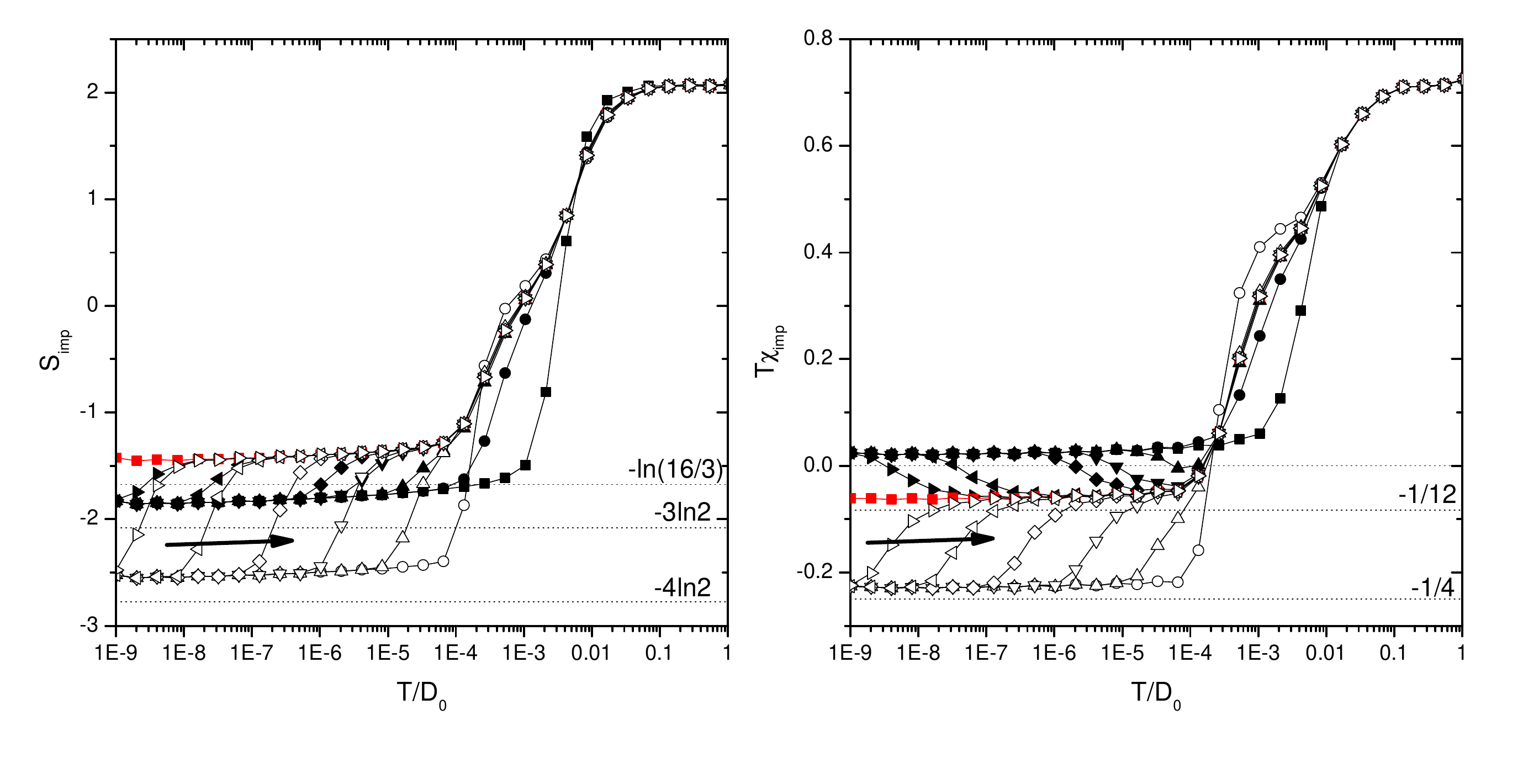} 
\caption{$S_{\text{imp}}$ and $T \chi_{\text{imp}}$ versus $T$ in the vicinity of the K-S/AF-ASC phase transition of the Kondo model Eq.~(\ref{3I2CK}) with a logarithmically divergent LDOS given by Eq.~(\ref{logLDOS}). $\Lambda=1.5D_{0}$; $J_{11}= J_{1\bar{1}} =0.1D_{0}/\sqrt{2}$, and different curves correspond to different values of $V_{1}$. The critical value $\left\vert V_{1}\right\vert /D_{0}=v_{c} \approx 0.126$ is shown in solid red squares; solid black symbols are in the K-S phase, and open black symbols are in the AF-ASC phase, with $\left| V_{1}/D_{0}-v_{c}\right| =10^{-7}$ (right-pointing triangles), $10^{-6}$ (left-pointing triangles), $10^{-5}$ (diamonds), $10^{-4}$ (down-pointing triangles), $10^{-3}$ (up-pointing triangles), $0.01$ (circles) and $0.1$ (squares, only for K-S) in the direction of the arrow. Data in these figures is not $z$-averaged and therefore contains spurious oscillations.}
\label{STchi3I2CKtr}
\end{figure}

\begin{figure}[!h]
\includegraphics[width=0.6\columnwidth]{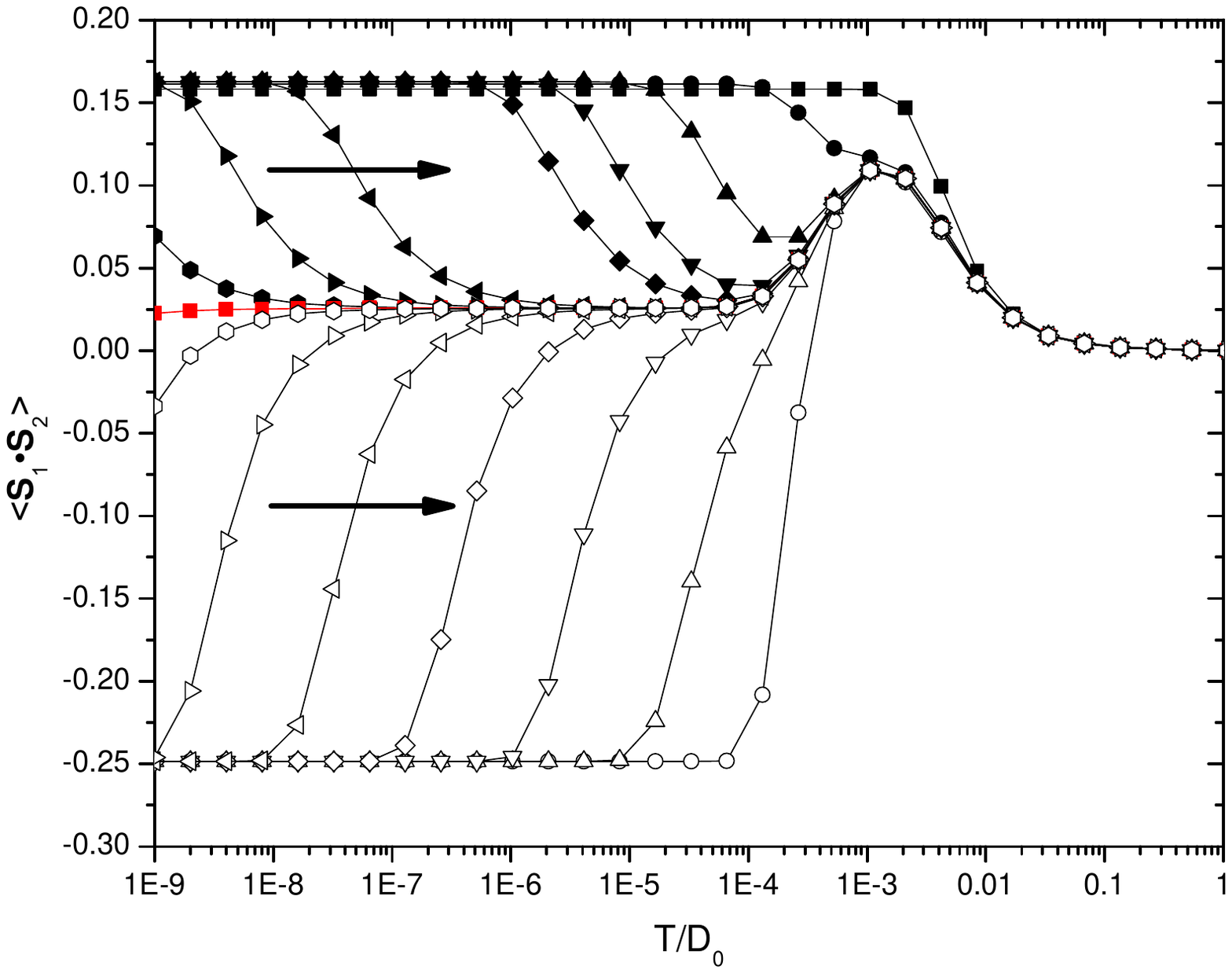} 
\caption{Equal-time impurity spin correlation $\left< \bm{S}_1 \cdot \bm{S}_2 \right>$ in the vicinity of the K-S/AF-ASC phase transition of the Kondo model Eq.~(\ref{3I2CK}) with a logarithmically divergent LDOS given by Eq.~(\ref{logLDOS}).
$\Lambda=1.5D_{0}$; $J_{11}= J_{1\bar{1}} =0.1D_{0}/\sqrt{2}$, and different curves correspond to different values of $V_{1}$. The critical value $V_{1}/D_{0}=v_{c} \approx 0.126$ is shown in solid red squares; solid black symbols are in the K-S phase, and open black symbols are in the AF-ASC phase, with $\left| V_{1}/D_{0}-v_{c}\right| =10^{-8}$ (hexagons), $10^{-7}$ (right-pointing triangles), $10^{-6}$ (left-pointing triangles), $10^{-5}$ (diamonds), $10^{-4}$ (down-pointing triangles), $10^{-3}$ (up-pointing triangles), $0.01$ (circles) and $0.1$ (squares, only for K-S) in the direction of the arrows.}
\label{spinspin3I2CKtr}
\end{figure}

We now examine the K-S/AF-ASC transition, motivated by the observation that
a transition of similar nature may exist in an Anderson model. For $J_{11}=J_{1\bar{1}}=0.1D_{0}/\sqrt{2}$, the K-S/AF-ASC transition takes place at $\left\vert V_{1}\right\vert /D_{0} = v_{c}\approx 0.126$. Fig.~\ref{STchi3I2CKtr} shows the high- to low-temperature crossover of $S_{\text{imp}}$ and $T\chi _{\text{imp}}$ as $V_{1}/D_{0}-v_{c}$ sweeps from $-0.01$ to $0.1$, and Fig.~\ref{spinspin3I2CKtr} shows
the corresponding behavior of $\left\langle \bm{S}_{1}\cdot \bm{S}%
_{2}\right\rangle $. We can explain the unstable critical point separating
the two phases as a simple level crossing of the spin-$1/2$ doublet ground
states in the K-S phase and the spin singlet ground state in the AF-ASC
phase. The doublet and the singlet do not mix, as they belong in different
sectors of the Hilbert space. At the critical point, $S_{\text{imp}}\left(
T=0\right) =-\ln \left( 16/3\right) $ relative to the 4-site-vacancy graphene,
which is greater than the AF-ASC value by $\ln 3$, a signature of the
accidental degeneracy. Moreover, the values of both $T\chi _{\text{imp}}$
and $\left\langle \bm{S}_{1}\cdot \bm{S}_{2}\right\rangle $ at the critical
point can be obtained as weighted averages of the K-S value (with weight $2/3
$) and the AF-ASC value (with weight $1/3$). We mention that this simple
level crossing picture also applies to the AF-CR critical point in the
single-channel case\cite{PhysRevB.88.075104}.

Finally we briefly discuss the effect of the helicity-0 channel. We assume that
the irrelevant couplings $J_{00}$, $J_{01}$ and $V_{0}$ are not too large
compared to the relevant couplings, so that the intermediate-coupling phase transition identified in Ref.~\onlinecite{PhysRevLett.64.1835} does not take place. In most cases, these irrelevant
couplings merely shift the phase boundaries without modifying the phase
diagram qualitatively. A notable exception is the K-I fixed point. In the
p-h symmetric strongly antiferromagnetic RKKY limit $J_{11}=V_{1}=0$ and $%
J_{1\bar{1}}\neq 0$, we find that $V_{0}$ by itself or the combination of $%
J_{00}$ and $J_{01}$ does not affect the low-energy K-I behavior. However,
the combination $J_{00}\neq 0$, $J_{01}=0$ and $V_{0}\neq 0$ drives the
system into the K-S phase, while the combination $J_{00}=0$, $J_{01}\neq 0$
and $V_{0}\neq 0$\ drives the system into the AF-ASC phase. Therefore, as
with its constant-LDOS analog\cite{cond-mat/9607190}, the K-I phase is
highly fragile against p-h symmetry breaking, and unlikely to be
experimentally observed.

We summarize our results on the fixed points of the 3-impurity 3-channel Kondo model Eq.~(\ref{3I3CK}) in Table~\ref{listSTchi}. The results for $S_{\text{imp}}\left( T=0\right) $ and $T\chi_{\text{imp}}$ are in full agreement with Figs.~\ref{STchi3I2CK} and \ref{STchiKI} upon changing the reference system to graphene with a 4-site vacancy, i.e. subtracting $2\ln 4$ from $S_{\text{imp}}$ and $1/4$ from $T\chi _{\text{imp}}$.

\begin{table}[htb]
\caption{Properties of various fixed points of the 3-impurity 3-channel Kondo model Eq.~(\ref{3I3CK}). The charge number is measured relative to half-filling; we assume a negative charge if the p-h symmetry is explicitly broken by potential scattering. Pristine graphene is chosen as the reference system for $S_{\text{imp}}$ and $\chi _{\text{imp}}$. Spin $\left( 1/2 \right) _{3}$ refers to three independent spin-$1/2$ impurities; $0< \left< \bm{S}_1 \cdot \bm{S}_2 \right> \leq 1/4$ in the K-S phase. The results for K-S and AF-ASC also apply to the 5-atom-cluster Anderson model (see Sec.~\ref{sec:Anderson}), with the exception that $-1/4\leq \left< \bm{S}_1 \cdot \bm{S}_2 \right> <0$ in the AF-ASC phase of the Anderson model.} \label{listSTchi} 
\begin{tabular}{C{4cm}C{1.5cm}C{2cm}C{1.5cm}C{1.5cm}C{2cm}C{1.5cm}C{1.5cm}C{1.5cm}}
\hline\hline \\[-1em]
Fixed point & Stability & Non-normalizable zero modes & Spin & Charge & Helicity degeneracy & $S_{\text{imp}}$ & $T\chi _{\text{imp}}$ & $\left< \bm{S}_1 \cdot \bm{S}_2 \right>$\\ \\[-1em]
\hline \\[-0.5em]
    free-spin symmetric local moment (LM) & unstable & $2$ & $\left(\frac{1}{2}\right)_{3}$ & $0$ & - & $7\ln 2$ & $1$ & $0$\\  \\[-0.5em]  
    free-spin asymmetric local moment (free-ALM) & unstable & $0$ & $\left(\frac{1}{2}\right)_{3}$ & $-2$ & - & $3\ln 2$ & $\frac{3}{4}$ & $0$\\  \\[-0.5em]  
    ferromagnetic symmetric Kondo (K-S) & stable & $0$ & $\frac{1}{2}$ & $0$ & - & $\ln 2$ & $\frac{1}{4}$ & $\in \left(0,\frac{1}{4}\right]$\\  \\[-0.5em]  
    ferromagnetic asymmetric local moment (F-ALM) & stable & $0$ & $\frac{3}{2}$ & $-2$ & - & $\ln 4$ & $\frac{5}{4}$ & $\frac{1}{4}$\\  \\[-0.5em]  
    antiferromagnetic symmetric isospin Kondo (K-I) & unstable & $0$ & $0$ & $\pm 1$ & - & $\ln 2$ & $0$ & $-\frac{1}{4}$\\  \\[-0.5em]  
    antiferromagnetic asymmetric strong-coupling (AF-ASC) & stable & $0$ & $0$ & $-1$ & - & $0$ & $0$ & $-\frac{1}{4}$\\  \\[-0.5em]  
    antiferromagnetic asymmetric local moment (AF-ALM) & stable & $0$ & $\frac{1}{2}$ & $-2$ & $2_{\text{spin}}$ & $\ln 4$ & $\frac{1}{4}$ & $-\frac{1}{4}$\\  \\[-0.5em]  
    ferromagnetic asymmetric strong-coupling (F-ASC) & stable & $0$ & $1$ & $-1$ & $2_{\text{channel}}$ & $\ln 6$ & $\frac{2}{3}$ & $\in \left(0,\frac{1}{4} \right)$\\  \\[-0.5em]  

\hline\hline
\end{tabular}
\end{table}

\subsection{Logarithmic LDOS with an infrared cutoff\label{sec:LDOScutoff}}

In a more realistic model of the graphene sheet, the small next-nearest-neighbor hopping $t^{\prime}$ between carbon atoms replaces the zero mode associated with a vacancy by a number of quasi-localized states shifted slightly away from the Dirac point. While the LDOS remains strongly enhanced near the energies of these quasi-localized states, it is no longer logarithmically divergent\cite{PhysRevB.77.115109}. Nevertheless, following Ref.~\onlinecite{PhysRevB.88.075104}, we can fine-tune the Fermi energy to the vacancy-induced peak of the LDOS, and heuristically model the effect of a next-nearest-neighbor hopping by imposing an infrared energy cutoff $X$ on the LDOS.

\begin{figure}[!h]
\includegraphics[width=1\columnwidth]{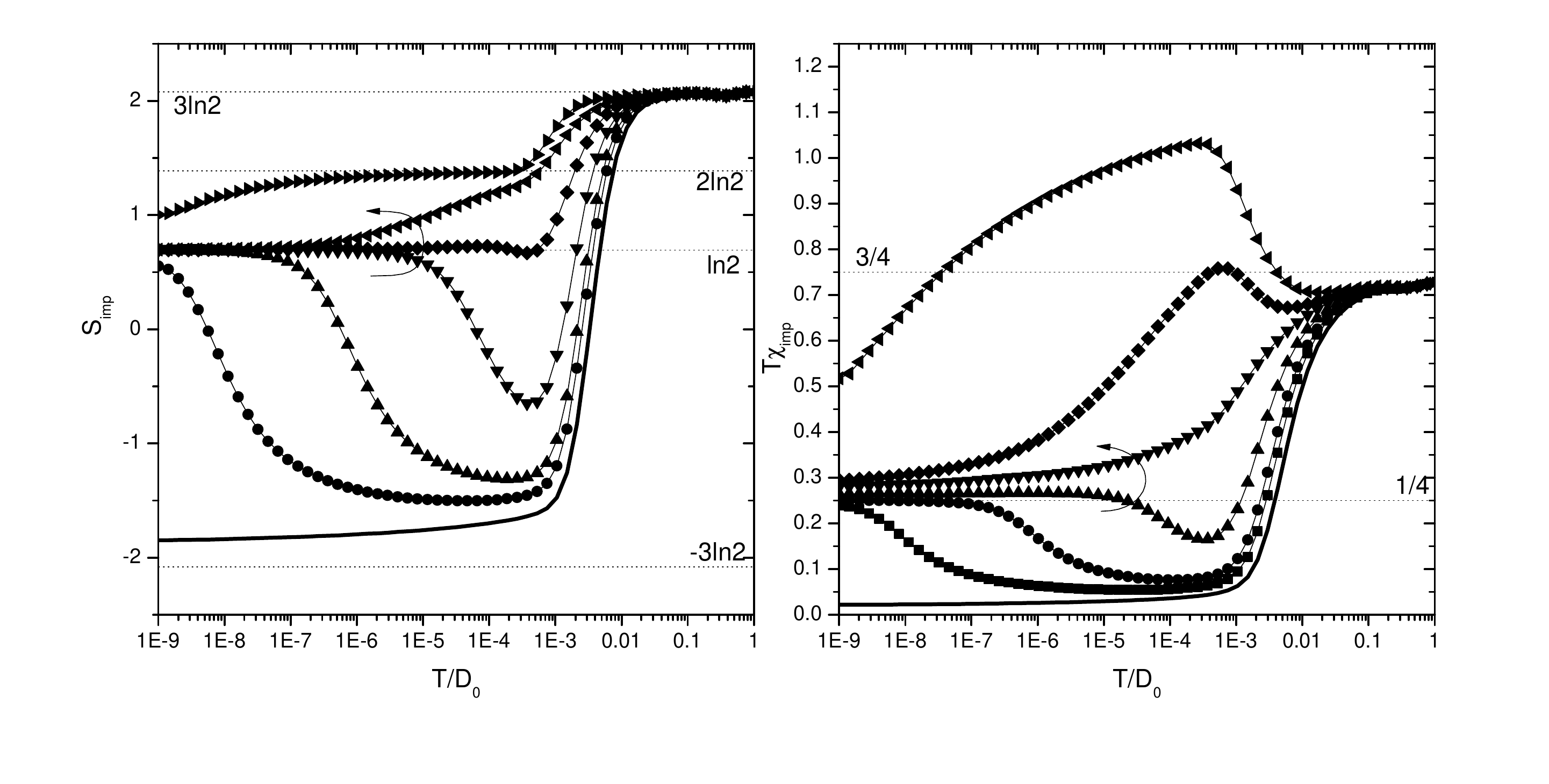} 
\caption{$S_{\text{imp}}$ and $T \chi_{\text{imp}}$ in the Kondo model Eq.~(\ref{3I2CK}) with different infrared cutoffs $X$ imposed on the logarithmically divergent LDOS; at energies below $X$ the original LDOS Eq.~(\ref{logLDOS}) is replaced by its value at $X$. $\Lambda=1.5D_{0}$; $\left( J_{11}, J_{1\bar{1}}, V_{1} \right)=\left( 0.1/\sqrt{2}, 0.1/\sqrt{2}, 0.02\right) D_{0}$. In each plot the thick line corresponds to $X=0$, and $X$ takes the following values along the direction of the arrow: $10^{-8}$, $10^{-6}$, $10^{-4}$, $10^{-3}$, $10^{-2.5}$ and $0.01$.}
\label{STchi3I2CKDOS}
\end{figure}

\begin{figure}[!h]
\includegraphics[width=0.6\columnwidth]{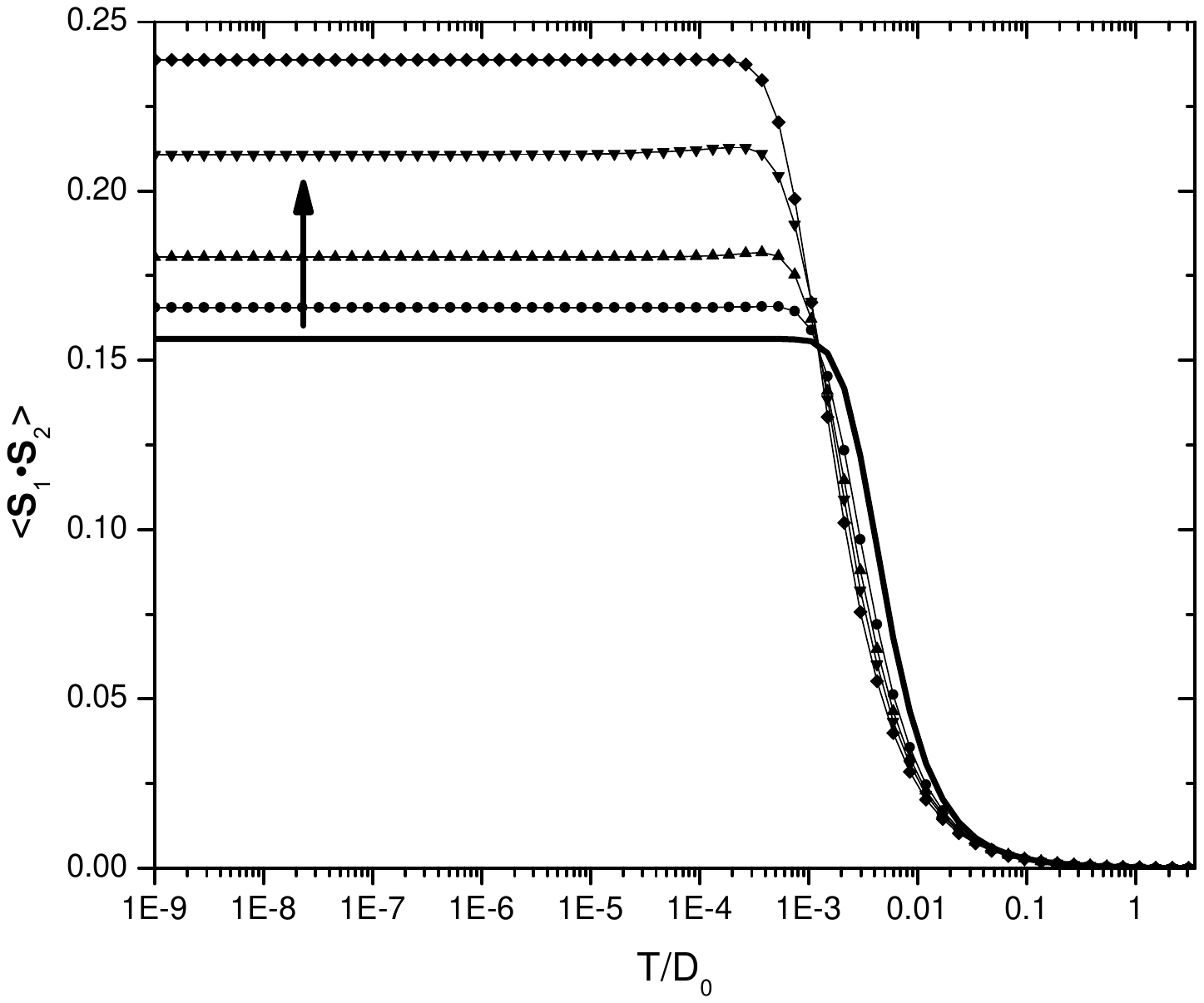} 
\caption{$\left< \bm{S}_1 \cdot \bm{S}_2 \right>$ in the Kondo model Eq.~(\ref{3I2CK}) with different infrared cutoffs $X$ on the logarithmically divergent LDOS Eq.~(\ref{logLDOS}). $\Lambda=1.5D_{0}$; $\left( J_{11}, J_{1\bar{1}}, V_{1} \right)=\left( 0.1/\sqrt{2}, 0.1/\sqrt{2}, 0.02\right) D_{0}$. The thick line corresponds to $X=0$, and $X$ takes the following values along the direction of the arrow: $10^{-4}$, $10^{-3}$, $10^{-2.5}$ and $0.01$.}
\label{spinspin3I2CKDOS}
\end{figure}

To be concrete, we replace $\rho \left( \omega \right) $ by a constant $\rho \left( X\right) $ for $\left\vert \omega \right\vert <X$ in Eq.~(\ref{logLDOS}), so that the LDOS becomes a large constant value at small energies. Whereas such a \textquotedblleft hard\textquotedblright\ cutoff scheme is slightly different from the \textquotedblleft soft\textquotedblright\ cutoff adopted in Ref.~\onlinecite{PhysRevB.88.075104}, we can verify that the two cutoff schemes do not lead to qualitatively different results. In the $X\rightarrow 0$ limit our LDOS recovers the logarithmic divergence in the $t^{\prime}=0$ case. For our choice of the ultraviolet energy cutoff in the LDOS $\Lambda=1.5D_{0}$, we find that $X/D_{0}\sim 0.01$ reproduces the LDOS peak height found by solving the tight-binding model\cite{PhysRevB.77.115109} with the experimentally estimated value $t^{\prime}\approx 0.1t$\cite{PhysRevB.88.165427}. Fig.~\ref{STchi3I2CKDOS} shows $S_{\text{imp}}$ and $T\chi _{\text{imp}}$ in the K-S phase of the 3-impurity 2-channel Kondo model Eq.~(\ref{3I2CK}) as we increase the infrared cutoff $X$ from $0$ to $10^{-2}D_{0}$, and Fig.~\ref{spinspin3I2CKDOS} shows the corresponding behavior of $\left\langle \bm{S}_{1}\cdot \bm{S}_{2}\right\rangle $.

At sufficiently low energies $T\ll X$, there is no longer any contribution
to $S_{\text{imp}}$ and $T\chi _{\text{imp}}$ from the non-normalizable zero
mode, so both $S_{\text{imp}}$ and $T\chi _{\text{imp}}$ recover their
values in the constant-LDOS spin-$3/2$ 2-channel Kondo problem in this
limit, namely $\ln 2$ and $1/4$. Also, when $X$ is far smaller than $%
T_{K}^{X=0}$ (the Kondo temperature at $X=0$), the RG flow is still towards
the K-S fixed point in the energy range $X\ll T\ll T_{K}^{X=0}$. These
features are also present in the single-channel case\cite{PhysRevB.88.075104}%
. On the other hand, as $X$ increases, Figs.~\ref{STchi3I2CKDOS} and \ref%
{spinspin3I2CKDOS} both show an increase of the total impurity spin at low
energies, which is particularly pronounced for larger $X$\ ($X\gtrsim
10^{-4}D_{0}$). When $X=10^{-2}D_{0}$, $\left\langle \bm{S}_{1}\cdot \bm{S}%
_{2}\right\rangle $ is close to $1/4$, so that the effective spin is almost
completely a spin-$3/2$; this effective spin controls the physics across a
wide range of energies between its formation around $T\sim 10^{-3}D_{0}$ and
the onset of screening below $T\sim 10^{-6}D_{0}$.

\section{Anderson model\label{sec:Anderson}}

Having discussed the 3-impurity Kondo model in great detail, in this section
we turn back to our initial approximation of ignoring the hydrogen impurity and the central A site.
This approximation is based on an infinite hydrogen-carbon coupling strength. Realistic estimates put the hydrogen-carbon coupling around twice the nearest-neighbor hopping between carbon atoms\cite{PhysRevLett.101.196803,PhysRevLett.105.056802,PhysRevLett.110.246602}; it is therefore important to check whether our intuitions from the
Kondo model carry over to the full 5-atom cluster Anderson model. This Anderson model is also represented by the
Hamiltonian

\begin{equation}
H=H_{\text{vac}}+H_{\text{hyb}}+H_{\text{imp}}^{\prime }\text{;}
\label{Anderson}
\end{equation}%
$H_{\text{vac}}$ and $H_{\text{hyb}}$ are already given in Eqs.~(\ref{bathH}%
) and (\ref{hybH}). The impurity Hamiltonian has additional terms:

\begin{eqnarray}
H_{\text{imp}}^{\prime } &=&\sum_{j=1}^{3}\left( \epsilon
_{b}n_{b,j}+Un_{b,j\uparrow }n_{b,j\downarrow }\right) +\epsilon
_{a}n_{a,0}+Un_{a,0\uparrow }n_{a,0\downarrow }+\epsilon
_{H}n_{H}+U_{H}n_{H\uparrow }n_{H\downarrow }  \notag \\
&&-\left[ \left( t_{H}g^{\dag }+t_{0}b_{1}^{\dag }+t_{0}b_{2}^{\dag
}+t_{0}b_{3}^{\dag }\right) a_{0}+\text{h.c.}\right] \text{.}
\end{eqnarray}%
Here we have labeled the hydrogen impurity as $g$, the central A site $a(%
\vec{0})$ as $a_{0}$, and defined $n_{a,0\alpha }\equiv a_{0\alpha }^{\dag
}a_{0\alpha }$, $n_{H\alpha }\equiv g_{\alpha }^{\dag }g_{\alpha }$, $%
n_{a,0}=n_{a,0\uparrow }+n_{a,0\downarrow }$ and $n_{H}=n_{H\uparrow
}+n_{H\downarrow }$. Compared to Eq.~(\ref{Anderson0}), Eq.~(\ref{Anderson})
has a number of new coupling constants: the on-site chemical potentials $%
\epsilon _{a}$ and $\epsilon _{H}$, the Hubbard interaction on the hydrogen
impurity $U_{H}$, the hydrogen-carbon coupling strength $t_{H}$, and the
nearest-neighbor hopping between the central A site and its nearest
neighbors $t_{0}$. $t_{0}$ is generally different from $t$ due to the
presence of the hydrogen impurity\cite{PhysRevB.85.115405}. We also note that, due to the two additional impurity sites, Eq.~(\ref{Anderson}) cannot be mapped to the simple Kondo model Eq.~(\ref{3I3CK}) even in the limit $U\sim \left\vert \epsilon _{b}\right\vert \gg t$.

Because of the immense size of the parameter space, we now focus on the
experimentally relevant regime where all parameters (including $U$ and $U_{H}
$) are of comparable magnitudes. We also continue to neglect the helicity-0
channel with a linear LDOS and keep only the helicity-$\pm 1$ channels with
a logarithmically divergent LDOS. Under these assumptions, quite generally, we find that
the ground state of the system is a charge-neutral spin doublet state when
the p-h symmetry breaking terms are weak, or a spin singlet state with
charge $+1$ (or $-1$) when the p-h symmetry breaking terms are strong. For
reasons that will become clear later we again call these two phases K-S and
AF-ASC respectively.

To be more concrete, we choose $U=t$, $U_{H}=2.8t$, $\epsilon _{b}=-U/2$, $%
t_{H}=2t$ and $t_{0}=0.6t$. When $\epsilon _{a}=-U/2-0.7t$
and $\epsilon _{H}=-U_{H}/2$, a Hartree-Fock calculation of the LDOS in
the full Anderson-Hubbard model (where the Hubbard interaction is also
included in $H_{\text{vac}}$) has been reported to agree qualitatively with
density-functional theory results\cite{PhysRevB.85.115405}. However, we
argue that the p-h symmetry breaking should be stronger on the hydrogen
impurity than on the central A site. In the following we therefore let $%
\epsilon _{a}=-U/2$ and vary $\epsilon _{H}$ instead, placing the p-h
symmetry breaking term on the hydrogen impurity. We nevertheless note that
our results below are not qualitatively modified by the presence of
additional p-h symmetry breaking terms on the central A site or its nearest-neighboring B sites, as long as these terms are not too large compared to $t$.

\begin{figure}[!h]
\includegraphics[width=1\columnwidth]{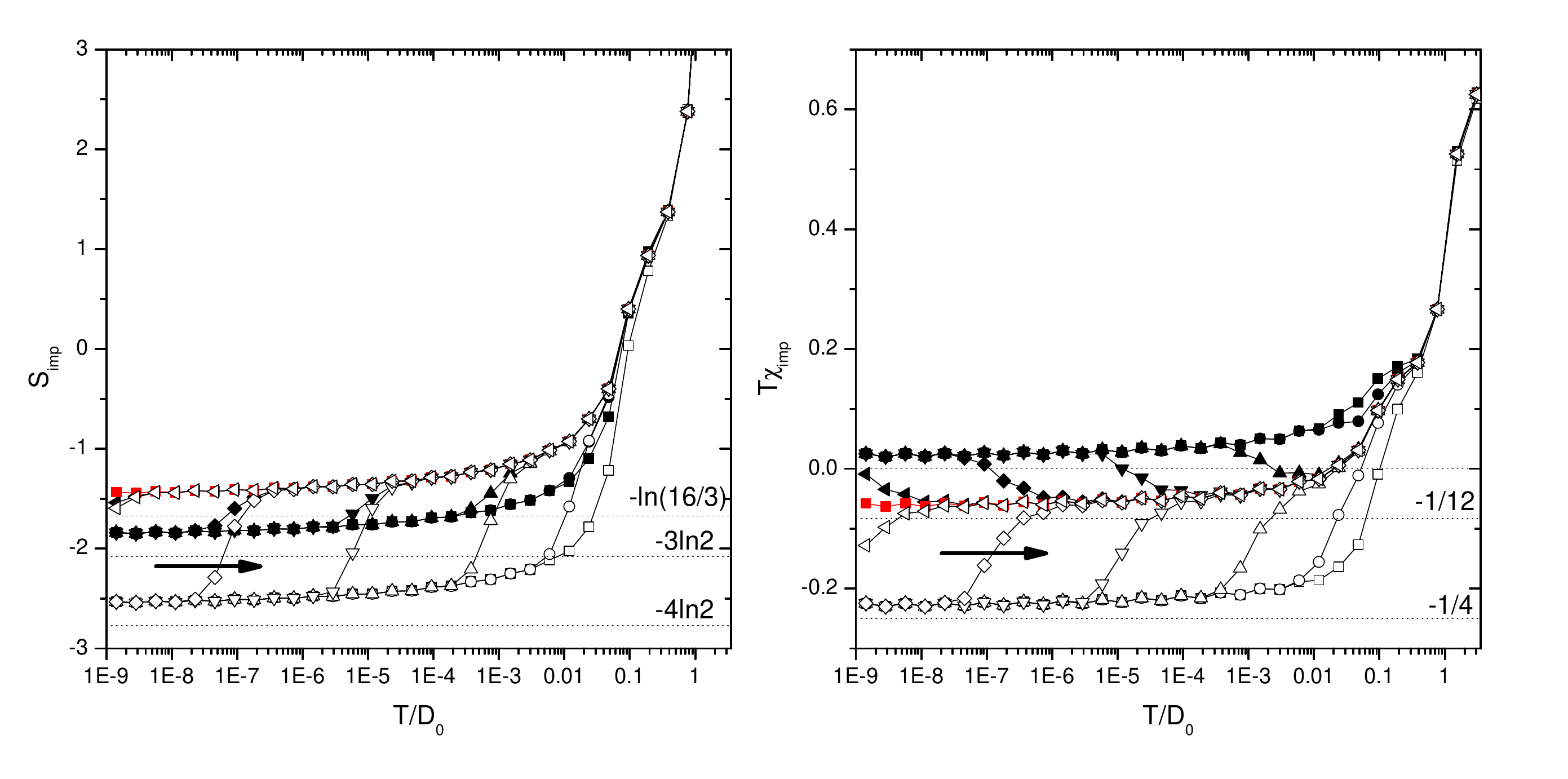} 
\caption{$S_{\text{imp}}$ and $T \chi_{\text{imp}}$ versus $T$ in the vicinity of the K-S/AF-ASC phase transition of the Anderson model Eq.~(\ref{Anderson}) with a logarithmically divergent LDOS given by Eq.~(\ref{logLDOS}). $\Lambda=1.5D_{0}$; $U=t$, $U_{H}=2.8t$, $\epsilon _{a}=\epsilon _{b}=-U/2$, $t_{H}=2t$, $t_{0}=0.6t$, and different curves correspond to different values of $\epsilon _{H}$. The critical value $\left( \epsilon _{H}+U_{H}/2\right) /t=\tilde{\epsilon }_{c} \approx -1.055$ is shown in solid red squares; solid black symbols are in the K-S phase, and open black symbols are in the AF-ASC phase. $\left( \epsilon _{H}+U_{H}/2\right) /t=0$ for solid black squares, $-0.75$ for solid circles, $-1.25$ for open circles and $-2$ for open squares. Closer to the transition, $\left| \left( \epsilon _{H}+U_{H}/2\right) /t-\tilde{\epsilon }_{c}\right| =10^{-8}$ (left-pointing triangles), $10^{-6}$ (diamonds), $10^{-4}$ (down-pointing triangles), $0.01$ (up-pointing triangles) in the direction of the arrows. Data in these figures is not $z$-averaged and therefore contains spurious oscillations.}
\label{STchi5I2CAtr}
\end{figure}

\begin{figure}[!h]
\includegraphics[width=1\columnwidth]{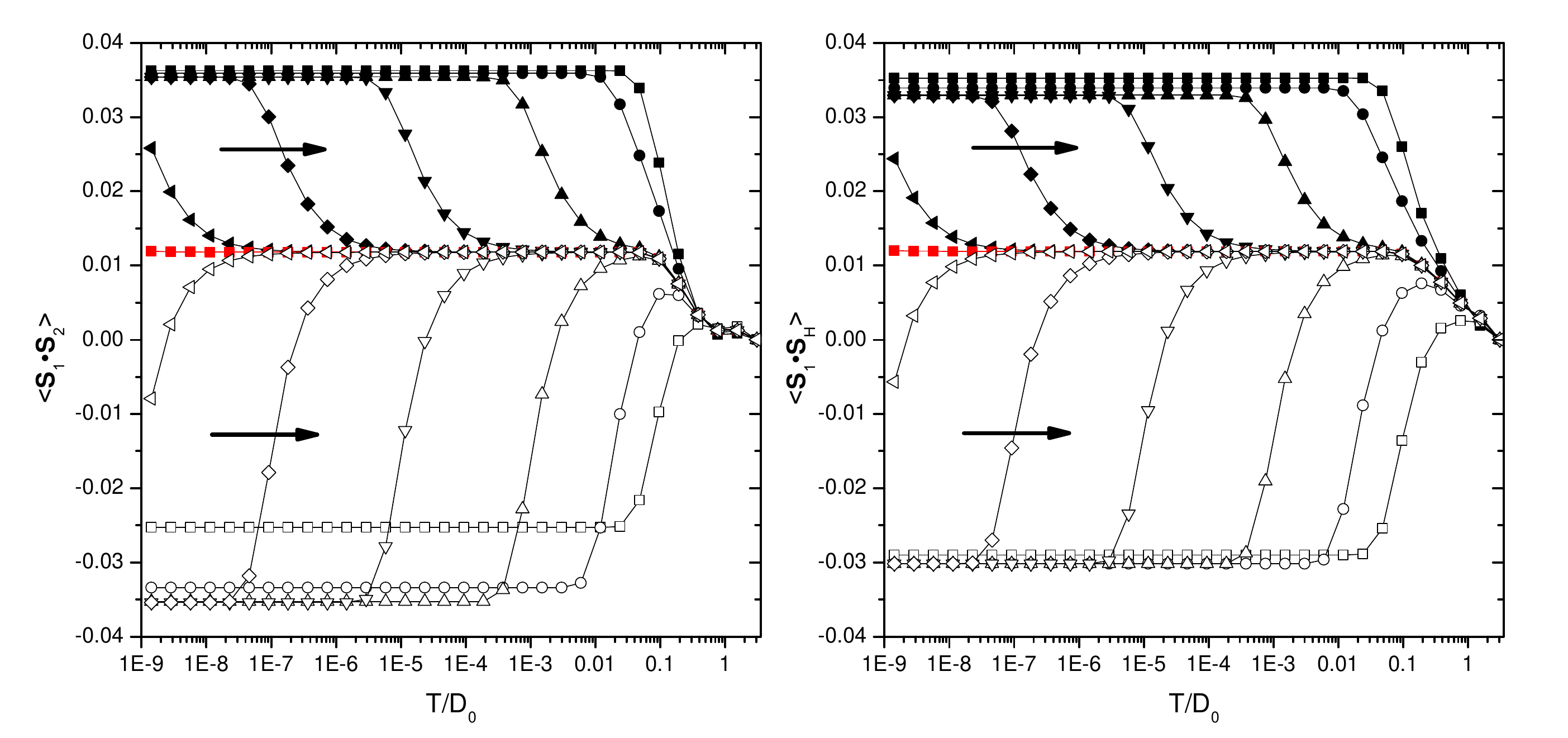} 
\caption{Equal-time impurity spin correlations $\left< \bm{S}_1 \cdot \bm{S}_2 \right>$ and $\left< \bm{S}_1 \cdot \bm{S}_H \right>$ in the vicinity of the K-S/AF-ASC phase transition of the Anderson model Eq.~(\ref{Anderson}) with a logarithmically divergent LDOS given by Eq.~(\ref{logLDOS}).
$\Lambda=1.5D_{0}$; $U=t$, $U_{H}=2.8t$, $\epsilon _{a}=\epsilon _{b}=-U/2$, $t_{H}=2t$, $t_{0}=0.6t$, and different curves correspond to different values of $\epsilon _{H}$. The critical value $\left( \epsilon _{H}+U_{H}/2\right) /t=\tilde{\epsilon }_{c} \approx -1.055$ is shown in solid red squares; solid black symbols are in the K-S phase, and open black symbols are in the AF-ASC phase. $\left( \epsilon _{H}+U_{H}/2\right) /t=0$ for solid black squares, $-0.75$ for solid circles, $-1.25$ for open circles and $-2$ for open squares. Closer to the transition, $\left| \left( \epsilon _{H}+U_{H}/2\right) /t-\tilde{\epsilon }_{c}\right| =10^{-8}$ (left-pointing triangles), $10^{-6}$ (diamonds), $10^{-4}$ (down-pointing triangles), $0.01$ (up-pointing triangles) in the direction of the arrows.}
\label{spinspin5I2CAtr}
\end{figure}

Fig.~\ref{STchi5I2CAtr} shows the typical behavior of $S_{\text{imp}}$ and $%
T\chi _{\text{imp}}$ in the K-S and the AF-ASC phases and across the phase transition in between. The K-S/AF-ASC transition occurs at $\left( \epsilon _{H}+U_{H}/2\right) /t=\tilde{\epsilon }_{c} \approx -1.055$, and we tune $\left( \epsilon _{H}+U_{H}/2\right) /t$ from $0$ to $-2$. The low-temperature behavior of $S_{\text{imp}}$ and $T\chi _{\text{imp}}$ is completely identical to that of their Kondo model counterparts, not only inside each
phase but also at the transition (cf. Figs.~\ref{STchi3I2CK} and \ref{STchi3I2CKtr}). The K-S/AF-ASC transition can again be explained as a simple level crossing of the spin-$1/2$ doublet ground states in the K-S phase and the spin singlet ground state in the AF-ASC phase. It is also interesting to consider the
equal-time spin correlations $\left\langle \bm{S}_{1}\cdot \bm{S}%
_{2}\right\rangle $ and $\left\langle \bm{S}_{1}\cdot \bm{S}%
_{H}\right\rangle $, where $\bm{S}_{1}$ is again the spin on the
nearest-neighbor B site $b_{1}$, and $\bm{S}_{H}\equiv \frac{1}{2}%
\sum_{\alpha \beta }g_{j\alpha }^{\dagger }\bm{\sigma}_{\alpha \beta
}g_{j\beta }$ is the spin on the hydrogen impurity; these are plotted in
Fig.~\ref{spinspin5I2CAtr} for different values of $\epsilon _{H}$. We see
that $\left\langle \bm{S}_{1}\cdot \bm{S}_{2}\right\rangle $ is
ferromagnetic in the K-S phase, but becomes antiferromagnetic in the AF-ASC phase (although now much smaller in magnitude than $-1/4$), changing sign
across the phase transition as in the Kondo model (cf. Figs.~\ref{spinspin3I2CK} and \ref{spinspin3I2CKtr}). A similar behavior is seen in $\left\langle \bm{S}_{1}\cdot \bm{S}_{H}\right\rangle $, i.e. the
spin correlation between the hydrogen impurity and the nearest-neighbor B sites
also goes from ferromagnetic to antiferromagnetic across the K-S/AF-ASC
transition. On the other hand, the spin correlations involving the spin on
the central A site, $\bm{S}_{0}\equiv \frac{1}{2}\sum_{\alpha \beta
}a_{0\alpha }^{\dagger }\bm{\sigma}_{\alpha \beta }a_{0\beta }$, are almost
unchanged at the transition.

Further information of the K-S and AF-ASC phases is given in Fig.~\ref%
{disc5I2CA}, where we plot the low-energy expectation values of the
following operators: the total spin on the nearest neighbor B sites $(\bm{S}%
_{1}+\bm{S}_{2}+\bm{S}_{3})^{2}$, the total spin on the hydrogen impurity
and the central A site $(\bm{S}_{0}+\bm{S}_{H})^{2}$, the total spin on the 5-atom cluster $\bm{S}_{\text{tot}}^{2}\equiv (\bm{S}_{1}+\bm{S}_{2}+\bm{S}_{3}+%
\bm{S}_{0}+\bm{S}_{H})^{2}$, the occupancy of a nearest-neighbor B site $n_{b,1}$, the occupancy of the central A site $n_{a,0}$, and the occupancy
of the hydrogen impurity $n_{H}$. $(\bm{S}_{0}+\bm{S}_{H})^{2}$ is almost
unchanged across the phase transition and remains small (less than $1/4$)
for $-2t\leq \epsilon _{H}+U_{H}/2\leq 0$, while both $(\bm{S}_{1}+\bm{S}%
_{2}+\bm{S}_{3})^{2}$ and $\bm{S}_{\text{tot}}^{2}$ fall dramatically when
the system goes from the K-S phase to the AF-ASC phase. While $(\bm{S}_{1}+\bm{S}%
_{2}+\bm{S}_{3})^{2}$ decreases from $1.6$ to $1.0$, $\bm{S}_{\text{%
tot}}^{2}$ undergoes a sharper drop from $1.7$ to $0.8$, which is a natural result of $\left\langle \bm{S}_{1}\cdot \bm{S}_{H}\right\rangle $ changing from ferromagnetic in the K-S phase to antiferromagnetic in the AF-ASC phase. Recalling that the K-S phase has total spin $1/2$ and the AF-ASC phase has total spin $0$, we conclude that the spin of the 5-atom cluster is Kondo-screened by the conducting channels in both the K-S phase and the AF-ASC phase. We stress that, in contrast to the linear-LDOS case\cite{PhysRevB.85.115405}, such a Kondo effect does not require a very strong coupling between the 5-atom cluster and the surrounding non-interacting bath. Meanwhile, $n_{b,1}$, $%
n_{a,0}$ and $n_{H}$ all experience a sudden increase across the transition as $%
\epsilon _{H}<-U_{H}/2$ decreases, which is consistent with the total charge
in the ground state increasing by one; $n_{b,1}$ increases by the largest
amount of the three, from $0.98$ to $1.2$.

\begin{figure}[!h]
\includegraphics[width=1\columnwidth]{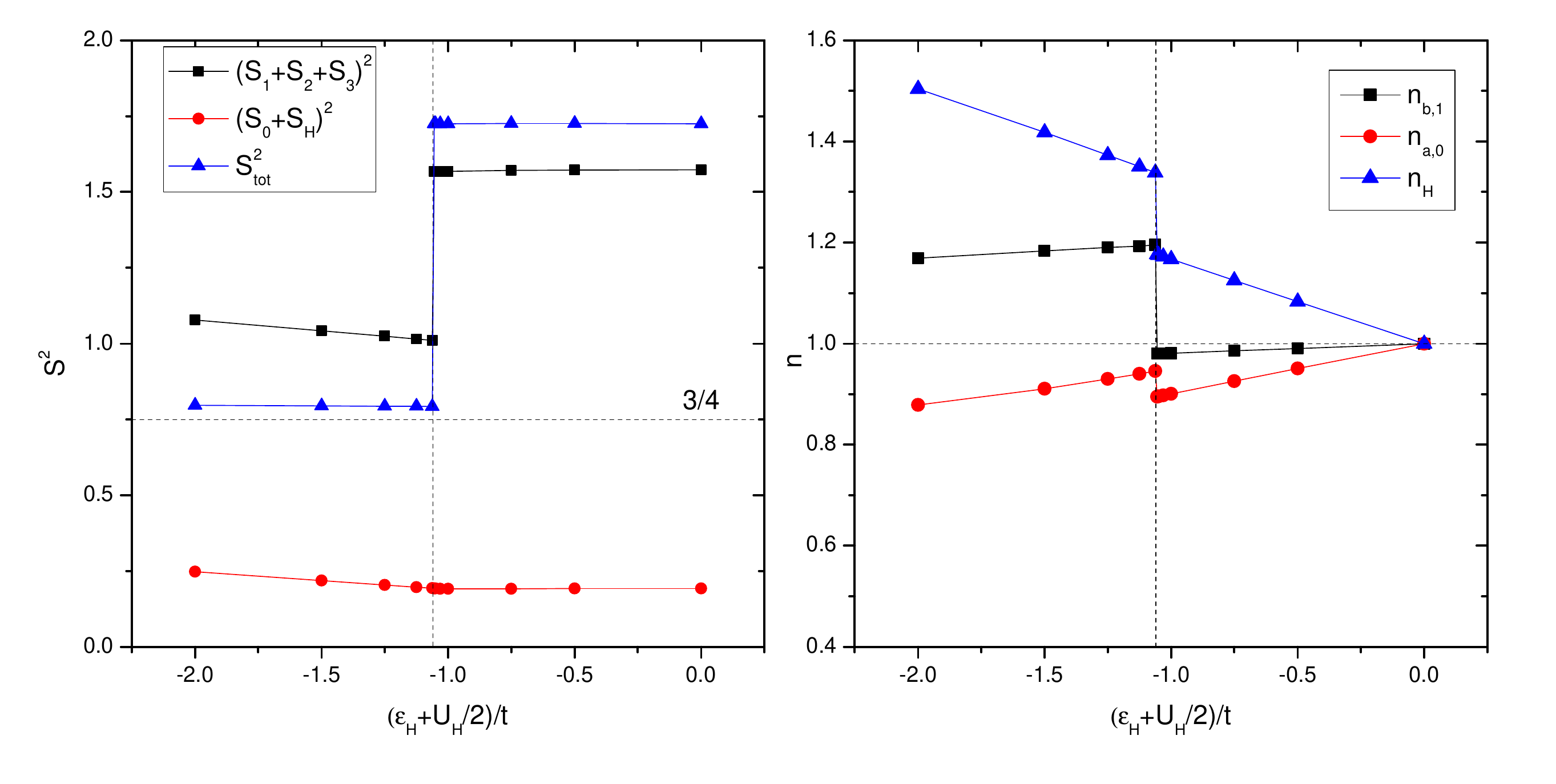} 
\caption{Low-energy total impurity spins and orbital occupancies as a function of $\epsilon _{H}$ in the Anderson model Eq.~(\ref{Anderson}) with a logarithmically divergent LDOS given by Eq.~(\ref{logLDOS}). See main text for an explanation of the plotted quantities. $\Lambda=1.5D_{0}$; $U=t$, $U_{H}=2.8t$, $\epsilon _{a}=\epsilon _{b}=-U/2$, $t_{H}=2t$, $t_{0}=0.6t$.}
\label{disc5I2CA}
\end{figure}

\begin{figure}[!h]
\includegraphics[width=1\columnwidth]{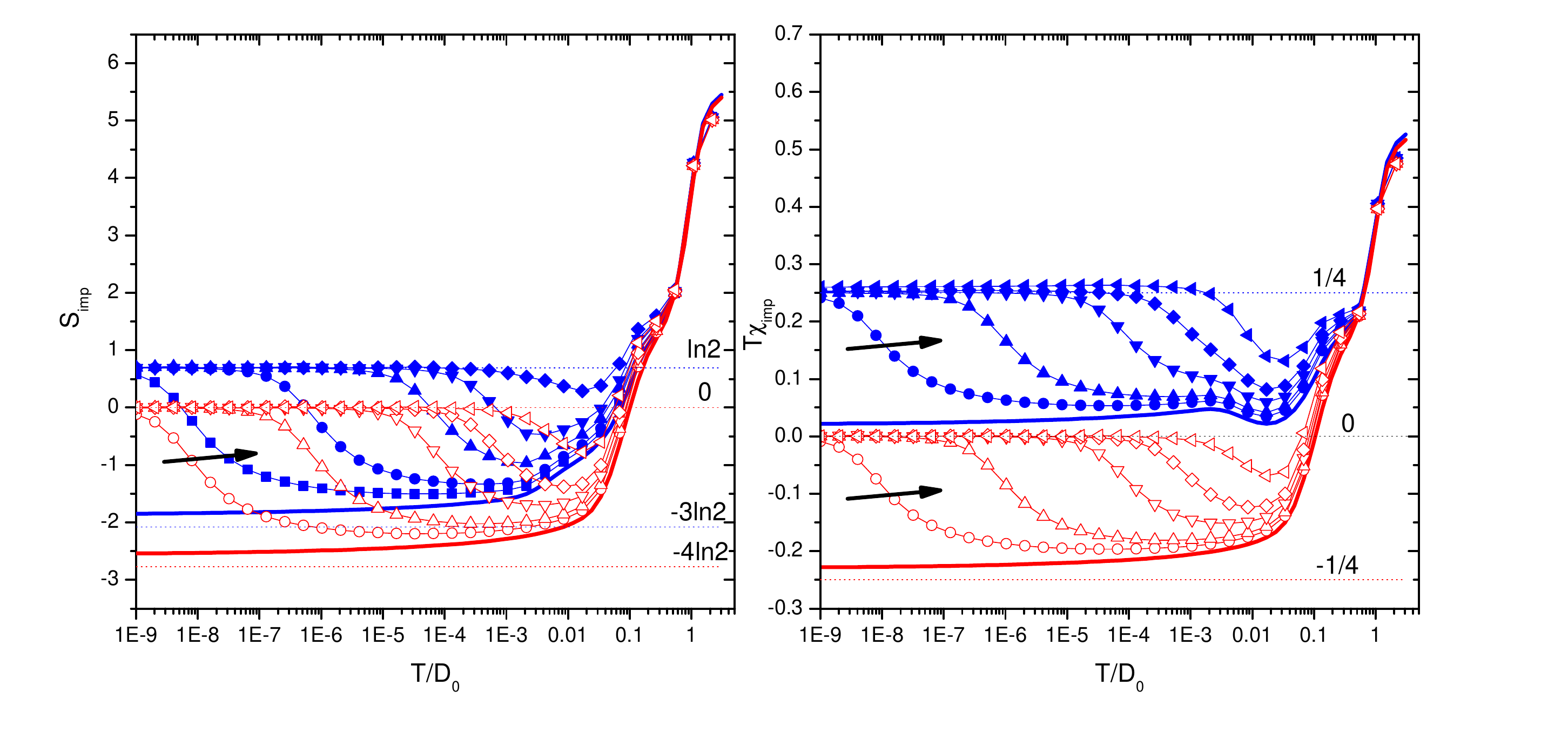} 
\caption{$S_{\text{imp}}$ and $T \chi_{\text{imp}}$ in the Anderson model Eq.~(\ref{Anderson}) with different infrared cutoffs $X$ on the logarithmically divergent LDOS Eq.~(\ref{logLDOS}). $\Lambda=1.5D_{0}$; $U=t$, $U_{H}=2.8t$, $\epsilon _{a}=\epsilon _{b}=-U/2$, $t_{H}=2t$, $t_{0}=0.6t$. The K-S phase with $\left( \epsilon _{H}+U_{H}/2\right) /t=-1$ is shown with solid blue symbols and the AF-ASC phase with $\left( \epsilon _{H}+U_{H}/2\right) /t=-2$ is shown with open red symbols. Thick lines correspond to $X=0$, and $X/D_{0}$ takes the following values along the direction of the arrows: $10^{-8}$, $10^{-6}$, $10^{-4}$, $10^{-3}$, and $0.01$.}
\label{STchi5I2CADOS}
\end{figure}

Finally, we briefly discuss the effect of an infrared energy
cutoff $X$ of the LDOS on the Anderson model. Fig.~\ref{STchi5I2CADOS} shows $S_{\text{imp}}$ and 
$T\chi _{\text{imp}}$ for different values of $X$ in both the K-S phase and
the AF-ASC phase. Not too surprisingly, the behavior of $S_{\text{imp}}$ and 
$T\chi _{\text{imp}}$ approaches the $X=0$ case at intermediate energy
scales $X\ll T\ll T_{K}^{X=0}$, while at lower energies $T\ll X$, $S_{\text{%
imp}}$ and $T\chi _{\text{imp}}$ return to their values in the constant-LDOS
version of the model where there is no contribution from non-normalizable
zero modes. While increasing $X$ also leads to an increase in magnitude for
the spin correlations $\left\langle \bm{S}_{1}\cdot \bm{S}_{2}\right\rangle $
and $\left\langle \bm{S}_{1}\cdot \bm{S}_{H}\right\rangle $, this is a tiny
effect in comparison with the spin correlation of the Kondo model shown in
Fig.~\ref{spinspin3I2CKDOS}.

To summarize this section, our results indicate that the behavior of the Anderson model which contains the hydrogen impurity and the four nearest carbon atoms is qualitatively captured by the 3-impurity Kondo model. There exist a p-h symmetric spin-$1/2$ K-S phase where the impurity spins align ferromagnetically, and a p-h asymmetric spin-singlet AF-ASC phase where the impurity spins align antiferromagnetically. It is possible that the K-S/AF-ASC transition picture is applicable to even more realistic models of the hydrogen impurity: weaker p-h symmetry breaking favors ferromagnetic spin correlation and leads to magnetic ground states, while stronger p-h symmetry breaking favors antiferromagnetic spin correlation and tends to suppress the ground state degeneracy.

\section{Conclusions and outlook\label{sec:conclusions}}

In this paper, we have studied the Kondo effect associated with a single hydrogen impurity on graphene. The hydrogen impurity is strongly coupled to the ``central'' carbon atom directly below it. First we consider the limit of infinite coupling, so that the hydrogen atom and the central carbon atom are effectively decoupled from the rest of the system, but the $C_3$ rotation symmetry of the system is preserved. To model the induced magnetization, we place a strong Hubbard interaction on the three nearest neighbor carbon atoms, creating three magnetic impurities. The remaining graphene sheet with four vacancy sites, approximated to be nearest-neighbor and non-interacting, supports two conduction channels which hybridizes with the three impurities with a local density of states diverging logarithmically as a function of energy near the Dirac point, in addition to a conduction channel whose LDOS vanishes linearly.

We study the resulting 3-impurity, 3-channel Kondo model with the numerical renormalization group method. Couplings to the conduction channel with a linear LDOS are irrelevant and usually negligible, and the phase diagram is controlled by the Kondo and potential scattering coupling constants associated with the two conduction channels with a logarithmically divergent LDOS. The regime where the potential scattering is not too strong sees the competition between a p-h symmetric Kondo phase (K-S) and a p-h asymmetric strong-coupling phase (AF-ASC). Ferromagnetic RKKY interactions between the magnetic impurities and weaker potential scattering favor the K-S phase, where the p-h symmetric ground state is a residual spin-$1/2$ after screening by the two conduction channels, and the impurity spins tend to align ferromagnetically. On the other hand, antiferromagnetic RKKY interactions and stronger potential scattering favor the AF-ASC phase, where the ground state is a spin singlet with one electron removed from or added to half filling, and the impurity spins align antiferromagnetically. In the strong potential scattering regime, the potential scattering coupling strength renormalizes to infinity, and the magnetic impurities decouple from the conduction channels, forming a local moment whose size depends on the RKKY interactions.

Relaxing the approximation of infinite hydrogen-carbon hybridization, we obtain an Anderson model with 5 impurity sites: the hydrogen atom, the central carbon atom and its three nearest neighbors in the tight-binding model. For realistic Hubbard interaction strengths on impurities, we find through NRG that the ground state is the p-h symmetric spin-$1/2$ K-S phase when the p-h symmetry breaking is not too strong, and the particle-hole asymmetric spin singlet AF-ASC phase otherwise. Kondo screening is shown to take place in both phases of the Anderson model. In the K-S phase, the spins of the nearest neighbor carbon atoms align ferromagnetically with each other and with the spin of the hydrogen atom, whereas in the AF-ASC phase they align antiferromagnetically with each other and with the hydrogen spin.These provide evidence that our 3-impurity Kondo model approximation is qualitatively reasonable.

Many open questions remain to be answered. First of all, we have assumed throughout this work that the bulk chemical potential is fine-tuned to the singularity of the vacancy-induced logarithmically divergent LDOS, which coincides with the zero point of the bulk density of states. Perturbations at the LM fixed point are thus strongly relevant for the helicity-1, \={1} channels with a logarithmically divergent LDOS, and strongly irrelevant for the helicity-0 channel with a linear LDOS. If the bulk chemical potential is shifted by an applied gate voltage, it is interesting to check whether the helicity-0 channel will have a progressively more important influence on the Kondo temperature and the transport properties of the system, as one may expect from results on the single-channel Kondo problem in gated or doped graphene\cite{PhysRevLett.102.046801,Europhys.Lett.90.27006,PhysRevB.89.195424,PhysRevB.95.115408,PhysRevB.97.155419}.

As a closely related point, in obtaining the 5-impurity Anderson model, we have neglected the next-nearest-neighbor hopping between carbon atoms. While imposing an infrared cutoff on the logarithmically divergent LDOS partially mimics its effects\cite{PhysRevB.77.115109}, the next-nearest-neighbor hopping will also change the wave functions of the bulk conduction electrons and their coupling to the hydrogen impurity. A more careful treatment of the next-nearest-neighbor hopping is thus necessary for a quantitative comparison with experiments.

The electron-electron interaction on carbon atoms farther away from the hydrogen impurity than the three nearest neighbors is another essential ingredient in a more realistic model, since experiments have shown that the spin-polarized state induced by the hydrogen impurity has a large spatial extension\cite{Science.352.437}. Such interactions have been taken into account in previous studies within the Hartree-Fock approximation\cite{JPSJ.76.064713,PhysRevB.85.115405} and using dynamical mean-field theory\cite{PhysRevB.83.241408,PhysRevB.91.035132}. To the best of our knowledge, it is not clear how the vacancy-induced non-normalizable zero modes behave in the presence of bulk electron-electron interactions when they are not strong enough to turn graphene into a Mott insulator. Addressing this issue will be useful for a theoretical understanding of the unusually long-ranged coupling between the magnetic moments induced by different hydrogen adatoms\cite{Science.352.437,PhysRevB.96.165411}.

\begin{acknowledgments}
We thank Eva Andrei, Josh Folk and Eran Sela for helpful discussions. The research of IA was supported by
NSERC Discovery Grant 04033-2016 and the Canadian Institute for Advanced Research.
\end{acknowledgments}

\appendix

\section{Conduction channels at low energies\label{sec:app4vscatbas}}

In this appendix we derive the low-energy behavior of the three conduction
channels of definite helicities, Eqs.~(\ref{c0tilde}), (\ref{c1tilde}) and (%
\ref{c1bartilde}). This is achieved by diagonalizing the non-interacting
Hamiltonian with a 4-site vacancy $H_{\text{vac}}$ and finding the scattering
state wave functions.

The vacancy can be implemented by strong potential scattering: $H_{\text{vac}%
}=H_{0}+V_{i}$, with $H_{0}$ describing the translationally invariant
pristine graphene,

\begin{equation}
H_{0}=-t\sum\nolimits_{\vec{R}}\left\{ \left[ b^{\dag }\left( \vec{R}\right)
+b^{\dag }\left( \vec{R}-\vec{a}_{2}\right) +b^{\dag }\left( \vec{R}-\vec{a}%
_{1}\right) \right] a\left( \vec{R}\right) +\text{h.c.}\right\} \text{,}
\end{equation}%
and $V_{i}$ simulating the vacancy,

\begin{equation}
V_{i}=U_{1}a^{\dag }\left( \vec{0}\right) a\left( \vec{0}\right) +U_{2}\left[
b^{\dag }\left( \vec{0}\right) b\left( \vec{0}\right) +b^{\dag }\left( -\vec{%
a}_{2}\right) b\left( -\vec{a}_{2}\right) +b^{\dag }\left( -\vec{a}%
_{1}\right) b\left( -\vec{a}_{1}\right) \right] \text{.}
\end{equation}%
The limit $U_{2}\rightarrow \pm \infty $ corresponds to the vacancy sites $a(\vec{0%
})$, $b(\vec{0})$, $b\left( -\vec{a}_{2}\right) $ and $b\left( -\vec{a}%
_{1}\right) $. Physically we do not expect the value of $U_{1}$ to affect
the scattering state wave functions in the $U_{2}\rightarrow \pm \infty $
limit, because $a(\vec{0})$ will be isolated from the other sites.

We will work in the basis of the $H_{0}$ eigenstates. These are given by

\begin{equation}
\left( 
\begin{array}{c}
\psi _{+\vec{k}} \\ 
\psi _{-\vec{k}}%
\end{array}%
\right) =\frac{1}{\sqrt{2}}\left( 
\begin{array}{cc}
1 & -\frac{1+e^{-i\vec{k}\cdot \vec{a}_{1}}+e^{-i\vec{k}\cdot \vec{a}_{2}}}{%
\left\vert 1+e^{i\vec{k}\cdot \vec{a}_{1}}+e^{i\vec{k}\cdot \vec{a}%
_{2}}\right\vert } \\ 
1 & \frac{1+e^{-i\vec{k}\cdot \vec{a}_{1}}+e^{-i\vec{k}\cdot \vec{a}_{2}}}{%
\left\vert 1+e^{i\vec{k}\cdot \vec{a}_{1}}+e^{i\vec{k}\cdot \vec{a}%
_{2}}\right\vert }%
\end{array}%
\right) \left( 
\begin{array}{c}
a_{\vec{k}} \\ 
b_{\vec{k}}%
\end{array}%
\right) \text{;}  \label{psiab}
\end{equation}%
here we have performed the Fourier transform

\begin{align}
a\left( \vec{R}\right) & =\int_{1BZ}\frac{d^{2}k}{\sqrt{S_{1BZ}}}e^{i\vec{k}%
\cdot \vec{R}}a_{\vec{k}}\text{,}  \notag \\
b\left( \vec{R}\right) & =\int_{1BZ}\frac{d^{2}k}{\sqrt{S_{1BZ}}}e^{i\vec{k}%
\cdot \vec{R}}b_{\vec{k}}\text{,}  \label{ABfourier}
\end{align}%
where $S_{1BZ}=8\pi ^{2}/(\sqrt{3}a^{2})$ is the area of the hexagonal first
Brillouin zone. It is straightforward to rewrite $H_{0}$\ and $V_{i}$ in
terms of $\psi _{\pm }$,

\begin{equation}
H_{0}=\int_{1BZ}d^{2}k\epsilon _{\vec{k}}\left( \psi _{+\vec{k}}^{\dag }\psi
_{+\vec{k}}-\psi _{-\vec{k}}^{\dag }\psi _{-\vec{k}}\right) \text{,}
\end{equation}%
with the dispersion

\begin{equation}
\epsilon _{\vec{k}}=t\left\vert 1+e^{i\vec{k}\cdot \vec{a}_{1}}+e^{i\vec{k}%
\cdot \vec{a}_{2}}\right\vert \text{,}
\end{equation}%
and

\begin{align}
V_{i}& =\frac{1}{2}\int_{1BZ}\frac{d^{2}kd^{2}k^{\prime }}{S_{1BZ}}\left\{ %
\left[ U_{1}\left( \psi _{+,\vec{k}}^{\dag }+\psi _{-,\vec{k}}^{\dag
}\right) \left( \psi _{+,\vec{k}^{\prime }}+\psi _{-,\vec{k}^{\prime
}}\right) \right] +U_{2}\left[ 1+e^{i\left( \vec{k}-\vec{k}^{\prime }\right)
\cdot \vec{a}_{1}}+e^{i\left( \vec{k}-\vec{k}^{\prime }\right) \cdot \vec{a}%
_{2}}\right] \right.  \notag \\
& \left. \times \frac{1+e^{-i\vec{k}\cdot \vec{a}_{1}}+e^{-i\vec{k}\cdot 
\vec{a}_{2}}}{\left\vert 1+e^{i\vec{k}\cdot \vec{a}_{1}}+e^{i\vec{k}\cdot 
\vec{a}_{2}}\right\vert }\frac{1+e^{i\vec{k}^{\prime }\cdot \vec{a}_{1}}+e^{i%
\vec{k}^{\prime }\cdot \vec{a}_{2}}}{\left\vert 1+e^{i\vec{k}^{\prime }\cdot 
\vec{a}_{1}}+e^{i\vec{k}^{\prime }\cdot \vec{a}_{2}}\right\vert }\left( \psi
_{+,\vec{k}}^{\dag }-\psi _{-,\vec{k}}^{\dag }\right) \left( \psi _{+,\vec{k}%
^{\prime }}-\psi _{-,\vec{k}^{\prime }}\right) \right\} \text{.}
\label{Vi4vsc}
\end{align}%
The scattering states are given by

\begin{equation}
\phi _{\pm ,\vec{p}}^{\dag }=\psi _{\pm ,\vec{p}}^{\dag
}+\int_{1BZ}d^{2}p^{\prime }\left[ G_{\pm ,+}\left( \vec{p},\vec{p}^{\prime
}\right) \psi _{+,\vec{p}^{\prime }}^{\dag }+G_{\pm ,-}\left( \vec{p},\vec{p}%
^{\prime }\right) \psi _{-,\vec{p}^{\prime }}^{\dag }\right] \text{,}
\label{ssp}
\end{equation}%
where for $\lambda $, $\lambda ^{\prime }=\pm 1$, $G_{\lambda ,\lambda
^{\prime }}\left( \vec{p},\vec{p}^{\prime }\right) \left( \lambda \epsilon _{%
\vec{p}}-\lambda ^{\prime }\epsilon _{\vec{p}^{\prime }}\right) $ is finite.
By definition $\phi $ diagonalizes $H_{\text{vac}}$:

\begin{equation}
\left[ H_{\text{vac}},\phi _{\pm ,\vec{p}}^{\dag }\right] =\pm \epsilon _{%
\vec{p}}\phi _{\pm ,\vec{p}}^{\dag }\text{.}
\end{equation}%
The energy eigenvalues are those of the $H_{0}$ eigenstates because the
impurity is localized. This equation is then solved for $G_{\lambda ,\lambda
^{\prime }}\left( \vec{p},\vec{p}^{\prime }\right) $.

In the limit of $U_{2}\rightarrow \pm \infty $, the scattering states are
indeed independent of $U_{1}$:

\begin{align}
\phi _{+,\vec{p}}^{\dag }& =\psi _{+,\vec{p}}^{\dag }+\int_{1BZ}\frac{d^{2}k%
}{S_{1BZ}}\left( -\frac{1}{\epsilon _{\vec{p}}-\epsilon _{\vec{k}}+i0}\psi
_{+,\vec{k}}^{\dag }+\frac{1}{\epsilon _{\vec{p}}+\epsilon _{\vec{k}}+i0}%
\psi _{-,\vec{k}}^{\dag }\right) \frac{1}{\frac{3}{2}L\left( \epsilon _{\vec{%
p}}+i0\right) -\frac{\epsilon _{\vec{p}}}{6t^{2}}\left[ -2+\epsilon _{\vec{p}%
}L\left( \epsilon _{\vec{p}}+i0\right) \right] }  \notag \\
& \times \left( \left[ 1+e^{i\left( \vec{k}-\vec{p}\right) \cdot \vec{a}%
_{1}}+e^{i\left( \vec{k}-\vec{p}\right) \cdot \vec{a}_{2}}\right] \frac{%
1+e^{i\vec{p}\cdot \vec{a}_{1}}+e^{i\vec{p}\cdot \vec{a}_{2}}}{\left\vert
1+e^{i\vec{p}\cdot \vec{a}_{1}}+e^{i\vec{p}\cdot \vec{a}_{2}}\right\vert }%
\frac{1+e^{-i\vec{k}\cdot \vec{a}_{1}}+e^{-i\vec{k}\cdot \vec{a}_{2}}}{%
\left\vert 1+e^{i\vec{k}\cdot \vec{a}_{1}}+e^{i\vec{k}\cdot \vec{a}%
_{2}}\right\vert }-\frac{1}{2}\left\{ 1-\frac{L\left( \epsilon _{\vec{p}%
}+i0\right) }{\frac{\epsilon _{\vec{p}}}{3t^{2}}\left[ -2+\epsilon _{\vec{p}%
}L\left( \epsilon _{\vec{p}}+i0\right) \right] }\right\} \frac{\epsilon _{%
\vec{p}}\epsilon _{\vec{k}}}{t^{2}}\right) \text{.}
\end{align}%
The negative-energy states are found by interchanging $+$ with $-$, and
inverting the signs of all absolute values. We have introduced the shorthand

\begin{equation}
L\left( z\right) \equiv \int_{1BZ}\frac{d^{2}q}{S_{1BZ}}\frac{2z}{%
z^{2}-\epsilon _{\vec{q}}^{2}}\text{.}  \label{Lz}
\end{equation}%
It is useful to give a low-energy asymptotic formula for $L\left( \omega
^{+}\right) \equiv L\left( \omega +i0\right) $, valid for $\left\vert \omega
\right\vert \ll \Lambda \sim t$, obtained by only keeping contributions
from near the two Dirac points:

\begin{equation}
L\left( \omega ^{+}\right) \approx 2\frac{\sqrt{3}a^{2}}{8\pi ^{2}}\int
d^{2}k\frac{2\omega ^{+}}{\left( \omega ^{+}\right) ^{2}-v_{F}^{2}k^{2}}%
\approx -\frac{2\omega }{\sqrt{3}\pi t^{2}}\left( \ln \frac{\Lambda
_{0}^{2}}{\omega ^{2}}+i\pi \operatorname{sgn}\omega \right) \text{,}  \label{Lomega}
\end{equation}%
where $\Lambda _{0}=\Lambda e^{\frac{\pi }{6\sqrt{3}}}$ is another
ultraviolet energy cutoff.

Using Eqs.~(\ref{ABfourier}) and (\ref{psiab}), we can rewrite $a$ in terms
of $\phi $:

\begin{align}
a\left( \vec{R}\right) & =\frac{1}{\sqrt{2}}\int_{1BZ}\frac{d^{2}p}{\sqrt{%
S_{1BZ}}}\left[ \left( e^{i\vec{p}\cdot \vec{R}}-\frac{1}{L\left( \epsilon _{%
\vec{p}}+i0\right) -L\left( \vec{a}_{1},\epsilon _{\vec{p}}+i0\right) }%
\left\{ \frac{1+e^{i\vec{p}\cdot \vec{a}_{1}}+e^{i\vec{p}\cdot \vec{a}_{2}}}{%
\left\vert 1+e^{i\vec{p}\cdot \vec{a}_{1}}+e^{i\vec{p}\cdot \vec{a}%
_{2}}\right\vert }\left[ \tilde{L}\left( -\vec{R},\epsilon _{\vec{p}%
}+i0\right) \right. \right. \right. \right.  \notag \\
& \left. +e^{-i\vec{p}\cdot \vec{a}_{1}}\tilde{L}\left( -\vec{R}-\vec{a}%
_{1},\epsilon _{\vec{p}}+i0\right) +e^{-i\vec{p}\cdot \vec{a}_{2}}\tilde{L}%
\left( -\vec{R}-\vec{a}_{2},\epsilon _{\vec{p}}+i0\right) \right] +\frac{1}{2%
}\left[ 1-\frac{3t^{2}}{\epsilon _{\vec{p}}}\frac{L\left( \epsilon _{\vec{p}%
}+i0\right) }{-2+\epsilon _{\vec{p}}L\left( \epsilon _{\vec{p}}+i0\right) }%
\right]  \notag \\
& \left. \left. \left. \times \left[ \frac{2\epsilon _{\vec{p}}}{t^{2}}%
\delta _{\vec{R}\vec{0}}-\frac{\epsilon _{\vec{p}}^{2}}{t^{2}}L\left( \vec{R}%
,\epsilon _{\vec{p}}+i0\right) \right] \right\} \right) \phi _{+,\vec{p}%
}+\left( \epsilon _{\vec{p}}\rightarrow -\epsilon _{\vec{p}}\right) \right] 
\text{,}  \label{aphi}
\end{align}%
where the $\epsilon _{\vec{p}}\rightarrow -\epsilon _{\vec{p}}$ part is the
contribution from the negative energy eigenstates $\phi _{-,\vec{p}}$, and
we have further defined

\begin{equation}
L\left( \vec{R},z\right) \equiv\int_{1BZ}\frac{d^{2}k}{S_{1BZ}}e^{i\vec {k}%
\cdot\vec{R}}\frac{2z}{z^{2}-\epsilon_{\vec{k}}^{2}}\text{,}  \label{LRz}
\end{equation}

\begin{equation}
\tilde{L}\left( \vec{R},z\right) \equiv \int_{1BZ}\frac{d^{2}k}{S_{1BZ}}e^{i%
\vec{k}\cdot \vec{R}}\frac{2t\left( 1+e^{i\vec{k}\cdot \vec{a}_{1}}+e^{i\vec{%
k}\cdot \vec{a}_{2}}\right) }{z^{2}-\epsilon _{\vec{k}}^{2}}\text{.}
\label{LtildeRz}
\end{equation}%
It is useful to note that $L\left( \vec{R},z\right) $ have all the
symmetries of the hexagonal lattice, and that $L\left( \vec{a}_{1},z\right) $
is related to $L\left( z\right) $ by

\begin{equation}
3\left[ L\left( z\right) +2L\left( \vec{a}_{1},z\right) \right] =\frac{1}{%
t^{2}}\left[ -2z+z^{2}L\left( z\right) \right] \text{.}  \label{L1L0}
\end{equation}

According to Eq.~(\ref{aphi}), the symmetric linear combinations $a_{1,2,3}$
have the form

\begin{subequations}
\begin{align}
a_{1}& =\frac{1}{2}\int_{1BZ}\frac{d^{2}p}{\sqrt{S_{1BZ}}}\left[ \left( e^{i%
\vec{p}\cdot \vec{a}_{1}}+e^{i\vec{p}\cdot \vec{a}_{2}}\right) -\frac{1}{%
L\left( \epsilon _{\vec{p}}+i0\right) -L\left( \vec{a}_{1},\epsilon _{\vec{p}%
}+i0\right) }\left( \frac{1+e^{i\vec{p}\cdot \vec{a}_{1}}+e^{i\vec{p}\cdot 
\vec{a}_{2}}}{\left\vert 1+e^{i\vec{p}\cdot \vec{a}_{1}}+e^{i\vec{p}\cdot 
\vec{a}_{2}}\right\vert }\right. \right.  \notag \\
& \left. \left. \times \left\{ -\frac{2}{t}+\frac{\epsilon _{\vec{p}}}{t}%
\left[ L\left( \epsilon _{\vec{p}}+i0\right) -L\left( \vec{a}_{1},\epsilon _{%
\vec{p}}+i0\right) \right] \right\} -\frac{\epsilon _{\vec{p}}}{3t^{2}}\left[
-2+\epsilon _{\vec{p}}L\left( \epsilon _{\vec{p}}+i0\right) \right] +\frac{%
\epsilon _{\vec{p}}}{3}\frac{L\left( \epsilon _{\vec{p}}+i0\right) L\left( 
\vec{a}_{1},\epsilon _{\vec{p}}+i0\right) }{-2+\epsilon _{\vec{p}}L\left(
\epsilon _{\vec{p}}+i0\right) }\right) \right]  \notag \\
& \times \phi _{+,\vec{p}}+\left( \epsilon _{\vec{p}}\rightarrow -\epsilon _{%
\vec{p}}\right) \text{,}
\end{align}

\begin{align}
& a_{2}=\frac{1}{2}\int_{1BZ}\frac{d^{2}p}{\sqrt{S_{1BZ}}}\left[ \left( e^{-i%
\vec{p}\cdot \vec{a}_{2}}+e^{i\vec{p}\cdot \left( \vec{a}_{1}-\vec{a}%
_{2}\right) }\right) -\frac{1}{L\left( \epsilon _{\vec{p}}+i0\right)
-L\left( \vec{a}_{1},\epsilon _{\vec{p}}+i0\right) }\left( \frac{1+e^{i\vec{p%
}\cdot \vec{a}_{1}}+e^{i\vec{p}\cdot \vec{a}_{2}}}{\left\vert 1+e^{i\vec{p}%
\cdot \vec{a}_{1}}+e^{i\vec{p}\cdot \vec{a}_{2}}\right\vert }e^{-i\vec{p}%
\cdot \vec{a}_{2}}\right. \right.  \notag \\
& \left. \left. \times \left\{ -\frac{2}{t}+\frac{\epsilon _{\vec{p}}}{t}%
\left[ L\left( \epsilon _{\vec{p}}+i0\right) -L\left( \vec{a}_{1},\epsilon _{%
\vec{p}}+i0\right) \right] \right\} -\frac{\epsilon _{\vec{p}}}{3t^{2}}\left[
-2+\epsilon _{\vec{p}}L\left( \epsilon _{\vec{p}}+i0\right) \right] +\frac{%
\epsilon _{\vec{p}}}{3}\frac{L\left( \epsilon _{\vec{p}}+i0\right) L\left( 
\vec{a}_{1},\epsilon _{\vec{p}}+i0\right) }{-2+\epsilon _{\vec{p}}L\left(
\epsilon _{\vec{p}}+i0\right) }\right) \right]  \notag \\
& \times \phi _{+,\vec{p}}+\left( \epsilon _{\vec{p}}\rightarrow -\epsilon _{%
\vec{p}}\right) \text{,}
\end{align}

\begin{align*}
& a_{3}=\frac{1}{2}\int_{1BZ}\frac{d^{2}p}{\sqrt{S_{1BZ}}}\left[ \left( e^{i%
\vec{p}\cdot \left( \vec{a}_{2}-\vec{a}_{1}\right) }+e^{-i\vec{p}\cdot \vec{a%
}_{1}}\right) -\frac{1}{L\left( \epsilon _{\vec{p}}+i0\right) -L\left( \vec{a%
}_{1},\epsilon _{\vec{p}}+i0\right) }\left( \frac{1+e^{i\vec{p}\cdot \vec{a}%
_{1}}+e^{i\vec{p}\cdot \vec{a}_{2}}}{\left\vert 1+e^{i\vec{p}\cdot \vec{a}%
_{1}}+e^{i\vec{p}\cdot \vec{a}_{2}}\right\vert }e^{-i\vec{p}\cdot \vec{a}%
_{1}}\right. \right. \\
& \left. \left. \times \left\{ -\frac{2}{t}+\frac{\epsilon _{\vec{p}}}{t}%
\left[ L\left( \epsilon _{\vec{p}}+i0\right) -L\left( \vec{a}_{1},\epsilon _{%
\vec{p}}+i0\right) \right] \right\} -\frac{\epsilon _{\vec{p}}}{3t^{2}}\left[
-2+\epsilon _{\vec{p}}L\left( \epsilon _{\vec{p}}+i0\right) \right] +\frac{%
\epsilon _{\vec{p}}}{3}\frac{L\left( \epsilon _{\vec{p}}+i0\right) L\left( 
\vec{a}_{1},\epsilon _{\vec{p}}+i0\right) }{-2+\epsilon _{\vec{p}}L\left(
\epsilon _{\vec{p}}+i0\right) }\right) \right] \\
& \times \phi _{+,\vec{p}}+\left( \epsilon _{\vec{p}}\rightarrow -\epsilon _{%
\vec{p}}\right) \text{.}
\end{align*}%
For the helicity-$\pm 1$ combinations $c_{h=1}$ and $c_{h=\bar{1}}$, at low
energies it is permissible to keep only the terms that are logarithmically
divergent at the Dirac points:

\end{subequations}
\begin{equation}
c_{h=1}=\frac{1}{\sqrt{3}}\left( a_{1}+e^{i\frac{2\pi }{3}}a_{2}+e^{-i\frac{%
2\pi }{3}}a_{3}\right) \approx \frac{1}{3^{\frac{1}{4}}\sqrt{2}}\int d^{2}k%
\left[ \frac{ie^{i\theta _{\vec{k}}}\phi _{\vec{K},+,\vec{k}}}{k\left( \ln 
\frac{\Lambda ^{2}}{v_{F}^{2} k^{2}}+i\pi \right) }+\frac{ie^{i\theta _{\vec{k}}}\phi
_{\vec{K},-,\vec{k}}}{-k\left( \ln \frac{\Lambda ^{2}}{v_{F}^{2} k^{2}}-i\pi \right) }%
\right] \text{,}  \label{c1}
\end{equation}

\begin{equation}
c_{h=\bar{1}}=\frac{1}{\sqrt{3}}\left( a_{1}+e^{-i\frac{2\pi }{3}}a_{2}+e^{i%
\frac{2\pi }{3}}a_{3}\right) \approx \frac{1}{3^{\frac{1}{4}}\sqrt{2}}\int
d^{2}k\left[ \frac{ie^{-i\theta _{\vec{k}}}\phi _{\vec{K}^{\prime },+,\vec{k}%
}}{k\left( \ln \frac{\Lambda ^{2}}{v_{F}^{2} k^{2}}+i\pi \right) }+\frac{ie^{-i\theta
_{\vec{k}}}\phi _{\vec{K}^{\prime },-,\vec{k}}}{-k\left( \ln \frac{\Lambda
^{2}}{v_{F}^{2} k^{2}}-i\pi \right) }\right] \text{.}  \label{c1bar}
\end{equation}%
Here $\phi _{\vec{K},\pm ,\vec{k}}\equiv \phi _{\pm ,\vec{K}+\vec{k}}$. On
the other hand, for the helicity-$0$ linear combination $c_{h=0}$, the
divergent terms are suppressed by $O\left( k^{2}\ln k\right) $ at low
energies, and the constant incident terms dominate instead:

\begin{align}
c_{h=0}& =\frac{1}{\sqrt{3}}\left( a_{1}+a_{2}+a_{3}\right)  \notag \\
& =\frac{1}{2\sqrt{3}}\int_{1BZ}\frac{d^{2}p}{\sqrt{S_{1BZ}}}\left( \left\{
-3+\frac{L\left( \epsilon _{\vec{p}}+i0\right) }{L\left( \epsilon _{\vec{p}%
}+i0\right) -L\left( \vec{a}_{1},\epsilon _{\vec{p}}+i0\right) }\left[ \frac{%
\epsilon _{\vec{p}}^{2}}{t^{2}}-\frac{\epsilon _{\vec{p}}L\left( \vec{a}%
_{1},\epsilon _{\vec{p}}+i0\right) }{-2+\epsilon _{\vec{p}}L\left( \epsilon
_{\vec{p}}+i0\right) }\right] \right\} \phi _{+,\vec{p}}+\left( \epsilon _{%
\vec{p}}\rightarrow -\epsilon _{\vec{p}}\right) \right)  \notag \\
& \approx -\frac{3^{\frac{3}{4}}a}{4\sqrt{2}\pi }\int d^{2}k\left( \phi _{%
\vec{K},+,\vec{k}}+\phi _{\vec{K}^{\prime },+,\vec{k}}+\phi _{\vec{K},-,\vec{%
k}}+\phi _{\vec{K}^{\prime },-,\vec{k}}\right) \text{.}  \label{c0}
\end{align}

We now take advantage of the rotational invariance at low energies, and
introduce angular momentum eigenmodes labeled by the quantum number $m$:

\begin{equation}
\phi _{\vec{K}/\vec{K}^{\prime },\pm ,\vec{k}}=\frac{1}{\sqrt{2\pi k}}%
\sum_{m=-\infty }^{\infty }e^{im\theta _{\vec{k}}}\tilde{\phi}_{\vec{K}/\vec{%
K}^{\prime },m,\pm \left\vert k\right\vert }\text{.}  \label{phismk}
\end{equation}%
These eigenmodes $\tilde{\phi}$ obey

\begin{equation}
\left\{ \tilde{\phi}_{\vec{K},m,k},\tilde{\phi}_{\vec{K},m^{\prime},k^{%
\prime}}^{\dag}\right\} =\delta_{mm^{\prime}}\delta\left( k-k^{\prime
}\right) \text{,}
\end{equation}
and in terms of $\tilde{\phi}$,

\begin{equation}
H_{\text{Vac}}=\int_{-\infty }^{\infty }dk\,v_{F}k\sum_{m}\left( \tilde{\phi}%
_{\vec{K},m,k}^{\dag }\tilde{\phi}_{\vec{K},m,k}+\tilde{\phi}_{\vec{K}%
^{\prime },m,k}^{\dag }\tilde{\phi}_{\vec{K}^{\prime },m,k}\right) \text{.}
\end{equation}%
Inserting Eq.~(\ref{phismk}) into Eqs.~(\ref{c1}), (\ref{c1bar}) and (\ref%
{c0}) then yields Eqs.~(\ref{c0tilde}), (\ref{c1tilde}) and (\ref{c1bartilde}%
).

\section{Non-normalizable zero modes\label{sec:appzero}}

This appendix elaborates on the zero modes of the infinite graphene sheet
with 4 vacancy sites. As discussed in Sec. \ref{sec:model}, these
non-normalizable zero modes are responsible for the logarithmic divergence
in the LDOS of our impurity models. We will solve the lattice Schr\"{o}%
dinger equation by generalizing the method of Ref.~%
\onlinecite{PhysRevLett.96.036801}, give the long-distance asymptotics of
the two solutions, and briefly discuss their fate in the strong-coupling
regime of the impurity models.

\begin{figure}[!h]
\includegraphics[width=0.5\columnwidth]{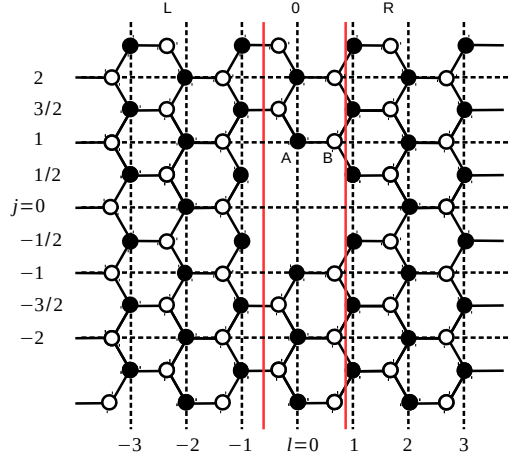} 
\caption{The alternative labeling scheme of 4-site-vacancy graphene lattice sites used in Appendix~\ref{sec:appzero}.
We divide the lattice into three parts: the left half-plane with a zigzag edge ($l<0$), the right half-plane with
a ``bearded'' edge ($l>0$), and a middle strip that contains the 4 vacancy sites ($l=0$).}
\label{zmlabel}
\end{figure}

It is convenient to relabel the lattice sites as in Fig.~\ref{zmlabel}, with
the two solid red lines dividing the plane into three parts: the left
half-plane with a zigzag\ edge, the right half-plane with a
\textquotedblleft bearded\textquotedblright\ edge, and the middle strip that
contains the four vacancy sites. The zero mode wave functions vanish on the
entire B sublattice, so we focus on the wave function on the A sublattice,
which we denote as $\phi _{l,j}$; here $l$ is an integer and $j$ is either
an integer or a half-integer, but $l+2j$ is always even.

Away from the vacancy, the Schr\"{o}dinger equation at zero energy reads

\begin{equation}
\phi _{l,j}+\phi _{l,j+1}+\phi _{l-1,j+\frac{1}{2}}=0\text{;}
\end{equation}%
this allows the expansion of the zero mode wave function on the left
half-plane in edge states of the zigzag edge\cite{PhysRevB.59.8271},

\begin{equation}
\phi _{l,j}=\int_{\frac{2\pi }{3a}}^{\frac{4\pi }{3a}}\frac{dk}{2\pi }\left(
-2\cos \frac{ka}{2}\right) ^{-l-1}e^{ikja}\phi _{k}^{L}\text{ }\left( l\leq
-1\right) \text{.}
\end{equation}%
as well as the expansion on the right half-plane in edge states of the
bearded edge,

\begin{equation}
\phi _{l,j}=\int_{-\frac{2\pi }{3a}}^{\frac{2\pi }{3a}}\frac{dk^{\prime }}{%
2\pi }\left( -2\cos \frac{k^{\prime }a}{2}\right) ^{-l+1}e^{ik^{\prime
}ja}\phi _{k^{\prime }}^{R}\text{ }\left( l\geq 1\right) \text{.}
\end{equation}%
Inserting these expansions into the $l=0$ and $l=1$ equations and
eliminating $\phi _{0,j}$, we have

\begin{equation}
\int_{-\frac{2\pi }{3a}}^{\frac{2\pi }{3a}}\frac{dk^{\prime }}{2\pi }%
e^{ik^{\prime }\left( j+\frac{1}{2}\right) a}\left( 2\cos \frac{k^{\prime }a%
}{2}\right) ^{2}\phi _{k^{\prime }}^{R}=\int_{\frac{2\pi }{3a}}^{\frac{4\pi 
}{3a}}\frac{dk}{2\pi }e^{ik\left( j+\frac{1}{2}\right) a}\phi _{k}^{L}\text{,%
}
\end{equation}%
which is true for any integer $j$ as long as $j\neq 0$ and $j\neq -1$. Using
the relation

\begin{equation}
\int_{-\frac{2\pi }{3a}}^{\frac{2\pi }{3a}}\frac{dk^{\prime }}{2\pi }%
e^{ik^{\prime }ja}=-\int_{\frac{2\pi }{3a}}^{\frac{4\pi }{3a}}\frac{dk}{2\pi 
}e^{ikja}
\end{equation}%
valid for nonzero integer $j$, we find two nontrivial solutions by
inspection:

\begin{equation}
\phi _{k}^{L,\left( 1\right) }=-e^{-i\frac{ka}{2}}\text{, }\phi _{k^{\prime
}}^{R,\left( 1\right) }=\frac{e^{-i\frac{k^{\prime }a}{2}}}{\left( 2\cos 
\frac{k^{\prime }a}{2}\right) ^{2}}
\end{equation}%
and

\begin{equation}
\phi _{k}^{L,\left( 2\right) }=-e^{i\frac{ka}{2}}\text{, }\phi _{k^{\prime
}}^{R,\left( 2\right) }=\frac{e^{i\frac{k^{\prime }a}{2}}}{\left( 2\cos 
\frac{k^{\prime }a}{2}\right) ^{2}}\text{.}
\end{equation}%
These solutions are linearly independent, and are therefore the only zero
energy solutions allowed\cite{PhysRevB.77.115109}.

We can show the long-distance asymptotic behavior of these solutions is
given by

\begin{equation}
\phi _{l,j}^{\left( 1\right) }\sim \left( -1\right) ^{l+1}\frac{1}{2\pi }%
\left( e^{i\frac{2\pi }{3}j}e^{-i\frac{\pi }{3}}\frac{1}{x+iy}+e^{-i\frac{%
2\pi }{3}j}e^{i\frac{\pi }{3}}\frac{1}{x-iy}\right) \text{,}
\label{zm1asymp}
\end{equation}%
and

\begin{equation}
\phi _{l,j}^{\left( 2\right) }\sim \left( -1\right) ^{l+1}\frac{1}{2\pi }%
\left( e^{i\frac{2\pi }{3}j}e^{i\frac{\pi }{3}}\frac{1}{x+iy}+e^{-i\frac{%
2\pi }{3}j}e^{-i\frac{\pi }{3}}\frac{1}{x-iy}\right) \text{,}
\label{zm2asymp}
\end{equation}%
where $x=\sqrt{3}la/2$, $y=ja$, and $r=\sqrt{x^{2}+y^{2}}\rightarrow \infty $%
. For instance, on the right half-plane, the first solution

\begin{equation}
\phi _{l,j}^{\left( 1\right) }=\int_{-\frac{2\pi }{3a}}^{\frac{2\pi }{3a}}%
\frac{dk^{\prime }}{2\pi }\left( -2\cos \frac{k^{\prime }a}{2}\right)
^{-l+1}e^{ik^{\prime }ja}\frac{e^{-i\frac{k^{\prime }a}{2}}}{\left( 2\cos 
\frac{k^{\prime }a}{2}\right) ^{2}}
\end{equation}%
is dominated by momenta near $k^{\prime }a=\pm 2\pi /3$ when\ $l\gg 1$:

\begin{eqnarray}
\phi _{l,j}^{\left( 1\right) } &\approx &\left( -1\right) ^{l+1}e^{i\frac{%
2\pi }{3}j}e^{-i\frac{\pi }{3}}\int_{\frac{2\pi }{3a}-\tilde{\Lambda} }^{\frac{2\pi 
}{3a}}\frac{dk^{\prime }}{2\pi }e^{\left[ \frac{\sqrt{3}}{2}\left(
l-1\right) +ij\right] \left( k^{\prime }a-\frac{2\pi }{3}\right) }  \notag \\
&&+\left( -1\right) ^{l+1}e^{-i\frac{2\pi }{3}j}e^{i\frac{\pi }{3}}\int_{-%
\frac{2\pi }{3a}}^{-\frac{2\pi }{3a}+\tilde{\Lambda} }\frac{dk^{\prime }}{2\pi }e^{%
\left[ -\frac{\sqrt{3}}{2}\left( l-1\right) +ij\right] \left( k^{\prime }a+%
\frac{2\pi }{3}\right) }
\end{eqnarray}%
where $\tilde{\Lambda} $ is a momentum cutoff of $O\left( 1/a\right) $. Performing
the integrals and taking the $\tilde{\Lambda} \rightarrow \infty $ limit, we
promptly obtain Eq.~(\ref{zm1asymp}). Eqs.~(\ref{zm1asymp}) and (\ref%
{zm2asymp}) suggest that the linear combinations $e^{-i\frac{\pi }{3}}\phi
^{\left( 1\right) }-e^{i\frac{\pi }{3}}\phi ^{\left( 2\right) }$ and $e^{i%
\frac{\pi }{3}}\phi ^{\left( 1\right) }-e^{-i\frac{\pi }{3}}\phi ^{\left(
2\right) }$ are eigenstates of the $C_{3}$ rotation, which is indeed the
case: the former has helicity 1 and the latter has helicity \={1}.

Since $\phi ^{\left( 1\right) }$ and $\phi ^{\left( 2\right) }$ are linearly
independent, we are unable to construct a normalizable zero mode whose wave
function drops to zero faster than $1/r$ as $r=\sqrt{x^{2}+y^{2}}\rightarrow
\infty $. However, if we consider removing even more sites from the graphene
lattice, more than two zero modes may be allowed. (The simplest example of
removing a site together with its three nearest neighbors and six next
nearest neighbors produces $\left\vert 1+6-3\right\vert =4$ zero modes.) In
such a situation, at most two zero modes decaying as $1/r$ are linearly
independent, which we can choose as $1/\left( x+iy\right) $ and $1/\left(
x-iy\right) $; any other zero mode can be combined with these two
non-normalizable modes to yield a wave function that decays faster than $1/r$%
. In other words, at most two conduction channels have a logarithmically
divergent LDOS.

We conclude this appendix by explaining why the zero modes cease to exist in
the \textquotedblleft strong-coupling lattice\textquotedblright . The
removal of $c_{1}$ and $c_{\bar{1}}$\ from the original lattice amounts to
the condition that the corresponding wave functions vanish. We can directly
calculate these wave functions for the two solutions:

\begin{equation}
c_{1}^{\left( 1\right) }=c_{\bar{1}}^{\left( 2\right) }=-\frac{e^{i\frac{%
2\pi }{3}}}{\sqrt{6}a},c_{\bar{1}}^{\left( 1\right) }=c_{1}^{\left( 2\right)
}=-\frac{e^{-i\frac{2\pi }{3}}}{\sqrt{6}a}\text{.}
\end{equation}%
It is easy to verify that $c_{\bar{h}}$ vanishes for the helicity-$h$ zero
mode $e^{-ih\frac{\pi }{3}}\phi ^{\left( 1\right) }-e^{ih\frac{\pi }{3}}\phi
^{\left( 2\right) }$. However, $c_{h}$ does not vanish for the helicity-$h$
solution, which means the electronic states $c_{1}$ and $c_{\bar{1}}$ cannot
be projected out without removing both zero modes.

\section{RKKY interaction in graphene with a 4-site vacancy\label{sec:appRKKY}}

In this appendix we calculate the RKKY interaction between magnetic
impurities in the Kondo model Eq.~(\ref{3I3CK}) to the second order in the Kondo couplings. We
show that the RKKY interaction at low temperatures is dominated by the
helicity $1$ and $\bar{1}$ channels, and remains ferromagnetic despite the
presence of the vacancy.

Following Ref.~\onlinecite{PhysRevB.84.115119}, to the second order in Kondo
couplings $J_{hh^{\prime }}$, we can write the RKKY interaction between $%
b_{1}$ and $b_{2}$ as

\begin{eqnarray}
H_{RKKY,12} &=&-\left[ J_{00}^{2}\chi _{00}+J_{11}^{2}\left( \chi _{11}+\chi
_{\bar{1}\bar{1}}\right) +J_{1\bar{1}}^{2}\left( e^{i\frac{2\pi }{3}}\chi _{1%
\bar{1}}+e^{-i\frac{2\pi }{3}}\chi _{\bar{1}1}\right) \right.  \notag \\
&&\left. +J_{01}^{2}\left( e^{i\frac{2\pi }{3}}\chi _{01}+e^{i\frac{2\pi }{3}%
}\chi _{\bar{1}0}+e^{-i\frac{2\pi }{3}}\chi _{0\bar{1}}+e^{-i\frac{2\pi }{3}%
}\chi _{10}\right) \right] \bm{S}_{1}\cdot \bm{S}_{2}\text{,}  \label{RKKY12}
\end{eqnarray}%
where the (isothermal) static spin susceptibilities $\chi _{hh^{\prime }}$\
are evaluated using Wick's theorem for the non-interacting Hamiltonian with
a 4-site vacancy $H_{\text{Vac}}$,

\begin{equation}
\chi _{hh^{\prime }}\equiv -\frac{1}{4}\int_{0}^{\beta }d\tau
G_{hh}^{c}\left( \tau \right) G_{h^{\prime }h^{\prime }}^{c}\left( -\tau
\right) \text{.}
\end{equation}%
The factor of $1/4$ comes from spin degrees of freedom, and $\beta =1/T$.
The imaginary time Green's function $G^{c}$ is defined by

\begin{equation}
G_{hh^{\prime }}^{c}\left( \tau \right) \equiv -\left\langle T_{\tau
}c_{h}\left( \tau \right) c_{h^{\prime }}^{\dag }\left( 0\right)
\right\rangle \text{.}
\end{equation}%
$G^{c}$ is diagonal in the helicity index, and may be expressed as linear
combinations of the real space Green's function $G_{aa}\left( \vec{R},\vec{R}%
^{\prime },\tau \right) \equiv -\left\langle T_{\tau }a\left( \vec{R},\tau
\right) a^{\dag }\left( \vec{R}^{\prime },0\right) \right\rangle $.

We proceed to find $G_{aa}$ by solving its equation of motion (coupled with
that of $G_{ba}\left( \tau \right) \equiv -\left\langle T_{\tau }b\left(
\tau \right) a^{\dag }\left( 0\right) \right\rangle $) in momentum space\cite%
{PhysRevB.79.155442,NewJPhys.14.083004}. The result is

\begin{subequations}
\label{GaaGbb}
\begin{align}
& G_{aa}\left( \vec{R},\vec{R}^{\prime },i\omega _{n}\right) =\frac{1}{2}%
L\left( \vec{R}-\vec{R}^{\prime },i\omega _{n}\right) -\frac{1}{2}\frac{1}{%
L\left( i\omega _{n}\right) -L\left( \vec{a}_{1},i\omega _{n}\right) }%
\left\{ \tilde{L}\left( -\vec{R},i\omega _{n}\right) \tilde{L}\left( -\vec{R}%
^{\prime },i\omega _{n}\right) \right.  \notag \\
& +\tilde{L}\left( -\vec{R}-\vec{a}_{1},i\omega _{n}\right) \tilde{L}\left( -%
\vec{R}^{\prime }-\vec{a}_{1},i\omega _{n}\right) +\tilde{L}\left( -\vec{R}-%
\vec{a}_{2},i\omega _{n}\right) \tilde{L}\left( -\vec{R}^{\prime }-\vec{a}%
_{2},i\omega _{n}\right) -\frac{1}{t^{2}}\left( i\omega _{n}\right)
^{2}L\left( \vec{R},i\omega _{n}\right)  \notag \\
& \left. \times L\left( \vec{R}^{\prime },i\omega _{n}\right) \frac{L\left( 
\vec{a}_{1},i\omega _{n}\right) }{L\left( i\omega _{n}\right) +2L\left( \vec{%
a}_{1},i\omega _{n}\right) }\right\} \text{,}
\end{align}%
where the fermionic Matsubara frequency $i\omega _{n}=\left( 2n+1\right) \pi
/\beta $. With the help of Eq.~(\ref{L1L0}) and the identity

\end{subequations}
\begin{equation}
\frac{z^{2}}{t^{2}}L\left( \vec{a}_{1},z\right) =\left[ L\left( z\right)
+5L\left( \vec{a}_{1},z\right) +2L\left( \vec{a}_{1}+\vec{a}_{2},z\right)
+L\left( 2\vec{a}_{1},z\right) \right] \text{,}
\end{equation}%
we find $G_{hh}^{c}$ in particularly simple forms:

\begin{equation}
G_{00}^{c}\left( \omega ^{+}\right) =\frac{\omega ^{+}}{2t^{2}}-\frac{3}{2}%
\frac{L\left( \omega ^{+}\right) }{-2+\omega ^{+}L\left( \omega ^{+}\right) }%
\approx \frac{\omega }{3t^{2}}-\frac{\sqrt{3}}{2\pi }\frac{\omega }{t^{2}}%
\left( \ln \frac{\Lambda ^{2}}{\omega ^{2}}+i\pi \operatorname{sgn}\omega
\right) \text{,}
\end{equation}

\begin{equation}
G_{11}^{c}\left( \omega ^{+}\right) =G_{\bar{1}\bar{1}}^{c}\left( \omega
^{+}\right) =\frac{\omega ^{+}}{2t^{2}}+\frac{6}{-2\omega ^{+}+\left[ \left(
\omega ^{+}\right) ^{2}-9t^{2}\right] L\left( \omega ^{+}\right) }\approx 
\frac{\pi }{\sqrt{3}}\frac{1}{\omega \left( \ln \frac{\Lambda ^{2}}{%
\omega ^{2}}+i\pi \operatorname{sgn}\omega \right) }\text{.}
\end{equation}%
The low-energy expressions of these Green's functions can also be found from
Eqs.~(\ref{c0tilde}), (\ref{c1tilde}) and (\ref{c1bartilde}). One can show,
term by term, that $G_{hh}^{c}\left( z\right) $ is analytic everywhere
except on the real axis.

We are ready to compute $\chi _{hh^{\prime }}$:

\begin{align}
\chi _{hh^{\prime }}& =-\frac{1}{4\beta }\sum_{i\omega _{n}}G_{hh}^{c}\left(
i\omega _{n}\right) G_{h^{\prime }h^{\prime }}^{c}\left( i\omega _{n}\right)
=\frac{1}{4}\int_{-\infty }^{\infty }\frac{d\omega }{2\pi i}n_{F}\left(
\omega \right) \left[ G_{hh}^{c}\left( \omega ^{+}\right) G_{h^{\prime
}h^{\prime }}^{c}\left( \omega ^{+}\right) -G_{hh}^{c}\left( \omega
^{-}\right) G_{h^{\prime }h^{\prime }}^{c}\left( \omega ^{-}\right) \right] 
\notag \\
& =\frac{1}{2}\int_{-\infty }^{\infty }\frac{d\omega }{2\pi }n_{F}\left(
\omega \right) \left[ \operatorname{Im}G_{hh}^{c}\left( \omega ^{+}\right) \operatorname{Re}%
G_{h^{\prime }h^{\prime }}^{c}\left( \omega ^{+}\right) +\operatorname{Re}%
G_{hh}^{c}\left( \omega ^{+}\right) \operatorname{Im}G_{h^{\prime }h^{\prime
}}^{c}\left( \omega ^{+}\right) \right] \text{,}
\end{align}%
where $n_{F}\left( \omega \right) =1/\left( e^{\beta \omega }+1\right) $, $%
\omega ^{\pm }\equiv \omega \pm i0$, and we have deformed the contour of
integration into two straight lines $\operatorname{Im}z=\pm 0^{+}$.

While $\chi _{00}$ and $\chi _{01}=\chi _{0\bar{1}}=\chi _{10}=\chi _{\bar{1}%
0}$ are finite, it turns out that $\chi _{11}=\chi _{1\bar{1}}=\chi _{\bar{1}%
1}=\chi _{\bar{1}\bar{1}}$ is divergent for temperatures $T\ll \Lambda $%
:

\begin{eqnarray}
\chi _{11} &\sim &\frac{\pi ^{2}}{3}\int_{-\Lambda }^{\Lambda }%
\frac{d\omega }{2\pi }n_{F}\left( \omega \right) \frac{-\pi \operatorname{sgn}\omega
\ln \frac{\Lambda ^{2}}{\omega ^{2}}}{\omega ^{2}\left( \ln ^{2}%
\frac{\Lambda ^{2}}{\omega ^{2}}+\pi ^{2}\right) ^{2}}  \notag \\
&=&\frac{\pi ^{2}}{6}\int_{-\Lambda }^{\Lambda }d\omega \left[
n_{F}\left( \omega \right) -\frac{1}{2}\right] \frac{-\operatorname{sgn}\omega \ln 
\frac{\Lambda ^{2}}{\omega ^{2}}}{\omega ^{2}\left( \ln ^{2}\frac{%
\Lambda ^{2}}{\omega ^{2}}+\pi ^{2}\right) ^{2}}  \notag \\
&\sim &\frac{\pi ^{2}}{3}\frac{1}{T\ln ^{3}\frac{\Lambda ^{2}}{T^{2}%
}}\text{.}
\end{eqnarray}%
Inserting this into Eq.~(\ref{RKKY12}), we find that at $T_{K}\ll T\ll
\Lambda $, the RKKY interaction can be approximated as

\begin{equation}
H_{RKKY,12}\sim \left[ J_{1\bar{1}}^{2}\left( T\right) -2J_{11}^{2}\left(
T\right) \right] \frac{\pi ^{2}}{3}\frac{1}{T\ln ^{3}\frac{\Lambda
^{2}}{T^{2}}}\bm{S}_{1}\cdot \bm{S}_{2}\text{,}
\end{equation}%
where $J_{hh^{\prime }}\left( T\right) $ are the renormalized Kondo
couplings at energy scale $T$.

\bibliography{KEG}

\end{document}